\pdfoutput=1
\documentclass{article}
\usepackage{arxiv}
\usepackage[T1]{fontenc}
\usepackage[utf8]{inputenc}
\usepackage{microtype}
\usepackage{amsmath,amssymb,graphicx,booktabs,array,tabularx,longtable,pdflscape,placeins}
\usepackage[round,authoryear]{natbib}
\usepackage[compatibility=false,hypcap=false]{caption}
\usepackage{xcolor,xurl}
\usepackage[colorlinks=true,allcolors=blue]{hyperref}
\graphicspath{{figures/}}


\newcommand{\tpms}{triply periodic minimal surface}
\newcommand{\rtol}{\log_{10}(1.02)}
\newcommand{\centerentry}[1]{\raisebox{-0.5\height}{#1}}
\newcommand{\tablestart}{\par\ifdim\dimexpr\pagegoal-\pagetotal\relax<0.23\textheight\newpage\fi}
\newcolumntype{L}[1]{>{\fontsize{7}{8.5}\selectfont\raggedright\arraybackslash}p{#1}}
\newcolumntype{C}[1]{>{\fontsize{7}{8.5}\selectfont\centering\arraybackslash}p{#1}}
\providecommand{\bibcommenthead}{}
\title{Separating learning from search and feedback in budgeted inverse design of periodic unit cells}
\renewcommand{\shorttitle}{Architecture selection and sizing}
\hypersetup{pdftitle={Separating learning from search and feedback in budgeted inverse design of periodic unit cells},pdfauthor={Shaoliang Yang and Jun Wang}}
\date{}
\author{Shaoliang Yang \\
  Santa Clara University\\
  Santa Clara, CA 95053, USA\\
  \texttt{syang11@scu.edu}
  \And
  Jun Wang\thanks{Corresponding author.} \\
  Santa Clara University\\
  Santa Clara, CA 95053, USA\\
  \texttt{jwang22@scu.edu}
}
\begin{document}
\maketitle
\begin{abstract}
To design a periodic material, engineers choose the shape and dimensions of a repeating cell to meet target properties. With few simulations, the benefit of a learned predictor can be hard to separate from the effect of which designs the search tries and how it uses simulator feedback. We compare two ways to predict properties in a library of 80 cell architectures, with up to 30 simulator calls per target. For each architecture, both searches use a table of predicted properties at different dimensions. One fills the table with predictions from neural networks. The other fits a simple model to the existing simulations for that architecture. Both searches use the difference between the predicted and simulated properties to adjust the next design. We generated 200 targets from simulated designs after fixing the analysis plan. The two searches showed no detectable difference in how often they matched all three properties within 2\% or in the mean error of their returned designs. The upper limit of the 95\% confidence interval for the gain in overall success rate was 4.5 percentage points. Follow-up controls on these targets changed success from 83.5\% to 36.0\% with a degraded table and 96.5\% with an oracle table. In separate tests, using the simulation result to adjust the next design increased success for both types of table. Results cover one library and a coarse \(32^3\) voxel simulator. Designers can use this comparison to judge what a learned predictor adds when the search, feedback and budget are the same.
\end{abstract}

\keywords{Inverse design \and Architected materials \and Periodic homogenisation \and Response tables \and Simulator feedback}

\section{Introduction}\label{sec:intro}

To design a periodic material for specified stiffness and conductivity, a designer chooses the geometry of its repeating unit cell and sets its dimensions. Inverse homogenisation uses cell simulations to work from desired material properties back to geometry \citep{sigmund1994}. With a library of candidate architectures and a limited budget, every simulation spent sizing one architecture leaves fewer calls to test others. A stored example may be far from the target even when changing its density or shape would give a good match.

Precomputed microstructure families give designers a way to interpolate from stored geometries \citep{schumacher2015}. Learned models can also predict a geometry for prescribed properties or generate designs across families \citep{bastek2022,zheng2023}. Both routes leave decisions for the search. It has to choose which architectures to simulate, where to start within each one and how to use the results of earlier calls.

We ask whether learning adds to a search that already makes these decisions with the aid of simple catalogue fits. A response table predicts the properties for a range of dimensions of one architecture. The search reads each table to find the dimensions that come closest to the target. After a simulation, it can adjust the predictions using the difference between the table and the observed properties, which we call the residual. We measure the effect of the full search by comparing it with local search from retrieved examples. To measure what learning adds, we replace the learned tables with catalogue fits and keep the search rules fixed.

Our benchmark uses one library of solid--void cells and a common catalogue of simulated examples. The library includes cells defined by periodic fields, strut lattices and blends of two periodic fields. The targets specify two measures of stiffness and one of conductivity. The main target sets come from simulated reference designs. We choose targets that none of the stored designs can meet at the chosen tolerance so the benchmark tests whether a search can improve on retrieval. The searches see the target properties and can return any architecture in the library. A common voxel simulator verifies the returned designs.

We address two research questions.
\begin{enumerate}
\item Under a fixed call budget, what do learned tables and simulator feedback each contribute to search performance?
\item How do the choice and starting dimensions of architectures affect the search, and how does table-guided search compare with bracketing, Bayesian optimisation and Nelder--Mead search (Sect.~\ref{sec:comparators})?
\end{enumerate}
The benchmark and matched comparisons give designers a way to decide whether to invest in a learned predictor or improve how they spend their simulation budget. The confirmatory comparison is the primary evidence about learning (Sect.~\ref{sec:learning}). Other tests examine feedback, the choice of search and the effects of changing the targets, tolerance or verification grid.

\section{Related work}\label{sec:related}

\subsection{Updating predictions with simulations}

Approximation-model management uses agreement between predicted and simulated improvement to decide where to trust an inexpensive model \citep{alexandrov1998}. Space mapping adjusts the inputs of the inexpensive model to bring its predictions closer to those of the detailed model \citep{bandler2004}. Engineering design uses metamodels to approximate expensive simulations \citep{viana2014}. Multifidelity methods combine predictions from models with different costs and levels of detail \citep{kennedy2000,peherstorfer2018}. These approaches motivate the local correction in Eq.~\ref{eq:residual}. Here the correction uses observations from a single simulator under a fixed allowance. The rounded voxel responses limit the applicability of smooth trust-region convergence results.

\subsection{Representing a library of unit cells}

Precomputed families let designers interpolate between microstructures to select compatible geometries \citep{schumacher2015}. Learning offers other ways to represent a library, for example by pooling information across microstructure classes \citep{wang2021}, predicting properties of cells that combine periodic surfaces \citep{chen2024}, or enforcing physical bounds in homogenisation \citep{keshav2026}. Our comparison holds the library fixed and tests the added contribution of learning when catalogue fits already support architecture ranking and sizing.

Other learned methods predict geometry directly or generate candidates in a continuous representation \citep{bastek2022,zheng2023,wang2020,regenwetter2022}. Separately, the way the catalogue is assembled can affect a design search \citep{lee2024}. Diversity-based selection and task-aware sampling provide ways to improve that coverage \citep{chan2021,lee2023}. Here both searches use the same catalogue so that table construction is the factor under test.

\subsection{Spending simulations to meet a specification}

Bayesian optimisation can use expected improvement to decide where to evaluate a design \citep{jones1998}. Related work uses predictions to shortlist candidates for detailed verification \citep{danesh2026} or to find designs likely to satisfy a property window \citep{jiang2026}. The utility of a surrogate also depends on the data and computation it needs \citep{alizadeh2020}. Our benchmark evaluates whether a design satisfies all target tolerances within the call allowance \citep{guo2024} and records the number of calls separately from the final error.

Fixed-budget ranking and selection allocates samples among competing alternatives \citep{chen2000,carpentier2016}. Here a call also supplies a property measurement that can guide the sizing of the same architecture. Work on acquiring information for engineering design \citep{panchal2009,chaudhari2020} and materials discovery \citep{lookman2019} motivates measuring what that observation contributes to later decisions. Shared tasks and baselines aid interpretation \citep{trabucco2022}. We use the same verification and accounting rules for all searches, and matched component comparisons to isolate learning and refinement.

\section{Design problem and benchmark}\label{sec:benchmark}

\subsection{Architectures and simulation}\label{sec:geometry}

\begin{figure}[!tbp]
\centering
\includegraphics[width=\textwidth]{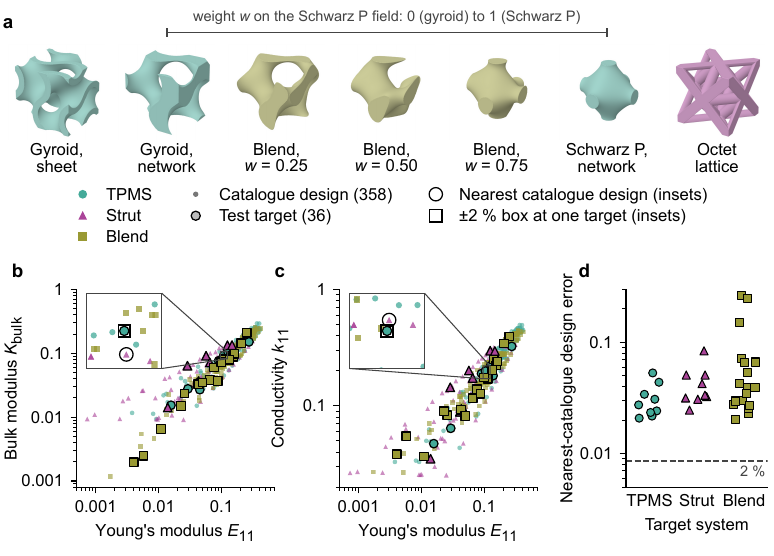}
\caption{Design space and target difficulty. \textbf{a} Gyroid sheet and network, three gyroid--Schwarz P network blends, Schwarz P network and octet lattice, drawn as smooth, closed isosurfaces of the underlying geometry evaluated on a \(128^3\) grid at nominal density 0.30. The gyroid sheet drawing omits two small corner pieces (0.3\% of its surface) connected to the body only through periodic cell faces; no other drawn cell has such pieces. Simulations use the \(32^3\) voxel model; after connected-component filtering its realised density is 0.300 for the level-set examples and 0.354 for the octet. The blend weight on Schwarz P increases from 0.25 to 0.75 between the pure limits. \textbf{b,c} The 358 catalogue designs and 36 test targets in two response planes. Young's modulus \(E_{11}\), Voigt bulk modulus \(K_{\mathrm{bulk}}\) and conductivity \(k_{11}\) are dimensionless relative to the solid. Insets show TPMS target 2 of Fig.~\ref{fig:target}, its 2\% box and its ringed nearest catalogue design. This target's retrieval error is closest to the median across targets. \textbf{d} Nearest-catalogue design error for every target: the largest \(\lvert\log_{10}(\mathrm{simulated}/\mathrm{target})\rvert\) of the three responses, in log units, with horizontal jitter within each system and the tolerance dashed. No catalogue design meets any target. TPMS denotes triply periodic minimal surface}
\label{fig:space}
\end{figure}

A discrete \emph{architecture} is denoted by \(c\). A design specifies the architecture and its continuous parameters \(\theta\), recorded to three decimal places. Table~\ref{tab:space} lists the architectures and parameter bounds. Figure~\ref{fig:space}a shows example unit cells from the three systems. In the \tpms{} (TPMS) system, a network thresholds a single level-set field and a sheet thresholds its absolute value. A blend applies either construction to the field \((1-w)f_a+w f_b\), where \(f_a\) and \(f_b\) are periodic fields scaled to unit standard deviation on the voxel grid. Each unordered pair has a fixed ordering and \(w\) is the weight on the second field. For a strut architecture, the field is the distance to the nearest periodically repeated member axis. For all three systems, the density parameter \(\rho\) sets the threshold through the rank of the sampled field values. The realised solid fraction can differ from \(\rho\) because of voxelisation, ties and removal of isolated material.

\begin{table}[tbp]
\centering
\small
\caption{Discrete architectures and continuous parameter bounds. TPMS denotes triply periodic minimal surface}
\label{tab:space}
\begin{tabular}{@{}llccl@{}}
\toprule
System & Discrete choice & Architectures & Density \(\rho\) & Blend weight \(w\) \\
\midrule
TPMS & 8 families, network or sheet & 16 & \(0.08\)--\(0.60\) & \\
Strut & 8 periodic strut topologies & 8 & \(0.06\)--\(0.50\) & \\
Blend & 28 family pairs, network or sheet & 56 & \(0.08\)--\(0.60\) & \(0.05\)--\(0.95\) \\
\bottomrule
\end{tabular}
\end{table}

The simulator estimates stiffness and conductivity on a \(32^3\) voxel grid. The solid has unit Young's modulus and conductivity and Poisson's ratio 0.3. Let \(C\), \(S=C^{-1}\) and \(k\) denote the effective stiffness, compliance and conductivity tensors, with \(C\) and \(S\) in Voigt notation, which writes each elasticity tensor as a matrix. The reported responses are \(E_{11}=1/S_{11}\), the Voigt bulk modulus \(K_{\mathrm{bulk}}=\sum_{i,j=1}^3 C_{ij}/9\), and axial conductivity \(k_{11}\). The moduli are in units of the solid Young's modulus and the conductivity is in units of the solid conductivity.

The solver checks that the design has a continuous load and heat path and that its simulated responses are valid. \hyperref[sec:supplement]{The Supplementary Information} specifies the geometry, solver and validity tests. The same simulator evaluates the catalogue, generates the principal target sets and verifies each returned design.

\subsection{Catalogue and target sets}

The shared offline catalogue contains 358 valid designs (128 TPMS, 80 strut and 150 blend). Candidate pure architectures were sampled at ten densities over each system's range. Candidate blends were sampled at up to four density and weight combinations per pair and mode. Duplicate and invalid designs were removed. The resulting blend architectures have one to four examples each.

To generate the targets, the simulator evaluated designs sampled independently of the catalogue. A candidate reference had to be valid and its response had to lie at least 0.02 in the maximum absolute log-ratio distance from every catalogue response. The target therefore has a feasible reference under the simulator. The gap of 0.02 in log-ratio distance exceeds the 2\% tolerance of about 0.00860 (Sect.~\ref{sec:scoring}), so catalogue retrieval alone fails. Figure~\ref{fig:space}b--d shows the catalogue, the test targets and each target's nearest-catalogue error. Retrieval alone thus scores zero by construction of the target filter. For a pure-architecture target (TPMS or strut), a search can match the reference by identifying its architecture and sizing its density parameter. Other architectures may also reach the target. The development split has 12 targets (3 TPMS, 3 strut and 6 blend), and the test split has 36 (9 TPMS, 9 strut and 18 blend). They share no reference design. The reference geometry is hidden from the search. The reference system is used for reporting, and a design from any system can meet a target.

Table~\ref{tab:target-sets} identifies which comparisons use each set. The planned learned-versus-catalogue comparison and the follow-up controls use the same 200 targets. Three further sets each have the test set's system counts. The \emph{fresh draw} uses a different reference-generator seed, the \emph{interpolated responses} combine two reference responses in log space, and the \emph{\(64^3\) responses} solve the fresh reference designs on that grid. The test set and fresh draw are reachable on the \(32^3\) search grid. Reachability on that grid is unknown for interpolated and \(64^3\) responses. A separate \emph{confirmatory set} contains 200 reachable targets from another generator seed (50 TPMS, 50 strut and 100 blend).

\begin{table}[!htbp]
\centering\small
\caption{Target sets and their role in the evaluation}
\label{tab:target-sets}
\begin{tabularx}{\textwidth}{@{}lclX@{}}
\toprule
Set & Targets & Reachable at \(32^3\) & Role \\
\midrule
Development & 12 & Yes & Select the learned configuration \\
Confirmatory & 200 & Yes & Planned table comparison and separate follow-up controls \\
Test & 36 & Yes & Benchmark search against retrieval and standard optimisers \\
Fresh draw & 36 & Yes & Repeat the benchmark on another draw and check resolution \\
Interpolated and \(64^3\) & 36 each & Unknown & Test performance when reachability is uncertain \\
\bottomrule
\end{tabularx}
\end{table}

\subsection{Scoring and call budget}\label{sec:scoring}

For target properties \(p_j^\star>0\) and a valid simulated response \(p_j(c,\theta)\), define the design error
\begin{equation}
  e(c,\theta)=\min\!\left\{1,\
    \max_{j=1,2,3}\left|\log_{10}\frac{p_j(c,\theta)}{p_j^\star}\right|\right\}.
  \label{eq:error}
\end{equation}
Design error measures the largest absolute log-ratio between a simulated property and its target, with a cap of 1. A returned design that fails the validity checks scores 1. A run succeeds if the error of its returned design is at most \(\tau=\rtol\approx 0.00860\). Each property ratio must then lie in \([1/1.02,1.02]\), the rule referred to here as the 2\% tolerance. A success is thus a match to the \(32^3\) voxel model that also generated the targets. It is not a statement about a physical property (Sect.~\ref{sec:target-sets}). We report success rate, mean error of the returned design, mean calls used and mean calls to success. Calls to success is the index of the first successful query and is set to 31 for a run with no such query, so its mean includes failures. The confirmatory decision uses success rate and design error, overall and on blends.

A call is charged per design in a batch, including invalid designs. Repeated queries are charged. The methods have unlimited access to the same catalogue and can use up to 30 simulator calls per target. The evaluation protocol independently re-solves the final returned design. This verification is separate from the search allowance and is applied to the retrieval control as well.

Each call can test another architecture or refine one that already has a valid observation. The searches use the same information and budget. They differ in how they rank and size architectures, how they adjust proposals after observing a response, and how many architectures they can test. \hyperref[sec:supplement]{The Supplementary Information} states the allocation problem formally.

The voxel simulator stands in for an expensive solver, and the call budget is imposed to study allocation. Direct timing on the confirmatory target generation gives 0.179 seconds per design in batches over 871 designs and a median of 0.75 seconds for a single design over 12 solves.

\section{Table inversion with simulator feedback}\label{sec:method}

\subsection{Precomputed response tables}\label{sec:tables}

The design catalogue stores simulated properties for existing designs. Each response table predicts three log properties at candidate dimensions within one architecture. The learned tables average the predictions of three feed-forward neural networks trained on the catalogue. The networks represent the same periodic field consistently whether it appears alone or in a blend.

The catalogue fits use the stored designs of each architecture. The pure-architecture fit interpolates piecewise linearly in log density. The blend fit estimates a plane over log density and blend weight. We call the resulting searches learned-table search and catalogue-table search. Both build their response tables before the first query and use the same grids, refinement rules and call budget. \hyperref[sec:supplement]{The Supplementary Information} gives the network settings, grids and fitting rules.

\subsection{Ranking and refinement}\label{sec:refine}

The search scores architectures with table predictions before testing them. Once an architecture has a valid observation, the search uses the observed mismatch to guide refinement within that architecture. Figure~\ref{fig:workflow}a--c shows the first batch, and panels d and e follow the later calls and verification.

\emph{First batch.} Let \(y^\star=(\log_{10}p_1^\star,\log_{10}p_2^\star,\log_{10}p_3^\star)\). For an architecture with no valid observation for the current target, the first proposal minimises the predicted error on its available grid \(\mathcal{G}_c\),
\begin{equation}
  m_c=\min_{\theta\in\mathcal{G}_c} \left\|T_c(\theta)-y^\star\right\|_\infty .
  \label{eq:score}
\end{equation}
Architectures are ordered by \(m_c\), with at most six from any system in the first batch of up to ten. Thus a promising architecture can be tested even if all its individual catalogue designs are far from the target.

\emph{Local refinement.} In a queried architecture, let \(\theta_b\) be the best observed design and \(z_b\) its simulated log responses. Let \(s_c\) contain the widths of the parameter ranges in Table~\ref{tab:space} and \(u=(\theta-\theta_b)/s_c\) denote the componentwise scaled offset. The corrected table is
\begin{equation}
  \widetilde T_c(\theta)=T_c(\theta)+r_b+G_c^\mathsf{T}u,\qquad
  r_b=z_b-T_c(\theta_b).
  \label{eq:residual}
\end{equation}
Here \(T_c(\theta)\) is interpolated linearly for a pure architecture and bilinearly for a blend. Each row of \(G_c\) gives the slope of the three residuals with respect to one scaled parameter. With one valid observation the slopes are zero. With more observations, we fit how the residual changes with the scaled design parameters. The fit gives more weight to designs closer to \(\theta_b\). \hyperref[sec:supplement]{The Supplementary Information} gives the fit. The residual shifts the table to the simulated properties at \(\theta_b\), and the fitted slopes adjust that shift at nearby designs.

To choose the next refinement, the search evaluates the corrected table in a window centred at \(\theta_b\). It queries the candidate with the smallest predicted error \(\|\widetilde T_c(\theta)-y^\star\|_\infty\). The window shrinks as the search collects valid observations. The search excludes catalogue designs, previously observed designs and a neighbourhood around each invalid query. An architecture with two invalid observations and no valid observation is dropped.

\emph{Later batches.} After the first batch, the usual allocation is three refinements and two proposals in architectures without valid observations, subject to available candidates and calls. During this stage, a proposed refinement must have a predicted error below the best observed error for that architecture. An architecture also yields its place after three consecutive queries without improvement.

To improve the ranking of untested architectures in a system, the search can apply a common correction to their predictions. After it observes valid designs in at least three architectures, it checks whether a shared shift would improve predictions for that system. \hyperref[sec:supplement]{The Supplementary Information} gives the rule for this shift.

\emph{Final refinement and stopping.} Once the best observed error drops below twice the tolerance (\(2\tau\)), the search instead refines up to two architectures ordered by their observed error. Search stops after a batch containing a successful query, when no proposal remains or when the budget is exhausted. The search returns the observed or catalogue design with the lowest error. \hyperref[sec:supplement]{The Supplementary Information} states the batch composition, the orderings and the tie rules in full.

\begin{figure}[!tbp]
\centering
\includegraphics[width=\textwidth,trim=0 435.969bp 0 0,clip]{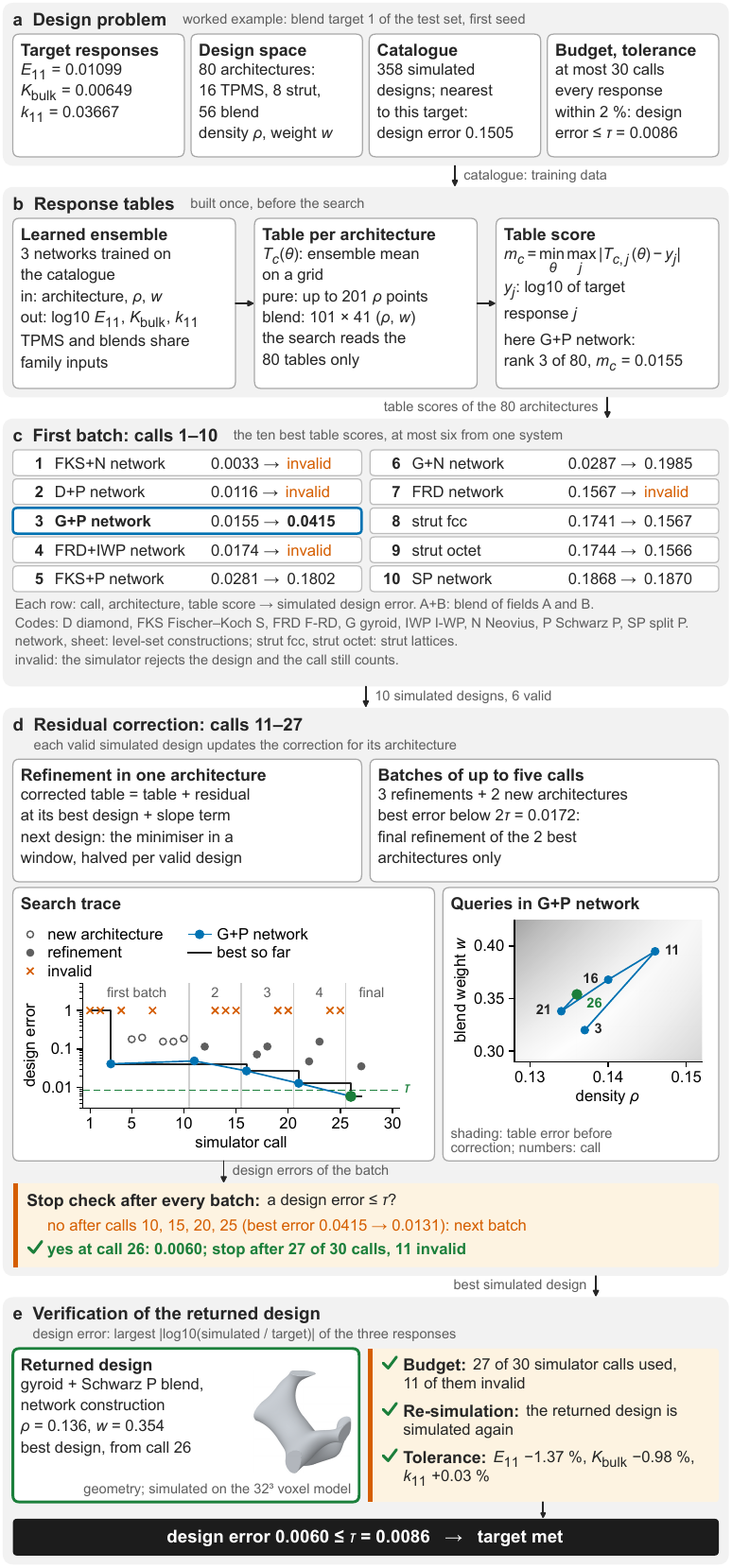}
\caption{Table inversion and residual correction for blend target 1 in Fig.~\ref{fig:target}, first seed. \textbf{a} Target, design space, catalogue and budget. \textbf{b} The ensemble supplies one response table per architecture. \textbf{c} Table scores and simulated design errors for the first batch of ten calls, with an architecture key (fcc denotes face-centred cubic). Four designs in this batch are invalid and still count against the budget. The continuation shows panels \textbf{d,e}}
\label{fig:workflow}
\end{figure}
\begin{figure}[!tbp]\ContinuedFloat
\centering
\includegraphics[width=\textwidth,trim=0 0 0 360bp,clip]{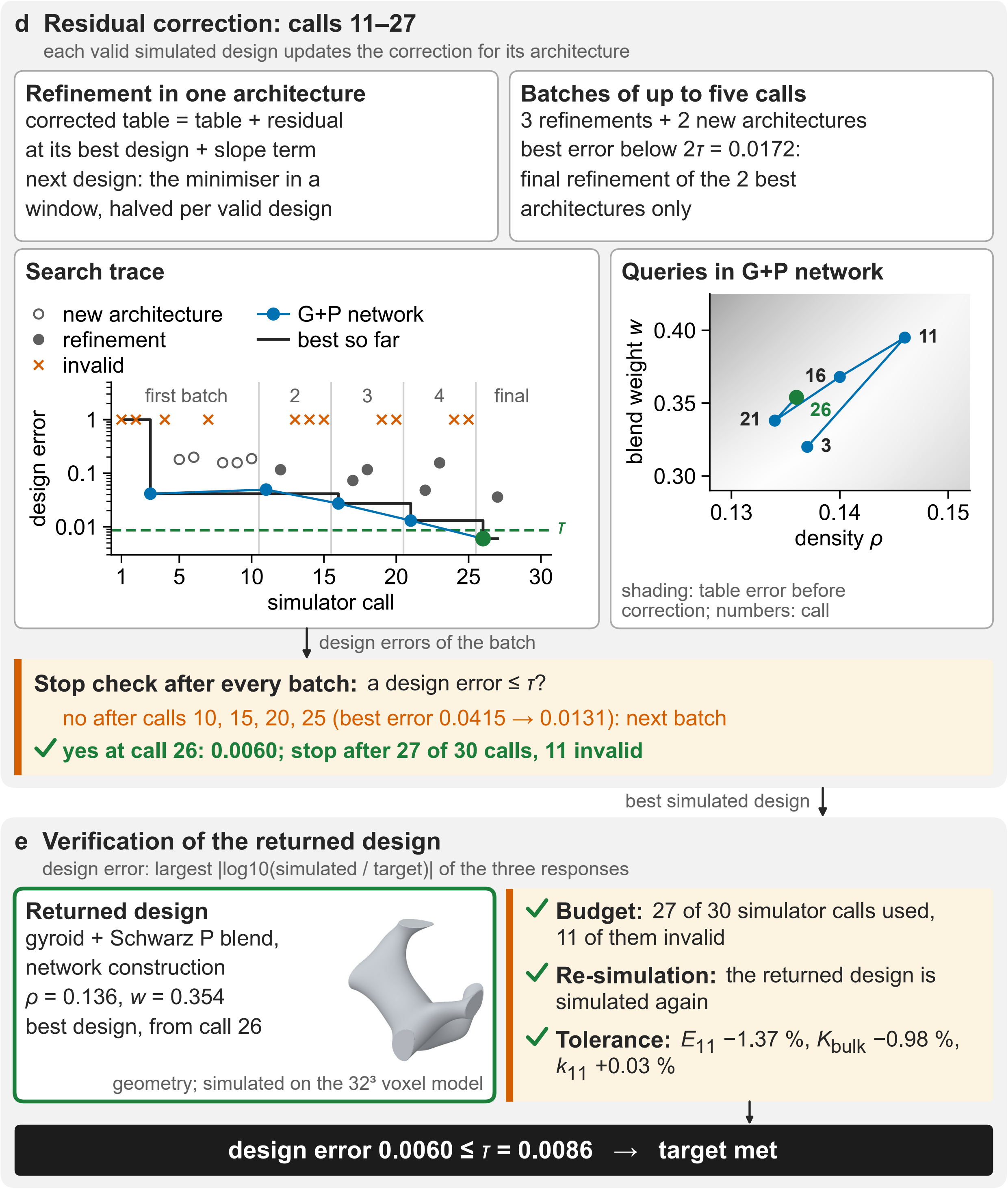}
\caption{(continued) \textbf{d} The query trace and five proposals in the gyroid--Schwarz P network. Each valid simulated design updates the correction for its architecture. Invalid queries are drawn at the error cap. Call 26 first meets the tolerance, and the last batch brings the total to 27 calls, 11 invalid. \textbf{e} The returned design is drawn as a smooth, closed isosurface of the underlying blended level-set field evaluated on a \(128^3\) grid at its nominal density. Simulations and re-simulation use the \(32^3\) voxel model of the same geometry}
\end{figure}

\subsection{Worked search trace}\label{sec:trace}

Figure~\ref{fig:workflow} follows a search for a blend cell with three specified properties, in order axial Young's modulus, Voigt bulk modulus and axial conductivity. The target values relative to the solid are approximately \((0.01099,0.00649,0.03667)\). The gyroid--Schwarz P network ranks third in the initial table ordering with a score of 0.0155. Its first simulated design has error 0.0415. Later refinements lower that error to 0.0131 at call 21 and 0.0060 at call 26. The returned design has \(\rho=0.136\) and \(w=0.354\) for the second field, Schwarz P. Re-simulation gives signed deviations of \(-1.37\%\), \(-0.98\%\) and \(+0.03\%\), respectively.

\subsection{Use of generative AI}\label{sec:ai}

Under the authors' direction, Claude Code helped implement the software and assisted in interpreting the comparisons. The inverse-design method and all benchmark and control runs made zero calls to language models.

\section{Experimental design}\label{sec:experiment}

The planned comparison isolates the effect of table construction. The other comparisons test how feedback, architecture selection and call allocation affect search. Table~\ref{tab:arms} identifies what each comparison changes and where it is evaluated.

\subsection{Confirmatory comparison}\label{sec:confirmatory-methods}

Before drawing the confirmatory targets, we fixed the target counts, generator seed (20261111), seven configurations, search seeds, primary contrast, endpoints and decision rule in a written analysis plan. The plan will be released with the study's data and code.

The learned-table search, Bayesian optimisation and retrieval with pattern search use ten seeds (0--9). The four deterministic configurations were run with two seeds, which gave identical outcomes, and count once per target.

The primary contrast subtracts the catalogue-table outcome from the learned-table outcome. It uses success rate and design error, each overall and on blends. For each endpoint the declared rule detects a difference when its paired 95\% target-bootstrap interval excludes zero. All four endpoints are reported together without multiplicity adjustment. Secondary contrasts include the standard optimisers, three-architecture catalogue search and calls. Target generation used 871 simulator calls outside the search budgets, plus one warm-up and 12 single-design solves for timing. No search receives reference designs.

\subsection{Benchmark comparisons}\label{sec:comparators}

Retrieval with pattern search ranks stored designs by their target error and picks the best catalogue design from each of the first three distinct architectures. It runs a coordinate pattern search within each chosen architecture, allotting up to ten calls to each. Initial steps are 0.04 in density and, for blends, 0.15 in weight. A sweep tests positive and negative coordinate moves, accepts the best improvement, and halves the steps if no move improves the error. The seed controls move order when the budget truncates a sweep. Retrieval alone, the retrieval control, returns the nearest stored design with no search calls. Random search draws 30 designs, choosing a system uniformly, then sampling its discrete choices and continuous parameters within their bounds. The common return and verification rules are given in Sect.~\ref{sec:scoring}.

The learned-table search uses tables from three neural networks and starts in up to ten architectures. Two candidate configurations were run on the 12 development targets. The first used a calibrated ensemble surrogate and trust-region refinement. It exceeded the preset compute-time limit, which stopped the run before a performance comparison. The second, a simplified variant of the first, is the reported method. It passed the development check below and is the only candidate run on the test targets. Neither candidate was run on the test targets before this selection.

The development check required at least two blend successes, at least five pure-system successes and mean design error at most 90\% of the error of catalogue-table search (Sect.~\ref{sec:controls}). The error criterion favoured configurations that performed better than the catalogue-table search on the development targets. Contrasts between the two on the test set can inherit this selection. The confirmatory targets were drawn after the configuration was fixed (Sect.~\ref{sec:confirmatory-methods}).

We fixed the batch schedule in Sect.~\ref{sec:refine} and the step sizes of retrieval with pattern search before the tests. Each configuration was evaluated on 36 test targets $\times$ three seeds (seeds 0, 1 and 2). The learned-table configuration was held fixed for the component comparisons. Variants and parameter sweeps evaluated on the same test targets are reported as attribution and sensitivity analyses. The additional sets assess the fixed configurations on other targets.

Three standard optimisers use the same catalogue and call allowance. These are our implementations of published algorithms. The catalogue-based starts and stopping rules described below are choices for this benchmark. Bracketing search orders architectures by catalogue-fitted table score and applies regula falsi with the Illinois update \citep{dowell1971} to density, with at most five calls per architecture. It brackets the midpoint of the largest and smallest signed log residuals and fixes a blend's weight at the table optimum. It ends refinement of an architecture when the absolute value of that midpoint is at most half the tolerance.

Nelder--Mead search \citep{nelder1965} uses the same table-ranked starts, optimises density and blend weight where present, and allots ten calls per start with an initial simplex edge of 5\% of each parameter range.

Bayesian optimisation fits a Gaussian process to predict properties with uncertainty. It updates that model using the simulated responses, then chooses the design with the largest expected improvement. It searches all architectures before refining within the chosen one. \hyperref[sec:supplement]{The Supplementary Information} gives its prediction and selection rules. Comparator settings were fixed before test evaluation. We did not tune them on the development runs.

\subsection{Component comparisons}\label{sec:controls}

The catalogue-table search keeps the search schedule and simulator feedback, replacing the learned predictions in both ranking and refinement. The categorical-encoding search replaces the shared field encoding (Sect.~\ref{sec:tables}) with an 80-entry indicator with one nonzero entry for the architecture, weight, density and log density. It uses the same hidden widths, training schedule and number of members. Two role swaps use learned tables either for ranking and starting designs or for local refinement, with catalogue-fitted tables in the other role.

The proposal control changes how the next refinement design is chosen. The corrected search minimises the predicted error from the corrected table. The uncorrected-proposal control minimises the error predicted by the original table. Both use the same rule to set the candidate window and use corrected predictions to decide which architecture to refine and when to stop refining it. They make identical initial queries. Their observations can diverge once they propose different designs.

The no-residual control uses uncorrected proposals and makes two further changes. It omits correction from architecture selection and drops the system-level shift.

We run the uncorrected-proposal control with learned tables and with catalogue-fitted tables on the test set, the fresh draw and the confirmatory set, with the targets, seeds and budget of the corresponding corrected searches. The learned-table control on the test set belongs to the original component comparisons. The other control runs were added after the confirmatory plan and are not among its declared configurations. A further control replaces the local refinement with pattern search from learned-table starts. The pattern-search control starts from the same proposals as learned-table search. We report it separately from the comparison of corrected and uncorrected table proposals.

To examine call allocation, we also compare two searches that rank architectures by table inversion, start in three architectures and open no further architectures. One uses learned tables, and the other uses catalogue-fitted tables. A control seeded with the three catalogue designs used by retrieval with pattern search uses their known properties as initial observations and applies catalogue-table residual correction with up to ten calls per start. Finally, we vary the first-batch size (3, 6 and 15 versus 10) and the ensemble size (1 and 9 versus 3).

\begin{table}[tbp]
\centering\small
\caption{Comparison arms, changes assessed and principal target sets}
\label{tab:arms}
\begin{tabularx}{\textwidth}{@{}>{\raggedright\arraybackslash}p{37mm}X>{\raggedright\arraybackslash}p{33mm}@{}}
\toprule
Comparison & Change assessed & Main evidence \\
\midrule
Learned / catalogue tables & Table construction under matched search and feedback & Planned comparison, confirmatory set \\
Corrected / uncorrected proposal & Residual use to choose the next design under matched decision rules & Test set and post-plan controls on the fresh and confirmatory sets \\
Degraded / catalogue / oracle tables & Response to larger changes in table quality & Post-plan controls, confirmatory set \\
Retrieval with and without pattern search & Charged local search within retrieved architectures & Test set \\
Catalogue-table search / retrieval with pattern search & Table-guided starts, refinement and allocation as a bundle & Test set \\
Broad / three-architecture search & Architecture allocation conditional on the table type & Test set and secondary confirmatory contrasts \\
Standard optimisers & Performance under alternative budgeted search policies & Test set and secondary confirmatory contrasts \\
\bottomrule
\end{tabularx}
\end{table}

\subsection{Paired analysis and offline diagnostic}\label{sec:paired}

The seed changes network initial weights, the move order of retrieval with pattern search, random-control draws and the Monte Carlo acquisition samples of Bayesian optimisation. Catalogue-table search, three-architecture catalogue search, bracketing search, Nelder--Mead search and retrieval control are deterministic. Their repeated seeds provide identical outcomes, and their seed-range entries are shown as a dash. For stochastic configurations we report the range of three seed-specific metric means. A difference larger than both ranges exceeds the spread observed across these seeds.

For paired comparisons, we first average over the seeds for each target. We then compare the two searches on the same targets. We report 95\% percentile bootstrap intervals \citep{efron1979}. We resample targets within each system 20,000 times, keeping all seeds of each target together. Two-sided exact sign tests use the same target-level differences and omit ties. \hyperref[sec:supplement]{The Supplementary Information} also reports Wilcoxon signed-rank tests \citep{wilcoxon1945} for design error. Intervals and tests are descriptive for these contrasts, with no adjustment for multiple comparisons.

A separate five-fold diagnostic holds out catalogue designs and ranks architectures for their simulated property vectors using tables fitted to the other folds. Recall at rank \(k\) is the fraction of held-out designs whose source architecture appears in the first \(k\) ranks. We report means and ranges over three model seeds and compare pooled and categorical encodings, catalogue-fitted tables and nearest-design retrieval. The diagnostic uses existing labels and spends no simulator calls. It measures recovery of a known source architecture on the catalogue grid.

\subsection{Additional targets and grid-resolution checks}\label{sec:additional-methods}

We evaluate the fixed configurations on three additional sets, each with 36 targets and three seeds. A fresh draw uses another seed of the same reference generator and the same catalogue-separation rule. An interpolated set uses log-space combinations of the responses of two reference designs from different architectures within one system. The \(64^3\) responses use the fresh reference geometries solved at \(64^3\), while searches continue to query the \(32^3\) simulator. Reachability on the search grid is unknown for the latter two sets. Generating these evaluation sets required 410 simulator calls at \(32^3\) and 36 at \(64^3\), outside the per-target search allowance. No search receives the reference geometries or these generation responses as training data.

For learned-table search, catalogue-table search and retrieval with pattern search, we also solve the returned test and fresh-draw designs at \(64^3\) without further search. On the test set, these responses are compared with the \(32^3\) targets because the reference geometries remain hidden. On the fresh draw, both the returned geometry and the reference geometry are evaluated at \(64^3\). Property shifts are summarised over distinct returned designs. Success counts retain each target and seed. These checks measure a change of resolution, and convergence of the finer grid is untested.

\subsection{Checks beyond the confirmatory plan}\label{sec:postplan}

Two coverage analyses were added after the plan. One splits the confirmatory blend targets by the number of catalogue designs of the reference architecture. The other repeats learned-table search (seeds 0--2) and catalogue-table search on the confirmatory targets with a thinned catalogue. This catalogue keeps every second density of each pure architecture, starting from the lowest, and all blend designs, which leaves 259 of the 358 designs. Both table types are rebuilt from it.

The other controls test how sensitive the search is to table quality and how it performs with incomplete architecture coverage. We first repeated the 2,400 original learned-table and catalogue-table runs and obtained the same outcomes. The controls keep the 30-call budget and 2\% stop rule. The degraded table uses one fit per system, shared by all its architectures. The oracle table uses extra simulator evaluations of individual architectures. The coverage check removes the catalogue designs for 14 randomly selected blend architectures. \hyperref[sec:supplement]{The Supplementary Information} gives the construction costs and the details of these controls, which we planned after seeing the confirmatory results.

On the test set and fresh draw, the seven configurations also ran at 3\% and 5\% tolerance. Each search with early stopping used the corresponding tolerance to decide when to stop. The switch to refining only the two best architectures remains at twice the task tolerance, and bracketing search keeps its signed-residual threshold at half the tolerance. All other settings stay fixed. Retrieval with pattern search has no early stop, so its results are re-scored. The paired analyses of the smaller sets are post hoc, averaging seeds within target and resampling within system and, for pooled comparisons, within set.

\section{Results}\label{sec:results}

The planned learning comparison uses the 200-target set. The proposal and table-quality controls on those targets address separate questions about feedback and prediction quality. The remaining results compare search policies and examine sensitivity to target construction, tolerance and resolution.

\subsection{Planned comparison detects no gain from learning}\label{sec:learning}

\begin{figure}[!tbp]
\centering
\includegraphics[width=\textwidth]{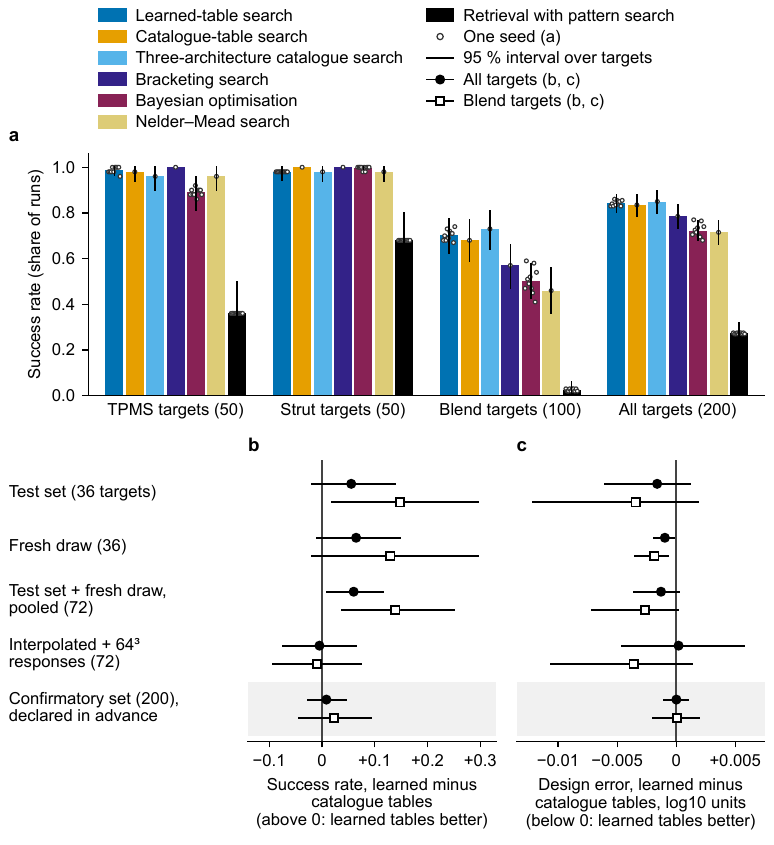}
\caption{Declared comparison and exploratory contrasts. \textbf{a} Success on 200 reachable targets (50 TPMS, 50 strut, 100 blend). Bars average seeds, dots show individual seeds and black lines give 95\% target-bootstrap intervals. Stochastic configurations use ten seeds, deterministic configurations count once. \textbf{b,c} Learned-table minus catalogue-table differences in success rate and design error. Filled circles denote all targets and open squares blends. Intervals resample targets 20,000 times within system and within set for pooled sets. The shaded confirmatory row uses the declared primary comparison. Smaller-set contrasts are post hoc and unadjusted for multiplicity. All searches use at most 30 simulator calls per target and zero language-model calls. TPMS denotes triply periodic minimal surface}
\label{fig:confirmatory}
\end{figure}

On the planned 200-target comparison, learned-table search succeeds in 1,687 of 2,000 runs with ten seeds per target. Deterministic catalogue-table search succeeds on 167 of 200 targets (Fig.~\ref{fig:confirmatory}). Search and feedback are matched. Overall success differs by 0.0085 \([-0.027,0.045]\), and blend success by 0.023 \([-0.044,0.093]\). Design-error differences are 0.00002 \([-0.00104,0.00099]\) overall and 0.00008 \([-0.00195,0.00195]\) on blends. All four declared intervals include zero under the decision rule in Sect.~\ref{sec:confirmatory-methods}.

The intervals measure variation across targets for the seeds and catalogue used in the experiment. The overall interval extends from a disadvantage of 2.7 percentage points to an advantage of 4.5. For blends, the interval extends from a disadvantage of 4.4 points to an advantage of 9.3. Smaller effects in either direction remain compatible with these data. The overall success sign test counts 18 targets better, 27 worse and 155 tied (\(p=0.23\)). This result governs the claim about the learned model.

\subsection{Correcting refinement proposals increases success}\label{sec:feedback}

The follow-up controls on the 200-target set show that corrected proposals improve success for both types of table. Learned-table search meets 1,687 of 2,000 runs with corrected proposals and 1,330 with uncorrected proposals, a difference of 0.179 \([0.142,0.217]\). Catalogue-table search meets 167 and 126 of the 200 targets, a difference of 0.205 \([0.150,0.260]\). These controls address the proposal rule and were added after the confirmatory plan.

\begin{figure}[!tbp]
\centering
\includegraphics[width=\textwidth]{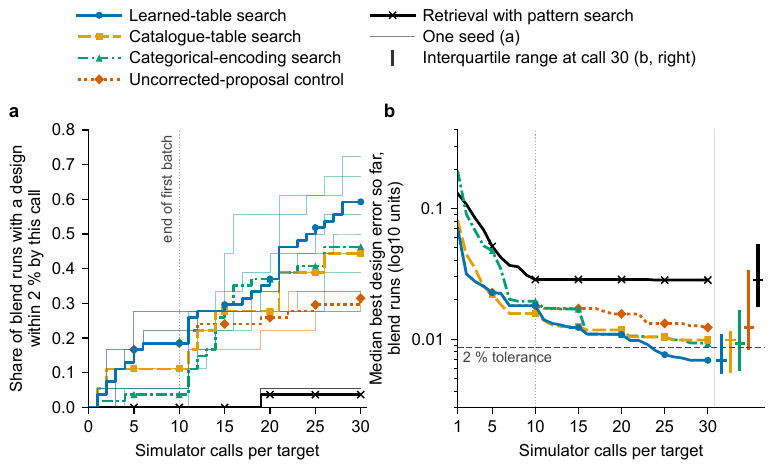}
\caption{Search progress on the 54 blend runs. \textbf{a} Share of runs met by each simulator call, with pooled-seed curves thick and single-seed curves thin. \textbf{b} Median best design error, with the last value carried forward after stopping. The vertical marker ends the first batch of ten calls. Bars to the right of call 30 give the interquartile range and median for every configuration, and the horizontal threshold is \(\tau=\log_{10}(1.02)\), the 2\% tolerance. These bars represent call 30. Line styles and markers identify configurations in greyscale. Learned-table search and the uncorrected-proposal control have identical first-batch traces}

\label{fig:diagnostic}
\end{figure}

The 36-target test set shows when the proposals start to diverge. The learned-table search and the uncorrected-proposal control make the same first ten queries for every run. On blend targets they each meet ten runs in that first batch. By the end of the search they meet 32 and 17 blend runs, respectively (Fig.~\ref{fig:diagnostic}a,b). Across all systems the corrected proposals meet 84 runs, compared with 65 under the uncorrected proposal rule. Twenty runs succeed only with correction, and one succeeds only with the uncorrected proposal. The mean error difference is \(-0.00796\) with interval \([-0.01338,-0.00355]\), and the success-rate difference is 0.176 with interval \([0.093,0.278]\). Both differences exceed the seed ranges. The control that removes all uses of the residual meets the same 65 runs as the uncorrected-proposal control. It issues the same queries as that control in 73 of the 108 runs and returns the same design in 100, so the further uses of the residual change some queries and no success outcome.

The proposal controls also improve success for catalogue fits on the test set and both table types on the fresh draw (Table~\ref{tab:correction_control} in \hyperref[sec:supplement]{the Supplementary Information}). With catalogue-fitted tables, corrected proposals meet 78 test runs and uncorrected proposals meet 57, a success-rate difference of 0.194 with interval \([0.083,0.333]\). On the fresh draw the corrected and uncorrected counts are 94 and 73 with learned tables and 87 and 69 with catalogue tables. Averaged over seeds within each target, no target in these six comparisons has a higher success rate with uncorrected proposals. The control runs on the fresh draw and the confirmatory set were added after the confirmatory plan, so these contrasts are not declared endpoints.

The uncorrected control makes smaller changes to its designs. On the test set, 50\% of its later queries in a given architecture stay within 0.003 in density and, for blends, 0.01 in weight of the first query in that architecture. The corresponding share for corrected search is 21\%. Measuring distance as the largest parameter change relative to its range, the median step from the best observed design is 0.007 for the control and 0.021 for the corrected search. The control therefore often samples again close to the optimum of the uncorrected table, and the contrast measures corrected proposals against that behaviour.

The contrast identifies an effect of using the residual to choose a refinement design under these search rules. The subsequent simulator observations need not agree after the first changed proposal. On the catalogue starts used by retrieval with pattern search, applying corrected table refinement meets 45 runs, compared with 23 for the retrieval with pattern search. The further pattern-search control from learned-table starts meets 59 runs, providing a second comparison with a coordinate search.

On the worked example in Fig.~\ref{fig:workflow}, catalogue-table search meets the target in the same architecture at call 26, and three-architecture catalogue search succeeds at call 7. Retrieval with pattern search uses 22 calls and returns its best catalogue design at error 0.1505.

\subsection{Search responds to large changes in table quality}\label{sec:table-quality}

\begin{figure}[!tbp]
\centering
\includegraphics[width=\textwidth]{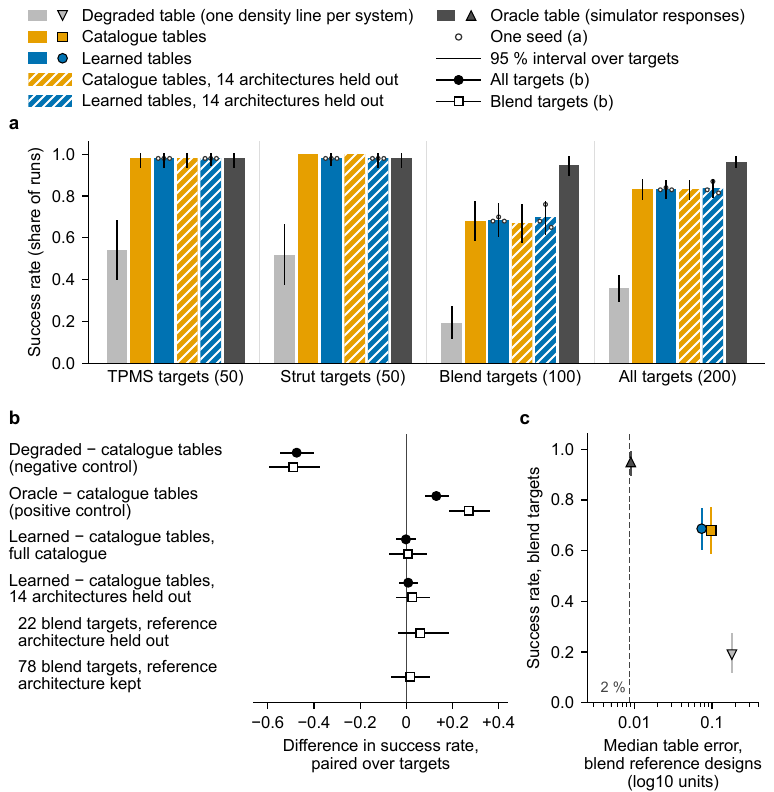}
\caption{Table-quality controls and held-out architectures on 200 confirmatory targets. \textbf{a} Success by system and overall; hatching marks held-out architectures. Bars average seeds, dots show seeds and black lines give 95\% target-bootstrap intervals. \textbf{b} Paired success differences for all targets (filled circles) and blends (open squares). \textbf{c} Blend success against median table error at reference designs; the dashed line marks the 2\% tolerance. Search rules and the \(32^3\) simulator are unchanged, with at most 30 calls and zero language-model calls. Learned arms use seeds 0--2; Fig.~\ref{fig:confirmatory} uses ten. Only the oracle uses extra simulator data. These post-plan controls are not declared confirmatory endpoints. Intervals use 20,000 target resamples without multiplicity adjustment. TPMS denotes triply periodic minimal surface}
\label{fig:table-controls}
\end{figure}

Success responds to both a poorer and a better response table (Fig.~\ref{fig:table-controls}, Tables~\ref{tab:table-controls} and \ref{tab:table-control-contrasts} in \hyperref[sec:supplement]{the Supplementary Information}). Relative to catalogue tables, the degraded table lowers overall success from 0.835 to 0.360, a difference of \(-0.475\) \([-0.545,-0.405]\), while the oracle table raises it to 0.965, a difference of 0.130 \([0.085,0.180]\). On blends the differences are \(-0.490\) and 0.270, respectively. The learned tables predict blend reference responses more accurately than catalogue tables (median maximum log error 0.073 against 0.097), but this improvement produces no detected success-rate gain. The larger changes to the degraded and oracle tables do affect success. The controls establish sensitivity to large differences in prediction quality.

When we remove the catalogue examples of 14 blend architectures, the paired intervals for learned versus catalogue tables include zero for success and design error, overall and on blends. Blend success differs by 0.027 \([-0.040,0.097]\). For the 22 targets whose reference architecture is held out, the success difference is 0.061 \([-0.030,0.182]\), leaving gains up to about 0.18 unresolved. Removing these architectures changes overall success little for either table type. This test uses one held-out subset and three learned-table seeds. The planned comparison holding out 28 architectures was omitted when earlier runs exceeded the plan's time limit. The oracle still misses seven targets. We tested one level of degradation.

\subsection{Table-guided searches outperform retrieval with pattern search on the test set}\label{sec:main-results}

\begin{figure}[!tbp]
\centering
\includegraphics[width=\textwidth]{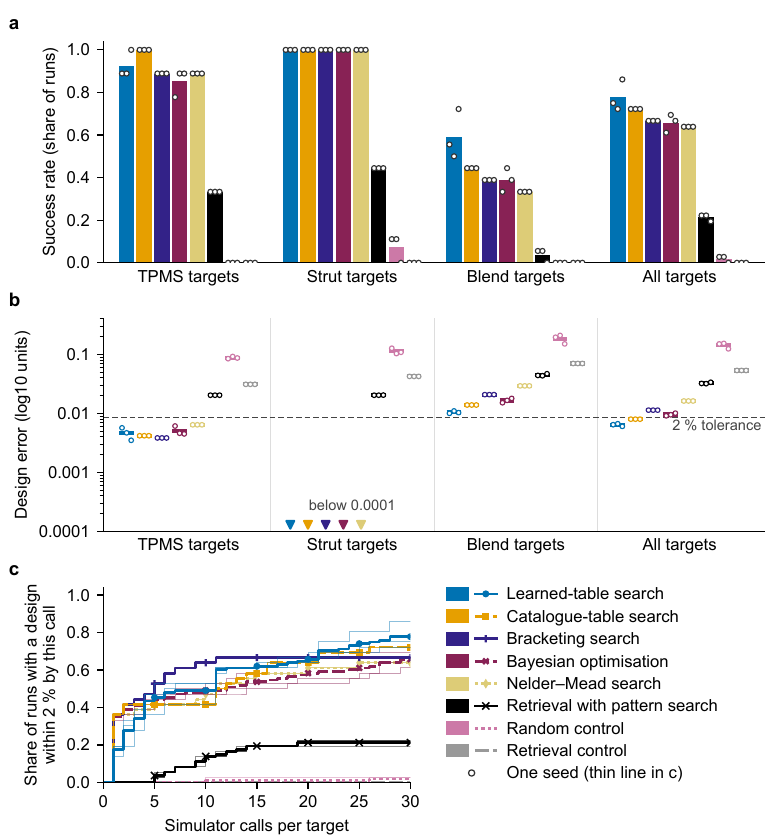}
\caption{Test results at a budget of 30 simulator calls.
\textbf{a} Success rate by target system and overall. Bars average three seeds, and dots show each seed.
\textbf{b} Mean design error on a log scale. The dashed line marks 2\%, and floor triangles mark means below \(10^{-4}\).
\textbf{c} Fraction of runs with a successful query by each call. Thick curves pool seeds, and thin curves show each seed. The retrieval control used zero calls and met zero targets. TPMS denotes triply periodic minimal surface}
\label{fig:main}
\end{figure}

On the 36-target test set, the learned-table search meets 84 of 108 runs, compared with 23 for retrieval with pattern search, none for retrieval alone and two for random search (Table~\ref{tab:main_comparison}, Fig.~\ref{fig:main}). Retrieval with pattern search uses step sizes fixed before test evaluation. The standard optimisers provide additional comparisons under the same call allowance. Mean design error is 0.00639 for the learned-table search and 0.03264 for retrieval with pattern search, a paired difference of \(-0.02625\) with 95\% interval \([-0.03751,-0.01697]\). The learned-table search has lower mean error on 32 targets and ties on four (two-sided sign test \(p<0.001\)). The success-rate difference is 0.565 with interval \([0.417,0.704]\). It meets all 23 runs met by retrieval with pattern search and another 61.

\begin{table}[tbp]\centering\begin{minipage}{\textwidth}
\centering
\fontsize{7}{8.5}\selectfont
\setlength{\tabcolsep}{2pt}
\caption{Test comparison over 36 targets $\times$ three seeds, with at most 30 simulator calls per target. Means are followed by seed ranges in parentheses. A dash denotes a deterministic configuration. Bracketed entries give 95\% target-bootstrap intervals. Deterministic counts correspond to 26/36 catalogue-table, 24/36 bracketing, 23/36 Nelder--Mead and 0/36 retrieval targets. Boundary intervals carry no information about new targets. All configurations use zero language-model calls. Calls to success includes failures at 31. Per-system results are in \hyperref[sec:supplement]{the Supplementary Information}}
\label{tab:main_comparison}
\begin{tabular*}{\linewidth}{@{\extracolsep{\fill}}lccccc@{}}
\toprule
Configuration & Runs met & Success rate & Design error & \shortstack{Calls to\\success} & \shortstack{Simulator calls\\per target} \\
\midrule
\shortstack[l]{Learned-table\\search} & 84/108 & \centerentry{\shortstack{0.778 (0.139)\\{[0.676, 0.880]}}} & \centerentry{\shortstack{0.00639 (0.00065)\\{[0.00471, 0.00848]}}} & 13.19 (0.22) & 17.03 \\
\addlinespace[2pt]
\shortstack[l]{Catalogue-table\\search} & 78/108 & \centerentry{\shortstack{0.722 (---)\\{[0.611, 0.833]}}} & \centerentry{\shortstack{0.00800 (---)\\{[0.00463, 0.01374]}}} & 13.86 (---) & 17.86 \\
\addlinespace[2pt]
Bracketing search & 72/108 & 0.667 (---) & 0.01138 (---) & 12.36 (---) & 12.03 \\
\addlinespace[2pt]
\shortstack[l]{Bayesian\\optimisation} & 71/108 & 0.657 (0.083) & 0.00958 (0.00100) & 14.99 (2.56) & 14.65 \\
\addlinespace[2pt]
Nelder--Mead search & 69/108 & 0.639 (---) & 0.01627 (---) & 14.97 (---) & 14.61 \\
\addlinespace[2pt]
\shortstack[l]{Retrieval with\\pattern search} & 23/108 & \centerentry{\shortstack{0.213 (0.028)\\{[0.102, 0.333]}}} & \centerentry{\shortstack{0.03264 (0.00180)\\{[0.02270, 0.04439]}}} & 26.47 (0.39) & 29.56 \\
\addlinespace[2pt]
Retrieval control & 0/108 & \centerentry{\shortstack{0.000 (---)\\{[0.000, 0.000]}}} & \centerentry{\shortstack{0.05352 (---)\\{[0.03850, 0.07252]}}} & 31.00 (---) & 0.00 \\
\addlinespace[2pt]
Random control & 2/108 & \centerentry{\shortstack{0.019 (0.028)\\{[0.000, 0.046]}}} & \centerentry{\shortstack{0.14299 (0.02949)\\{[0.09814, 0.19672]}}} & 30.76 (0.58) & 30.00 \\
\addlinespace[2pt]
\bottomrule
\end{tabular*}
\end{minipage}\end{table}

The learned-table search uses 17.03 simulator calls per target on average, compared with 29.56 for retrieval with pattern search. Mean calls to success, with a value of 31 assigned to failures, are 13.19 and 26.47, respectively. Within ten calls, the learned-table search has met 53 runs and retrieval with pattern search 15. The learned-table search improves the success rate in each system by 0.56 to 0.59. It meets 25 of 27 TPMS runs, 27 of 27 strut runs and 32 of 54 blend runs, compared with 9, 12 and 2 for retrieval with pattern search. The blend runs remain the least frequently solved.

Bracketing search, Bayesian optimisation and Nelder--Mead search meet 72, 71 and 69 runs, respectively, compared with 23 for retrieval with pattern search. Their blend successes are 21, 21 and 18 of 54, and each meets all 27 strut runs. The learned-table search meets 12 more runs than bracketing search, the most successful standard optimiser, a success-rate difference of 0.111 that is smaller than the learned search's 0.139 spread across its three seeds. Bayesian optimisation has the lowest mean design error among the standard optimisers (0.00958), compared with 0.00639 for the learned-table search. Bracketing search meets 69 runs within ten calls and has mean calls to success of 12.36, compared with 53 and 13.19 for learned-table search. Standard optimisers therefore close much of the difference from retrieval with pattern search. On the separate development set, the learned-table search meets 10 of 12 targets and retrieval with pattern search meets five.

\subsection{Standard optimisers use fewer calls at the 2\% tolerance}\label{sec:optimisers}

On the 200-target confirmatory set, the standard optimisers use fewer simulator calls per target at the 2\% tolerance. Learned-table search uses 15.27 calls, compared with 9.31 for bracketing search, 12.02 for Bayesian optimisation and 13.87 for Nelder--Mead search. Both table-guided searches meet more specifications. Overall success-rate differences for learned-table search against bracketing, Bayesian optimisation and Nelder--Mead are 0.059 \([0.019,0.099]\), 0.121 \([0.083,0.158]\) and 0.128 \([0.079,0.179]\). Catalogue-table search also has positive intervals against all three. These differences arise mainly on blends, where they are 0.133 \([0.059,0.212]\), 0.200 \([0.133,0.266]\) and 0.243 \([0.152,0.337]\). On the 100 pure targets bracketing search meets every target, and learned-table search meets 984 of its 1,000 runs. Bayesian optimisation is also behind on TPMS targets, by 0.098 \([0.040,0.168]\). Bracketing search holds blend weight at the table optimum, leaving density as its only way to refine a blend. The matched-stop tests in Sect.~\ref{sec:exploratory-learning} show how the comparison changes at 5\% tolerance.

\subsection{Remaining misses concentrate on blend targets}\label{sec:target-results}

\begin{figure}[!tbp]
\centering
\includegraphics[width=\textwidth]{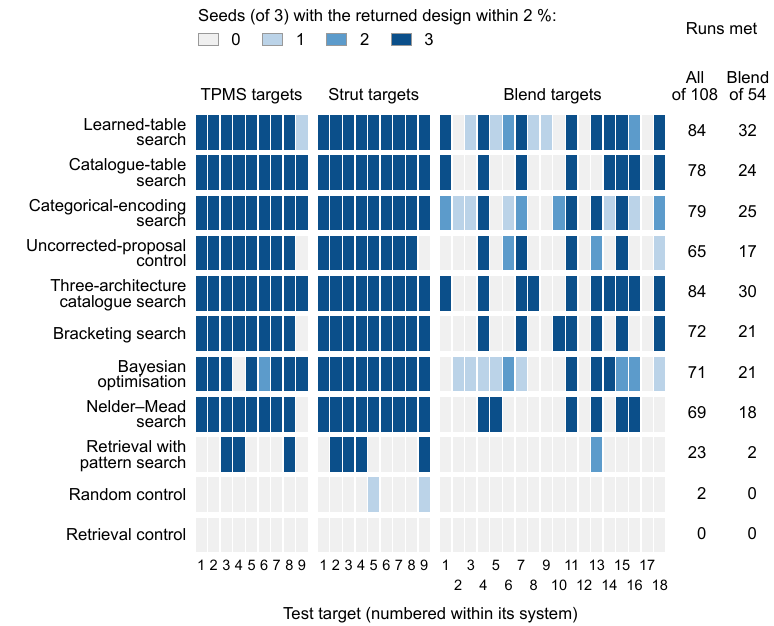}
\caption{Verified successes by test target. Columns follow benchmark order within each system. Each entry shows how many of three seeds met the tolerance after re-simulation of the returned design. The right columns total successes over all 108 runs and over the 54 blend runs. Deterministic variants can show only zero or three successes for a target. TPMS denotes triply periodic minimal surface}
\label{fig:target}
\end{figure}

On the test set, the learned-table search misses blend targets more often than pure targets. The learned-table search meets 14 of 18 blend targets for at least one seed and eight for all three seeds (Fig.~\ref{fig:target}). Retrieval with pattern search meets one blend target for two seeds. On the pure systems, the only learned-table-search misses are two seeds on the same TPMS target. Both the learned-table search and the catalogue-table search reproduce the strut targets to rounding precision on the voxel grid. In one successful TPMS run, the returned design belongs to a different system. All returned designs from the learned-table search pass the simulator's validity checks, including those that miss the tolerance.

\subsection{Three-architecture catalogue search uses fewer calls}\label{sec:allocation}

To examine the choice between testing another architecture and sizing a queried one, we restrict the architectures available after the first batch. On the 36-target test set, a search that uses catalogue-fitted tables and stays in its three best table-ranked architectures reaches 84 successes, with five successes unique to each of this variant and the learned-table search. It uses 10.81 calls per target and has mean calls to success of 9.94, compared with 17.03 and 13.19 for the learned-table search. The paired difference in calls to success is 3.24 for the learned-table search minus the three-architecture catalogue search (interval \([0.96,5.67]\)). The three-architecture catalogue search has higher mean design error (0.01010 versus 0.00639), with a paired interval for the error difference that includes zero.

\begin{figure}[!tbp]
\centering
\includegraphics[width=\textwidth]{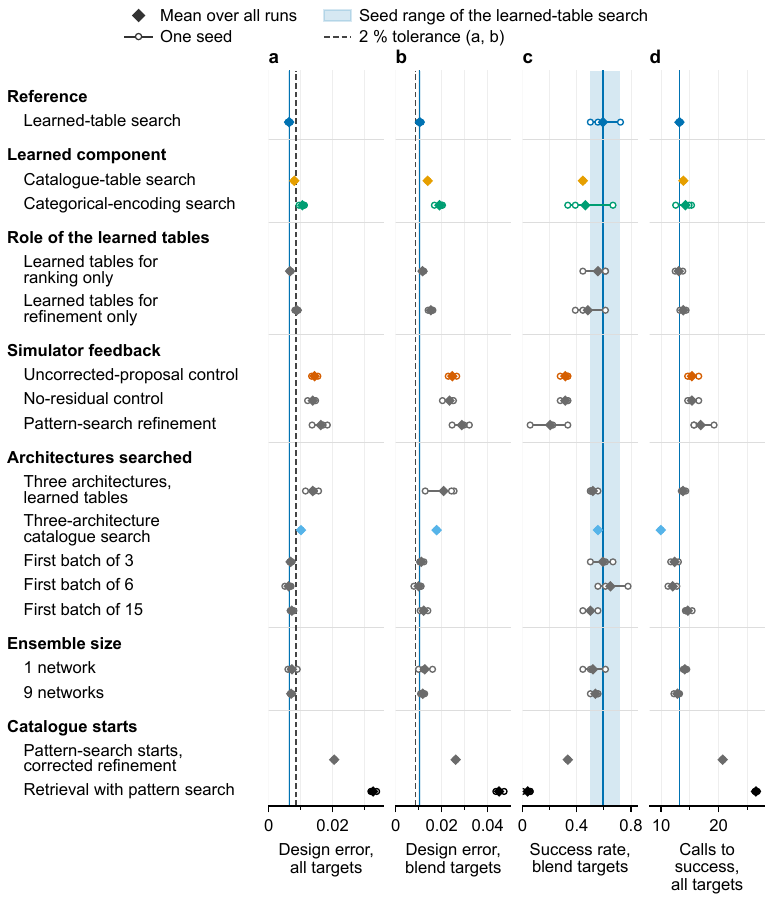}
\caption{Ablations and sensitivity analyses under the common call budget.
\textbf{a,b} Mean design error over all runs and over blend runs.
\textbf{c} Blend success rate.
\textbf{d} Mean calls to success over all runs. Diamonds show means, connected points show seed results, and the vertical band and line show the learned-table search's seed range and mean. Deterministic variants have a single point. The threshold in \textbf{a,b} is \(\tau\), the 2\% tolerance. These displays show seed variation. Target-level intervals are reported in \hyperref[sec:supplement]{the Supplementary Information}}
\label{fig:attribution}
\end{figure}

The same restriction to three architectures with learned tables meets 71 runs. Initial batches of 3, 6, 10 and 15 architectures in the broader learned-table search give 83, 87, 84 and 79 successes, respectively. The six-architecture setting was evaluated on these test targets and its 87 successes are an exploratory result. Using one, three or nine neural networks gives 80, 84 and 82 successes, with no detected paired difference in mean error. Figure~\ref{fig:attribution} summarises the component results and seed variation.

The three-architecture catalogue search also improves the success rate by 0.361 relative to using the three catalogue starts of retrieval with pattern search with corrected refinement (interval \([0.194,0.528]\)). Among the 63 runs that miss with corrected refinement from those starts, at least one of learned-table search, catalogue-table search, three-architecture catalogue search or retrieval with pattern search succeeds in 45. In each of those runs, a successful alternative uses an architecture outside those three starts. The comparison changes which architectures are tried, where they are started and how calls are allocated, so it cannot isolate a ranking-only effect.

These comparisons estimate how two sets of search choices contribute to the gain over retrieval with pattern search. From the three retrieved starts, pattern search meets 23 runs and corrected refinement meets 45. Under the table-ranked schedule, pattern search as the refinement rule meets 59 runs, and corrected refinement meets 84 with learned tables and 78 with catalogue tables. Changing the ranking, the starting designs and the allocation together therefore adds 36 runs under pattern search and 33 to 39 runs under corrected refinement. Changing the refinement rule adds 22 runs from the retrieved starts and 25 under the learned-table schedule. These arms assess the bundle of table-based ranking, starts and allocation separately from the refinement rule. The pattern-search arm under the table-ranked schedule was run with learned tables only, so its comparison with the 78 catalogue-table runs also changes the table type.

On the confirmatory set, three-architecture catalogue search meets 170 of 200 targets (85.0\%), compared with 84.4\% success for learned-table search, averaging its ten seeds for each target. Its 8.40 calls to success and 9.05 simulator calls per target compare with 10.82 and 15.27 for learned-table search. The paired calls-to-success difference is 2.42 \([1.40,3.41]\). These secondary contrasts are descriptive.

\subsection{Smaller target sets and catalogue coverage}\label{sec:exploratory-learning}

\begin{figure}[!tbp]
\centering
\includegraphics[width=\textwidth]{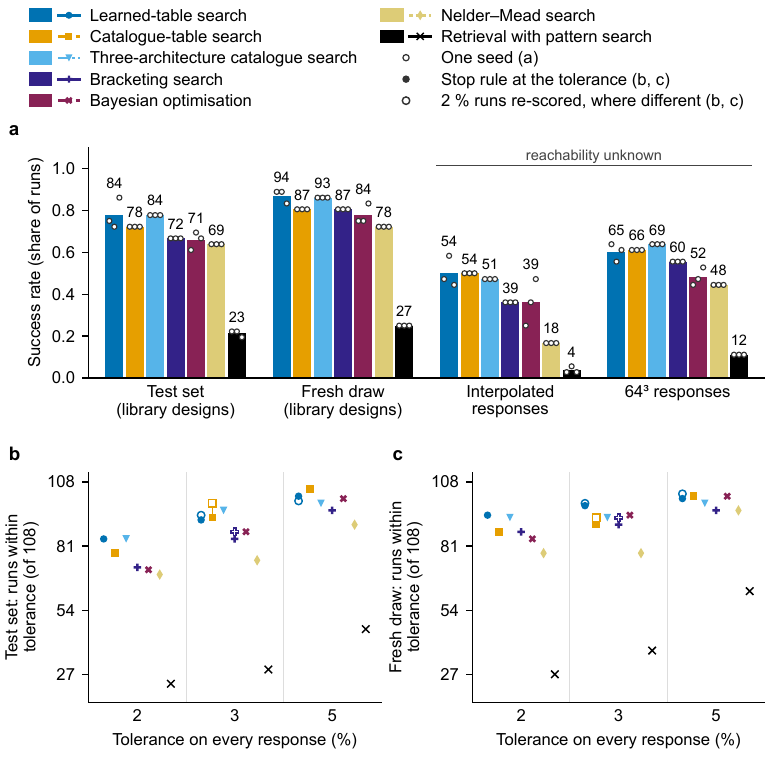}
\caption{Target sets and matched stop rules. \textbf{a} Success rates for the seven fixed configurations on the test set, fresh draw, interpolated responses and \(64^3\) responses. Each has 36 targets and three seeds, at most 30 simulator calls per target and a \(32^3\) search simulator. Bars give means, dots give seeds and numbers give runs met. Reachability is unknown for the last two sets. \textbf{b,c} Test-set and fresh-draw successes at 2\%, 3\% and 5\%. Filled markers use the matched stop rule. Larger open markers show 2\%-stop-rule returns re-scored where the counts differ. The vertical axes start at 15 runs. Retrieval with pattern search uses re-scoring because it has no early stop. TPMS denotes triply periodic minimal surface}
\label{fig:target-sets}
\end{figure}

The 36-target test set and fresh draw give exploratory estimates. We analysed these comparisons post hoc, after the development screen. Figure~\ref{fig:target-sets}a compares four smaller sets. Panels b and c show what changes when the test and fresh-draw searches use wider tolerances.

On the test set, learned tables meet 84 runs and catalogue tables meet 78. The paired success-rate difference is 0.056 with 95\% interval \([-0.019,0.139]\), and the blend difference is 0.148 with interval \([0.019,0.296]\). Mean design error differs by \(-0.00161\), with interval \([-0.00605,0.00122]\). These intervals describe target variation, while the three-seed ranges are 0.139 for overall success and 0.222 for blend success. On the development set, both catalogue variants meet 11 of 12 targets and learned-table search meets ten (Table~\ref{tab:dev_subset} in \hyperref[sec:supplement]{the Supplementary Information}).

Pooling the test set and fresh draw gives an exploratory advantage for learned tables. Across 72 reachable targets, the success rate is 6.0 percentage points higher, with a paired 95\% interval from 0.9 to 11.6 points. The intervals for design error and calls to success include zero. The learned configuration had passed a development check against catalogue tables, and the larger confirmatory comparison does not reproduce the pooled advantage. Tables~\ref{tab:paired_additional}--\ref{tab:paired_confirmatory} in \hyperref[sec:supplement]{the Supplementary Information} give the paired outcomes and contrasts.

The coverage checks ask whether learning helps more when fewer catalogue designs are available. We detect no advantage when grouping blend targets by catalogue coverage or when repeating the comparison with fewer pure-architecture examples. The latter check retains all blend designs. Table~\ref{tab:coverage} in \hyperref[sec:supplement]{the Supplementary Information} gives the counts and intervals.

With early stopping set to the tolerance being tested, at 3\% and 5\%, learned-table and catalogue-table searches meet 92 versus 93 and 102 versus 105 test runs. On the fresh draw they meet 98 versus 90 and 101 versus 102 (Fig.~\ref{fig:target-sets}b,c). The fresh-draw 3\% success difference is 0.074 \([0.019,0.148]\), the only one of these four paired intervals that excludes zero. The ordering varies with tolerance and set. Table~\ref{tab:tolerance_levels} reports these runs alongside the 2\%-stop-rule designs re-scored at wider tolerances. At 5\%, Bayesian optimisation meets 101 test runs and 102 fresh-draw runs with 4.25 and 4.34 simulator calls per target, respectively.

Among the learned variants, the shared encoding lowers mean error relative to the categorical-encoding search by 0.00414 overall (interval \([-0.00996,-0.00023]\) for shared minus categorical encoding) and by 0.00857 on blends (interval \([-0.02044,-0.00096]\)). Success-rate intervals for those comparisons include zero. The shared encoding therefore helps the design-error metric in this comparison with an untuned categorical control. The role swap with learned tables for ranking and catalogue refinement meets 84 runs. The reverse swap meets 80. Compared with using catalogue-fitted tables throughout, learned tables for ranking lowers mean blend error by 0.00224, with an interval of \([-0.01325,0.00536]\) for the difference.

The offline diagnostic gives little support for explaining the shared encoding's lower design error by better architecture ranking. The shared encoding recovers fewer blend architectures in the first ten ranks and more in the first twenty than the categorical encoding (Fig.~\ref{fig:offline} and Table~\ref{tab:offline_recall} in \hyperref[sec:supplement]{the Supplementary Information}). The verified search can meet the target through any architecture.

\subsection{Target construction and resolution change the ordering}\label{sec:target-sets}

The additional sets test how the test-set ordering depends on guaranteed reachability and the particular target draw (Fig.~\ref{fig:target-sets}a). On the fresh draw, learned-table search meets 94 runs, catalogue-table search 87 and three-architecture catalogue search 93. The corresponding mean design errors are 0.01230, 0.01326 and 0.01389. Thus the higher success count than on the test set accompanies a higher mean error. Bracketing search, Bayesian optimisation and Nelder--Mead search meet 87, 84 and 78 runs, and retrieval with pattern search meets 27. The learned-versus-catalogue success difference of 0.065 exceeds the learned seed range of 0.056 on this draw. The paired pooled contrast and confirmatory comparison (Sects.~\ref{sec:exploratory-learning} and \ref{sec:learning}) distinguish this estimate from evidence on a further draw.

The interpolated targets yield 54 successes each for learned-table and catalogue-table search, compared with 51 for three-architecture catalogue search and four for retrieval with pattern search. For targets defined by the fresh references' \(64^3\) responses, the counts are 65, 66, 69 and 12. The learned-table search's advantage over catalogue-table search is therefore inconsistent across sets. For all seven configurations in Fig.~\ref{fig:target-sets}a, success falls relative to the fresh draw when the simulator may have no design that matches a target. On the fresh blend targets, 40 of the learned-table search's 43 successes use an architecture different from the reference, and none returns the reference design. Matching properties can therefore succeed through an alternative architecture. The pure reachable targets largely reduce to architecture identification and one-parameter matching.

\begin{figure}[!tbp]
\centering
\includegraphics[width=\textwidth]{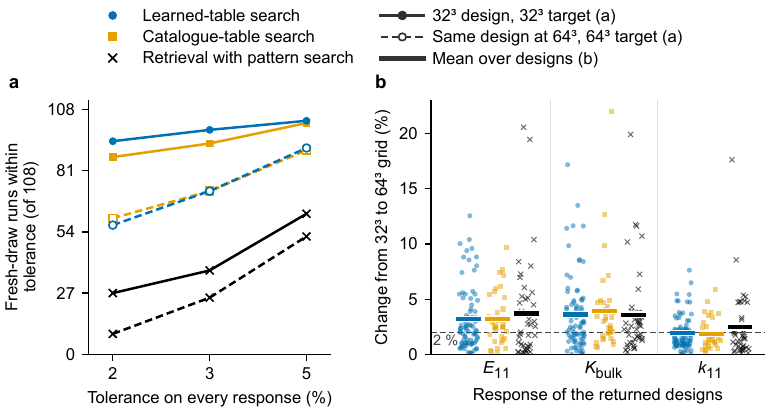}
\caption{Grid-resolution checks. \textbf{a} Fresh-draw designs returned under the 2\% stop rule, assessed at 2\%, 3\% and 5\%. Solid lines compare designs and targets on \(32^3\), dashed lines compare the same designs and reference responses on \(64^3\). \textbf{b} Absolute response changes over distinct returned test-set designs. Points show designs with horizontal jitter, bars show means and the dashed line is 2\%. Marker shapes and colours identify configurations. These panels measure the 32-to-64 shift, with finer-grid convergence untested}
\label{fig:grid-shift}
\end{figure}

A design that meets its target on the search grid can miss at finer resolution. The grid checks show property shifts comparable to or larger than the matching tolerance (Fig.~\ref{fig:grid-shift}a,b). Over distinct returned test designs, the mean absolute \(32^3\)-to-\(64^3\) changes in \(E_{11}\), \(K_{\mathrm{bulk}}\) and \(k_{11}\) are 3.19\%, 3.61\% and 1.94\% for learned-table search, and 3.22\%, 3.93\% and 1.87\% for catalogue-table search. Against the \(32^3\) targets, their success counts fall from 84 and 78 to 12 and 18. This comparison mixes resolutions on purpose to measure the change from the verified match on the search grid. On the fresh draw, solving both the returned and reference designs at \(64^3\) reduces counts from 94 and 87 to 57 and 60. The catalogue-table search is ahead in this finer-grid comparison. No returned design fails validity at \(64^3\). The finer-grid comparison shows that the \(32^3\) match can change by more than the tolerance with grid resolution. Convergence at \(64^3\) remains untested.

\section{Discussion}\label{sec:discussion}

\subsection{Implications for allocating simulations}

For a designer who must meet a tight tolerance, a simulation that misses the target can still improve the next proposal. The correction uses the observed properties at the best design to update the table locally. In the matched proposal tests this change improved success for both learned and catalogue-fitted tables (Sect.~\ref{sec:feedback}). The result connects model management and surrogate alignment \citep{alexandrov1998,bandler2004} with the value of a simulation for later design decisions \citep{panchal2009,chaudhari2020}.

The choice of search depends on the designer's tolerance and the cost of each call. At 2\%, the table-guided searches met more confirmatory targets than the standard optimisers and generally used more calls. At 5\%, Bayesian optimisation nearly matched the learned-table search on the test set and fresh draw using fewer than five calls per target (Sects.~\ref{sec:optimisers} and \ref{sec:exploratory-learning}). The three-architecture comparison shows that the choice of starting designs also matters when calls are scarce. Because the experiments change ranking, starting designs and allocation together, they estimate the contribution of those decisions as a group (Sect.~\ref{sec:allocation}). The refinement comparison leaves the separate effects of the two parts of the correction unresolved (Eq.~\ref{eq:residual}).

These trade-offs arise when the aim is to meet a specification under an allowance \citep{guo2024}. In a finite library with examples for each architecture and a common verifier, designers can test what a learned model adds by replacing it with catalogue fits while keeping the search and feedback fixed. The comparison gives the designer a basis for choosing between improving the predictor and changing where the search spends calls.

\subsection{Limits of the benchmark}

The test set and fresh-draw targets are generated by library members on the same parameter grid as the search. This construction tests inversion beyond retrieval while excluding impossible targets. For pure architectures it largely tests architecture identification plus one-parameter matching, which helps explain the near-exact strut recovery. Interpolated and finer-grid targets weaken that assumption, but their reachability by the \(32^3\) design library is unknown. Their lower success combines search difficulty with potential infeasibility. The test set has 36 targets and three seeds, while the confirmatory set has 200 targets and ten seeds for stochastic arms. The latter still represents one reachable-target generator and one catalogue. The component and sensitivity comparisons on the test set are exploratory. Tests of the same configurations on additional sets show how the results vary with the targets. The comparison arms use the same catalogue, so these comparisons concern search within the common library and information allowance. Learned generators and language-model-guided designers were outside the numerical comparison.

The verifier uses a coarse voxel grid. Targets, catalogue and verifier share one discretisation, so the study leaves search against responses computed by another model untested. Changes in a continuous parameter may leave the geometry unchanged. The finer-grid checks show that a 2\% match at \(32^3\) cannot be read as convergence or manufactured-property accuracy. The numerical results come from one material system with a fixed base Poisson ratio. Over the 358 catalogue designs, the three log responses have pairwise correlations of 0.92 to 0.98. One principal component accounts for 96\% of their variance. The strong correlations in the catalogue limit what this study can say about specifications that trade one property against another. Manufacturing constraints, shear, anisotropy and density requirements were outside the study. The allowance is 30 simulator calls. Training and table construction are paid once per seed for the shared catalogue and excluded from those counts. \hyperref[sec:supplement]{The Supplementary Information} lists the hand-set constants and distinguishes the two varied settings from fixed settings whose values were neither tuned nor sensitivity-tested.

Transfer to other simulators, property sets and libraries remains untested. We recorded the selection rule and its outcome. Individual development runs were incompletely logged, which limits reconstruction of configuration selection. We recorded the confirmatory plan before drawing the targets. It had no external registration. Roughly halving the pure-architecture catalogue did not produce a learned-table advantage (Sect.~\ref{sec:exploratory-learning}), and removing one subset of blend architectures detected no advantage, but more extensive blend hold-outs, larger catalogues and other libraries were not tested. Future work should vary catalogue size and coverage to determine when shared learning adds to per-architecture fitting. A property set with stronger trade-offs and a converged verifier would show whether the same comparison can guide the choice of predictor in a wider range of designs.

\section{Conclusion}\label{sec:conclusion}

With the same search, simulator feedback and call budget, the learned tables and catalogue fits showed no detectable difference in success or design error on the declared comparison. The paired intervals leave smaller benefits or costs unresolved. In separate controls, using simulator feedback to choose the refinement design increased success with both kinds of table. This gives designers with catalogue examples for each architecture and a common verifier a baseline for judging whether a learned model justifies its cost. The comparison can be repeated with other catalogues and property targets to find when learning contributes enough to change that decision.

\section*{Supplementary information}

\hyperref[sec:supplement]{The Supplementary Information} reports the full search schedule, the Bayesian-optimisation specification, the full ablations, paired contrasts, the correction control for both table types, the catalogue-coverage analyses, per-seed and development outcomes, per-system optimiser comparisons, additional target sets, grid-resolution checks, search constants and offline diagnostics.

\section*{Statements and Declarations}

\paragraph{Funding} This research was supported by the National Science Foundation (NSF) under the Computational and Data-Enabled Science and Engineering (CDS\&E) funding opportunity (program description PD~22-8084) [award number 2245299].

\paragraph{Competing interests} The authors declare no competing interests.

\paragraph{Author contributions} S.Y. designed the study, developed the methods, directed the implementation, experiments and analysis, and wrote and edited the manuscript. J.W. conceptualised the ideas, provided funding, and reviewed and edited the manuscript.

\paragraph{Data availability} The offline catalogue, target sets and per-run benchmark results will be available at \url{https://github.com/nbbllxx0/Budgeted-inverse-design-of-periodic-unit-cells}.

\paragraph{Code availability} The simulator, evaluation protocol, search implementations and analysis scripts will be available at \url{https://github.com/nbbllxx0/Budgeted-inverse-design-of-periodic-unit-cells}.

\paragraph{Use of generative AI} The authors conceived the study, wrote the manuscript, and take full responsibility for its content and conclusions. OpenAI Codex assisted with editing and formatting. Section~\ref{sec:ai} describes assistance with software and analysis. The authors reviewed and revised the AI-assisted work before adopting it.

\FloatBarrier
\appendix
\section*{Supplementary Information}\label{sec:supplement}
\setcounter{figure}{0}\setcounter{table}{0}
\renewcommand{\thetable}{S\arabic{table}}\renewcommand{\thefigure}{S\arabic{figure}}
\renewcommand{\theHtable}{S\arabic{table}}\renewcommand{\theHfigure}{S\arabic{figure}}

\section*{Benchmark construction}

The triply periodic minimal surface (TPMS) architectures use eight periodic fields from gyroid, Schwarz P, diamond, I-WP, Neovius, Fischer--Koch S, F-RD and split P. Each has network and sheet constructions. The strut set consists of simple cubic, body-centred cubic, face-centred cubic, octet, body-centred cubic with an added axial member, face-centred cubic with an added axial member, combined simple and body-centred cubic, and diamond truss. The added axial members run along \(z\). Each of the 28 level-set pairs has network and sheet constructions.

For the offline catalogue, the pure-architecture grid has ten evenly spaced densities over each system's design range. Four blend candidates per pair and mode are drawn using weights \(\{0.25,0.50,0.75\}\) and densities \(\{0.10,0.25,0.40,0.55\}\). After canonicalising the pair order, the generator removes repeated inputs and keeps valid simulator outputs. The catalogue generator uses seed 20261109. The test reference generator uses seed 977. The additional fresh draw uses seed 20261110. Reference densities are drawn uniformly over \([0.10,0.58]\) for pure level-set architectures, \([0.08,0.48]\) for strut architectures and \([0.12,0.55]\) for blends. Reference blend weights are drawn over \([0.15,0.85]\) before canonicalisation. Discrete choices are drawn uniformly within each reference system. Reference designs must be distinct, valid and at least 0.02 in maximum absolute log-ratio error from all catalogue designs.

\section*{Table conventions}

\emph{Learned-table search} uses predictions from three feed-forward neural networks and tests up to ten architectures in its first batch. \emph{Catalogue-table search} uses catalogue-fitted tables under the same schedule. The three-architecture variants open no further architectures after their initial batch. Learned tables for ranking include selection of the starting design, and learned tables for refinement use the model only in queried architectures. Counts pool three seeds per test target. Deterministic configurations give identical repeats, so a dash replaces their seed ranges. Target-level intervals use independent targets as the resampling unit. Calls to success is the index of the first successful query, or 31 if the run never succeeds. The simulator-call count is the number of designs the search queries. Design error is the largest absolute base-10 log-ratio between a returned property and its target, capped at 1. Success means all three simulated properties meet the tolerance stated in the article.

Tables~\ref{tab:ablation} and \ref{tab:ablation_pure} report the variant results by system. Tables~\ref{tab:contrasts_design_error} and \ref{tab:contrasts_success_rate} compare paired outcomes. A boundary interval such as [0.000, 0.000] or [1.000, 1.000] carries no information about performance on new targets. The paired intervals resample targets within their system 20,000 times using seed 0. Both variants and all seeds for a target use the same draw. The tests are two-sided and the reported \(p\) values are unadjusted. Tables~\ref{tab:per_seed}--\ref{tab:offline_recall} give seed-level results, alternate tolerances, development results and the offline diagnostic. Tables~\ref{tab:system_comparison}--\ref{tab:paired_confirmatory} give results by system, search settings, the additional target sets, grid sensitivity and paired comparisons. Table~\ref{tab:correction_control} compares corrected and uncorrected refinement proposals for learned and catalogue-fitted tables on the test set, the fresh draw and the confirmatory set. Table~\ref{tab:coverage} compares the two table types under reduced catalogue coverage. Tables~\ref{tab:table-controls} and \ref{tab:table-control-contrasts} report the degraded and oracle controls and the held-out-architecture test, including simulator calls and zero language-model calls.

\section*{Geometry and discretisation}

The unit cube is sampled at voxel centres \(X_i=(i+1/2)/n\), for \(i=0,\ldots,n-1\) on each axis, with \(n=32\) for search. Write \(x=2\pi X\), \(y=2\pi Y\), \(z=2\pi Z\), \(s_x=\sin x\), \(c_x=\cos x\), and similarly \(s_{2x}=\sin(2x)\), \(c_{2x}=\cos(2x)\). The sampled fields \(f_{\mathrm{G}}, f_{\mathrm{P}}, f_{\mathrm{D}}, f_{\mathrm{IWP}}, f_{\mathrm{N}}, f_{\mathrm{FKS}}, f_{\mathrm{FRD}}\) and \(f_{\mathrm{SP}}\) correspond in order to gyroid, Schwarz P, diamond, I-WP, Neovius, Fischer--Koch S, F-RD and split P.
\begin{align*}
f_{\mathrm{G}}={}&s_xc_y+s_yc_z+s_zc_x,\\
f_{\mathrm{P}}={}&c_x+c_y+c_z,\\
f_{\mathrm{D}}={}&s_xs_ys_z+s_xc_yc_z+c_xs_yc_z+c_xc_ys_z,\\
f_{\mathrm{IWP}}={}&2(c_xc_y+c_yc_z+c_zc_x)-(c_{2x}+c_{2y}+c_{2z}),\\
f_{\mathrm{N}}={}&3(c_x+c_y+c_z)+4c_xc_yc_z,\\
f_{\mathrm{FKS}}={}&c_{2x}s_yc_z+c_xc_{2y}s_z+s_xc_yc_{2z},\\
f_{\mathrm{FRD}}={}&4c_xc_yc_z-(c_{2x}c_{2y}+c_{2y}c_{2z}+c_{2z}c_{2x}),\\
f_{\mathrm{SP}}={}&1.1(s_{2x}s_zc_y+s_{2y}s_xc_z+s_{2z}s_yc_x)\\
&-0.2(c_{2x}c_{2y}+c_{2y}c_{2z}+c_{2z}c_{2x}).
\end{align*}
Each field is divided by its standard deviation over the sampled grid. A blend uses \((1-w)f_a/\operatorname{std}(f_a)+wf_b/\operatorname{std}(f_b)\), where \(f_a\) and \(f_b\) are the two fields of the blend and \(w\) is the weight on the second field. Network construction uses this signed field, while sheet construction uses its absolute value.

Strut axes are periodically repeated line segments. Denote a segment by \([a,b]\), let \(O=(0,0,0)\), \(H=(1/2,1/2,1/2)\), \(e_1,e_2,e_3\) be the Cartesian unit vectors, and \(V=\{0,1\}^3\). Define the following member sets.
\begin{align*}
\mathcal S={}&\{[O,e_1],[O,e_2],[O,e_3]\},\\
\mathcal B={}&\{[H,v]\mid v\in V\},\\
\mathcal F={}&\{[O,(1,1,0)],[(1,0,0),(0,1,0)],\\
&[O,(1,0,1)],[(1,0,0),(0,0,1)],\\
&[O,(0,1,1)],[(0,1,0),(0,0,1)]\},\\
\mathcal Z={}&\{[O,e_3]\}.
\end{align*}
Let \(Q\) be the six face centres, given by
\begin{align*}Q=\{&(1/2,1/2,0),(1/2,1/2,1),(1/2,0,1/2),\\
&(1/2,1,1/2),(0,1/2,1/2),(1,1/2,1/2)\}.\end{align*} Let \(\mathcal I\) contain every unordered pair \([p,q]\) in \(Q\) with length \(\sqrt{1/2}\). For the diamond truss, let
\begin{align*}
N_{\mathrm d}={}&\{(1/4,1/4,1/4),(3/4,3/4,1/4),\\
&(3/4,1/4,3/4),(1/4,3/4,3/4)\},\\
D={}&\{(1,1,1),(1,-1,-1),(-1,1,-1),(-1,-1,1)\},\\
\mathcal E={}&\{[b,b-d/4]\mid b\in N_{\mathrm d},\ d\in D\}.
\end{align*}
The member sets of the eight topologies are \(\mathcal S\) for simple cubic, \(\mathcal B\) for body-centred cubic, \(\mathcal F\) for face-centred cubic, \(\mathcal F\cup\mathcal I\) for octet, \(\mathcal B\cup\mathcal Z\) for body-centred cubic with an added axial member, \(\mathcal F\cup\mathcal Z\) for face-centred cubic with an added axial member, \(\mathcal S\cup\mathcal B\) for combined simple and body-centred cubic, and \(\mathcal E\) for the diamond truss. At each voxel centre \(P\), the strut field is the minimum Euclidean distance to a member segment translated by each vector in \(\{-1,0,1\}^3\). The closest point on a segment uses the clipped axial coordinate \(u=\min\{1,\max\{0,(P-a)\cdot(b-a)/\|b-a\|^2\}\}\).

For either a level-set or strut field \(g\), sort its \(M=n^3\) samples as \(g_{(0)},\ldots,g_{(M-1)}\). Let \(k=\min\{\max\{\operatorname{round}(\rho M),1\},M-1\}\), where \(\rho\) is the density parameter of the design. Solid voxels satisfy \(g\leq[g_{(k-1)}+g_{(k)}]/2\). This threshold is the strut radius for distance fields. Tied samples can change realised density. Continuous design parameters are first rounded to three decimals. Periodic six-neighbour connectivity joins opposite faces, and only the largest component is solved. More than 1\% disconnected solid rejects the design.

Periodic trilinear hexahedral elements use two-point Gauss quadrature in each direction. Opposite-face nodal degrees of freedom are identified through periodic indexing. The binary solid mask multiplies the element stiffness and conductivity matrices, giving void zero stiffness and conductivity. Nodes incident to solid elements define the solid-node mask. The solver projects out the mean solid-node displacement in each component, or the mean scalar conduction field, to remove rigid translation or the constant null mode. The Jacobi preconditioner is the reciprocal assembled diagonal where positive and zero elsewhere. The double-precision conjugate gradient solver checks the relative residual every ten iterations at tolerance \(10^{-8}\), with a maximum of 4,000 iterations. The implementation returns the current projected iterate at that cap. A re-solve of the 358 catalogue designs, the 36 fresh-draw and 200 confirmatory reference designs, and the 1,232 distinct designs returned by learned-table and catalogue-table search on the test set, the fresh draw and the confirmatory set comprised 16,434 linear systems. None reached the cap. The largest iteration count was 1,460, and every final relative residual was below \(10^{-8}\). The searches run for Table~\ref{tab:correction_control} and their checks solved 101,754 systems. Two of them reached the cap, with final relative residuals of \(5.5\times10^{-8}\) and \(3.8\times10^{-8}\). The axial screen requires the axial stiffness component \(C_{11}\), axial Young's modulus \(E_{11}\) and axial conductivity \(k_{11}\) to be at least \(10^{-6}\) in the article's normalisation. Designs with nonfinite responses or singular stiffness matrices are rejected.

\section*{Allocation and response tables}

Let \(\tau=\log_{10}(1.02)\) be the 2\% tolerance on design error. The search is a sequential allocation problem. Let \(\mathcal{C}\) be the set of 80 architectures, \(\Theta_c\) the rounded parameter set of architecture \(c\), and \(\mathcal{D}\) the offline catalogue. After \(t\) calls the history is \(h_t=\{(c_i,\theta_i,o_i)\}_{i=1}^{t}\), where \(o_i\) is the simulated response of design \((c_i,\theta_i)\) or an invalidity flag. \(p^\star\) is the vector of the three target properties. A search policy \(\pi\) maps \((\mathcal{D},p^\star,h_t)\) either to the next batch of designs or to a stop, subject to \(t\le B\) with \(B=30\). The policy then returns one design \(x_\pi\) from its queries and the catalogue. \(e\) is the design error defined in the article. We compare policies by the success rate \(\Pr[e(x_\pi)\le\tau]\) and the mean error \(\mathbb{E}[e(x_\pi)]\) over targets and seeds, and by the calls they use. Each call either tests an architecture with no valid observation or refines one with a valid observation. This choice determines how the policy allocates calls. The searches in the article are policies of this form.

Each architecture has a table \(T_c\) of predicted \(\log_{10}\) responses. We train three feed-forward neural networks (multilayer perceptrons, or MLPs) on the offline catalogue. Each table records their average predictions in log space. The shared encoding uses an eight-entry family vector shared across pure TPMS and blends. It assigns one unit entry to a pure family and entries \(1-w\) and \(w\) to the two blend families. A sheet flag, eight strut indicators, three system indicators, \(\rho\) and \(\log_{10}\rho\) give 22 inputs. Each MLP has three hidden layers of width 128 with the sigmoid linear activation \(x/(1+e^{-x})\). The density inputs and the three response outputs are standardised. Full-batch Adam training \citep{kingma2015} uses 3,000 steps, initial learning rate \(0.003\) with cosine decay and weight decay \(10^{-5}\) on the weights. These network settings were fixed without tuning.

The pure-architecture grid uses up to 201 density points. It spans the catalogue densities for that architecture plus half the original catalogue spacing on either side, clipped to the design bounds. A blend grid has 101 density points and 41 weight points across the design bounds. These grids are rounded to the design precision. All 80 tables are built before target queries and remain read-only during search.

For the search with catalogue-fitted tables, a pure-architecture table is piecewise linear in \(\log_{10}\rho\) with end-segment extrapolation. A blend table is a least-squares plane in \((\log_{10}\rho,w)\) fit to that architecture's catalogue responses. With one design it is constant, with two it has the minimum-norm slope fitting their response difference, and with three or more it uses the least-squares fit, also taking the minimum-norm slope if the fit is underdetermined. An architecture without catalogue observations has no prediction in this variant. The catalogue-table search shares the same table grids, refinement rules and call budget as the learned-table search.

\section*{Search schedule of the table-guided searches}

This section states the decision rules of the learned-table and catalogue-table searches in full.

The ranking can shift for an entire system after valid observations in three or more architectures. For each property, the shift is the median of the first residuals of those architectures. We omit each architecture in turn and predict its residuals from the others. The shift is used if this prediction reduces the mean of the largest absolute residual across the three properties.

Here \(\tau=\log_{10}(1.02)\) is the 2\% tolerance on design error, and Table~\ref{tab:constants} lists the search constants.

\emph{First proposals and ranking.} For an architecture without a valid observation, the first proposal is the point of its table grid with the lowest predicted error, excluding catalogue designs and the excluded neighbourhoods of invalid queries. Among tied grid points, the one nearest in scaled maximum norm to the architecture's catalogue design with the lowest target error is taken. Architectures are ordered by this predicted error and then by their position in the library list. The first batch takes the first ten architectures in this order, with at most six from one system.

\emph{State.} For every architecture the search keeps its valid observations, its invalid queries and a stall counter. The first valid observation in an architecture sets the counter to zero, and so does a valid observation whose error is below the best error observed so far in that architecture. Every other query in the architecture, valid or invalid, adds one. An architecture with two invalid queries and no valid observation receives no further proposal.

\emph{Later batches.} While the lowest observed error over all architectures is at least \(2\tau\), a batch has five designs, or as many as the remaining budget allows. Up to two places open architectures without a valid observation, in the ranking order and without the limit per system. Their tables include the system shift when the leave-one-out check of the article accepts it. The other places hold refinements, one per architecture. A queried architecture is a refinement candidate when it has a proposal, the predicted error of that proposal is below its best observed error, and its stall counter is below three. Candidates are ordered by the predicted error of the proposal when the stall counter is below two, and otherwise by the larger of the predicted and the best observed error. Ties are resolved by the best observed error and then by the library position. Places that one kind of proposal cannot fill pass to the other kind.

\emph{Close to the target.} Once the lowest observed error is below \(2\tau\), every batch has two designs. They are refinements in the two architectures with the lowest best observed errors that still have a proposal. The predicted-improvement test and the stall test do not apply, and no further architecture is opened.

The correction slope is fitted by minimum-norm weighted least squares on the offsets from the current best design, scaled by the parameter ranges. For scaled offset \(u\), the least-squares row multiplier is \(1/\sqrt{\|u\|_2^2+10^{-8}}\). The relative singular-value cutoff is \(10^{-6}\).

\emph{Refinement proposal.} Let \(m\) be the number of valid observations in the architecture. On each parameter axis the window is centred at the best observed design. Its half-width is the larger of \(0.25\cdot0.5^{\,m-1}\) times the parameter range and the floor (0.003 in density, 0.01 in blend weight), and it is clipped to the table axis. The window is sampled with the smallest step that is a multiple of 0.001 and gives at most 200 intervals. Catalogue designs and valid observations are excluded. Each invalid query excludes points within a scaled maximum-norm radius of 0.10. When the architecture has valid observations, the radius is capped at half the distance to the nearest one. The proposal minimises the predicted error of the corrected table over the remaining points. A tie goes to the lowest density and then to the lowest weight.

\emph{Stop and return.} The search stops after a batch that contains a successful query, when no proposal remains or when the budget is spent. It returns the valid observation with the lowest error, or the nearest catalogue design when that design has a strictly lower error.

\emph{Controls.} The uncorrected-proposal control keeps every rule above, including the candidate tests and orderings that use the corrected prediction. It changes only the queried design, which minimises the uncorrected table over the same window points. The no-residual control takes the same uncorrected proposal, orders refinement candidates by their best observed error, applies only the stall test and uses no system shift.

\section*{Bayesian-optimisation comparator}

The Gaussian process models the three \(\log_{10}\) responses over the architecture \(c\), \(\log_{10}\rho\) and the blend weight \(w\), with \(w=0.5\) for pure architectures. The continuous parameters use a Mat\'ern-5/2 covariance function \citep[Section 4.2]{rasmussen2006}. For each system and response, the mean is a least-squares line in \(\log_{10}\rho\) fitted to the catalogue. For response \(j\), the covariance of two designs is
\begin{align*}
&\sigma_j^2\,\gamma(c,c')\,\kappa\!\left(\frac{\log_{10}\rho-\log_{10}\rho'}{\ell_\rho}\right)\kappa\!\left(\frac{w-w'}{\ell_w}\right),\\
&\kappa(d)=\Bigl(1+\sqrt5\,|d|+\tfrac{5}{3}d^2\Bigr)e^{-\sqrt5|d|},
\end{align*}
with \(\gamma=1\) for the same architecture, \(\gamma=g_1\) for two architectures of one system and \(\gamma=g_1q\) otherwise. A nugget \(\eta\) is added to the variance of every observation. The three responses share this correlation and are modelled independently, and \(\sigma_j^2\) is the maximum-likelihood variance of response \(j\). The hyperparameters maximise the marginal likelihood of the 358 catalogue designs using a limited-memory method for bound-constrained optimisation (L-BFGS-B) \citep{byrd1995}, with \(\ell_\rho,\ell_w\in[0.03,3]\), \(g_1\in[0,0.99]\), \(q\in[0,1]\) and \(\eta\in[10^{-6},10^{-1}]\), started at \(0.3\), \(0.3\), \(0.5\), \(0.5\) and \(10^{-4}\). They are fitted once and held fixed. During a search, the process is conditioned on the catalogue and on the valid observations of the current target.

The incumbent is the lowest error among the catalogue designs and the valid observations. For a candidate with posterior mean \(\mu\) and standard deviations \(\sigma\), the acquisition is the mean of \(\max\{0,\ \text{incumbent}-\|\mu+\sigma\odot z_i-y^\star\|_\infty\}\) over 32 standard normal vectors \(z_i\). Here \(y^\star\) is the vector of log target properties, \(\odot\) is elementwise multiplication and \(\|\cdot\|_\infty\) takes the largest absolute component. The seed fixes these vectors for all candidates and calls in a run. Each call has two stages. The first stage evaluates every fourth density point and every second weight point of each architecture's table grid, together with the last point of each axis. The second stage evaluates all designs of the chosen architecture on the 0.001 grid within four density-grid steps and two weight-grid steps of the first-stage choice. Catalogue designs and queried designs are excluded, and an invalid query excludes the designs of its architecture within 0.05 of the parameter range on every axis. If the largest acquisition value is at most \(10^{-12}\), the candidate with the lowest posterior-mean error is taken. Each call queries one design. The search stops at the first successful query or when the budget is spent, and it returns a design by the rule of the table-guided searches.

\begingroup\scriptsize
\setlength{\tabcolsep}{2pt}
\setlength{\LTleft}{0pt}
\setlength{\LTright}{0pt plus 1fill}
\setlength{\LTcapwidth}{\textwidth}
\begin{longtable}{@{}L{24mm}C{8mm}C{9mm}C{21mm}C{21mm}C{12mm}C{12mm}C{12mm}@{}}
\caption{Component comparisons and sensitivity analyses on the test set. Each variant uses at most 30 simulator calls per target. Metric means are followed by seed ranges in parentheses where shown. A dash denotes a deterministic configuration.}\label{tab:ablation}\\

\toprule
 & \multicolumn{2}{c}{Runs met} & \multicolumn{2}{c}{Design error} & \shortstack{Success\\rate} & \shortstack{Calls to\\success} & \shortstack{Simulator\\calls} \\
\cmidrule(lr){2-3}\cmidrule(lr){4-5}
Variant & all (108) & blend (54) & all & blend & blend & all & per target \\
\midrule\endfirsthead
\multicolumn{8}{l}{Table~\thetable\ (continued)}\\
\toprule
 & \multicolumn{2}{c}{Runs met} & \multicolumn{2}{c}{Design error} & \shortstack{Success\\rate} & \shortstack{Calls to\\success} & \shortstack{Simulator\\calls} \\
\cmidrule(lr){2-3}\cmidrule(lr){4-5}
Variant & all (108) & blend (54) & all & blend & blend & all & per target \\
\midrule\endhead
\midrule\multicolumn{8}{r}{Continued on next page}\\\endfoot
\bottomrule\endlastfoot

\multicolumn{8}{l}{\textit{Reference}} \\
Learned-table search & 84 & 32 & 0.00639 (0.00065) & 0.01046 (0.00098) & 0.593 (0.222) & 13.19 (0.22) & 17.03 \\
\addlinespace[2pt]
\multicolumn{8}{l}{\textit{Learned component}} \\
Catalogue-table search & 78 & 24 & 0.00800 (---) & 0.01391 (---) & 0.444 (---) & 13.86 (---) & 17.86 \\
Categorical-encoding search & 79 & 25 & 0.01053 (0.00168) & 0.01904 (0.00350) & 0.463 (0.333) & 14.22 (2.75) & 17.48 \\
\addlinespace[2pt]
\multicolumn{8}{l}{\textit{Role of the learned tables}} \\
Learned tables for ranking only & 84 & 30 & 0.00668 (0.00039) & 0.01167 (0.00085) & 0.556 (0.167) & 13.07 (1.31) & 17.09 \\
Learned tables for refinement only & 80 & 26 & 0.00876 (0.00100) & 0.01526 (0.00208) & 0.481 (0.222) & 13.84 (1.08) & 17.86 \\
\addlinespace[2pt]
\multicolumn{8}{l}{\textit{Simulator feedback}} \\
Uncorrected-proposal control & 65 & 17 & 0.01435 (0.00194) & 0.02457 (0.00366) & 0.315 (0.056) & 15.35 (1.92) & 18.68 \\
No-residual control & 65 & 17 & 0.01375 (0.00249) & 0.02338 (0.00477) & 0.315 (0.056) & 15.35 (1.92) & 18.68 \\
Pattern-search refinement & 59 & 11 & 0.01630 (0.00476) & 0.02872 (0.00749) & 0.204 (0.278) & 16.87 (3.50) & 20.11 \\
\addlinespace[2pt]
\multicolumn{8}{l}{\textit{Architectures searched}} \\
Three architectures with learned tables & 71 & 28 & 0.01377 (0.00408) & 0.02080 (0.01252) & 0.519 (0.056) & 13.81 (0.75) & 13.49 \\
Three-architecture catalogue search & 84 & 30 & 0.01010 (---) & 0.01779 (---) & 0.556 (---) & 9.94 (---) & 10.81 \\
First batch of 3 architectures & 83 & 32 & 0.00686 (0.00063) & 0.01113 (0.00177) & 0.593 (0.167) & 12.34 (1.33) & 13.32 \\
First batch of 6 architectures & 87 & 35 & 0.00625 (0.00193) & 0.00995 (0.00311) & 0.648 (0.222) & 12.01 (1.53) & 14.02 \\
First batch of 15 architectures & 79 & 27 & 0.00724 (0.00113) & 0.01216 (0.00336) & 0.500 (0.111) & 14.63 (1.17) & 20.84 \\
\addlinespace[2pt]
\multicolumn{8}{l}{\textit{Ensemble size}} \\
1 MLP & 80 & 28 & 0.00727 (0.00292) & 0.01266 (0.00609) & 0.519 (0.167) & 14.12 (0.58) & 17.75 \\
9 MLPs & 82 & 29 & 0.00700 (0.00058) & 0.01175 (0.00177) & 0.537 (0.056) & 12.83 (0.97) & 16.84 \\
\addlinespace[2pt]
\multicolumn{8}{l}{\textit{Catalogue starts}} \\
Pattern-search starts with corrected refinement & 45 & 18 & 0.02049 (---) & 0.02604 (---) & 0.333 (---) & 20.69 (---) & 17.47 \\
Retrieval with pattern search & 23 & 2 & 0.03264 (0.00180) & 0.04489 (0.00357) & 0.037 (0.056) & 26.47 (0.39) & 29.56 \\
\addlinespace[2pt]
\multicolumn{8}{l}{\textit{Controls}} \\
Retrieval control & 0 & 0 & 0.05352 (---) & 0.07027 (---) & 0.000 (---) & 31.00 (---) & 0.00 \\
Random control & 2 & 0 & 0.14299 (0.02949) & 0.18540 (0.05899) & 0.000 (0.000) & 30.76 (0.58) & 30.00 \\

\end{longtable}
\endgroup

\begingroup\scriptsize\setlength{\tabcolsep}{1pt}
\setlength{\LTleft}{0pt}\setlength{\LTright}{0pt plus 1fill}\setlength{\LTcapwidth}{\textwidth}
\begin{longtable}{@{}L{29mm}C{12mm}C{20mm}C{14mm}C{12mm}C{20mm}C{14mm}@{}}
\caption{Component comparisons and sensitivity analyses on TPMS and strut targets. Each system has 27 test runs. TPMS denotes triply periodic minimal surface}
\label{tab:ablation_pure}\\
\toprule
 & \multicolumn{3}{c}{TPMS targets (27 runs)} & \multicolumn{3}{c}{Strut targets (27 runs)} \\
\cmidrule(lr){2-4}\cmidrule(lr){5-7}
Variant & Runs met & Design error & \shortstack{Calls to\\success} & Runs met & Design error & \shortstack{Calls to\\success} \\
\midrule\endfirsthead
\multicolumn{7}{@{}l}{Table~\thetable\ (continued)}\\
\toprule
 & \multicolumn{3}{c}{TPMS targets (27 runs)} & \multicolumn{3}{c}{Strut targets (27 runs)} \\
\cmidrule(lr){2-4}\cmidrule(lr){5-7}
Variant & Runs met & Design error & \shortstack{Calls to\\success} & Runs met & Design error & \shortstack{Calls to\\success} \\
\midrule\endhead
\bottomrule\endfoot
\multicolumn{7}{l}{\textit{Reference}} \\*
\shortstack[l]{Learned-table\\search} & 25 & 0.00463 & 7.37 & 27 & $<10^{-4}$ & 2.96 \\
\addlinespace[2pt]
\multicolumn{7}{l}{\textit{Learned component}} \\*
\shortstack[l]{Catalogue-table\\search} & 27 & 0.00419 & 4.67 & 27 & $<10^{-4}$ & 4.33 \\
\shortstack[l]{Categorical-encoding\\search} & 27 & 0.00405 & 7.56 & 27 & $<10^{-4}$ & 2.19 \\
\addlinespace[2pt]
\multicolumn{7}{l}{\textit{Role of the learned tables}} \\*
\shortstack[l]{Learned tables for\\ranking only} & 27 & 0.00338 & 5.78 & 27 & $<10^{-4}$ & 3.52 \\
\shortstack[l]{Learned tables for\\refinement only} & 27 & 0.00453 & 6.00 & 27 & $<10^{-4}$ & 3.78 \\
\addlinespace[2pt]
\multicolumn{7}{l}{\textit{Simulator feedback}} \\*
\shortstack[l]{Uncorrected-proposal\\control} & 24 & 0.00542 & 7.74 & 24 & 0.00283 & 5.19 \\
No-residual control & 24 & 0.00542 & 7.74 & 24 & 0.00283 & 5.19 \\
Pattern-search refinement & 23 & 0.00586 & 10.81 & 25 & 0.00189 & 5.04 \\
\addlinespace[2pt]
\multicolumn{7}{l}{\textit{Architectures searched}} \\*
\shortstack[l]{Three architectures\\with learned tables} & 20 & 0.00886 & 11.37 & 23 & 0.00464 & 6.04 \\
\shortstack[l]{Three-architecture\\catalogue search} & 27 & 0.00482 & 2.22 & 27 & $<10^{-4}$ & 2.33 \\
First batch of 3 architectures & 24 & 0.00516 & 8.44 & 27 & $<10^{-4}$ & 2.48 \\
First batch of 6 architectures & 25 & 0.00508 & 7.52 & 27 & $<10^{-4}$ & 2.52 \\
First batch of 15 architectures & 25 & 0.00463 & 8.48 & 27 & $<10^{-4}$ & 3.52 \\
\addlinespace[2pt]
\multicolumn{7}{l}{\textit{Ensemble size}} \\*
1 MLP & 25 & 0.00378 & 8.00 & 27 & $<10^{-4}$ & 2.78 \\
9 MLPs & 26 & 0.00449 & 6.48 & 27 & $<10^{-4}$ & 3.04 \\
\addlinespace[2pt]
\multicolumn{7}{l}{\textit{Catalogue starts}} \\*
\shortstack[l]{Pattern-search\\starts with corrected\\refinement} & 15 & 0.01356 & 15.11 & 12 & 0.01633 & 19.78 \\
\shortstack[l]{Retrieval with\\pattern search} & 9 & 0.02038 & 23.56 & 12 & 0.02040 & 21.22 \\
\addlinespace[2pt]
\multicolumn{7}{l}{\textit{Controls}} \\*
Retrieval control & 0 & 0.03111 & 31.00 & 0 & 0.04244 & 31.00 \\
Random control & 0 & 0.08799 & 31.00 & 2 & 0.11318 & 30.04 \\
\end{longtable}\endgroup

\begingroup\fontsize{7}{8.5}\selectfont
\setlength{\tabcolsep}{2pt}
\setlength{\LTleft}{0pt}\setlength{\LTright}{0pt plus 1fill}
\setlength{\LTcapwidth}{\textwidth}

\begin{longtable}{@{}L{32mm}C{16mm}C{29mm}C{24mm}C{20mm}@{}}
\caption{Paired design-error contrasts over all test targets and over blends. Differences subtract the mean of \(b\) from that of \(a\). \emph{Exceeds both seed ranges} indicates that the absolute difference exceeds both seed ranges. The 95\% intervals bootstrap paired target means. Sign and Wilcoxon signed-rank tests \citep{wilcoxon1945} use paired target means and omit ties. Boundary intervals carry no information about new targets.}\label{tab:contrasts_design_error}\\

\toprule
Contrast (\(a-b\)) & Difference & 95\% interval & Targets better / worse / tied & Sign / Wilcoxon \(p\) \\ \midrule\endfirsthead
\multicolumn{5}{l}{Table \thetable\ (continued)}\\
\toprule
Contrast (\(a-b\)) & Difference & 95\% interval & Targets better / worse / tied & Sign / Wilcoxon \(p\) \\ \midrule\endhead
\midrule\multicolumn{5}{r}{Continued on next page}\\\endfoot
\bottomrule\endlastfoot
Learned-table search $-$ retrieval with pattern search & $-$0.02625 & [$-$0.03751, $-$0.01697] & 32 / 0 / 4 & $<$0.001 / $<$0.001 \\*
\multicolumn{5}{@{}p{\textwidth}@{}}{All targets have runs met 84 versus 23, with 61 versus 0 met by one variant only. Seed ranges are 0.00065 and 0.00180. The difference exceeds both seed ranges.} \\ \addlinespace[2pt]
Learned-table search $-$ retrieval with pattern search & $-$0.03443 & [$-$0.05509, $-$0.01820] & 18 / 0 / 0 & $<$0.001 / $<$0.001 \\*
\multicolumn{5}{@{}p{\textwidth}@{}}{Blend targets have runs met 32 versus 2, with 30 versus 0 met by one variant only. Seed ranges are 0.00098 and 0.00357. The difference exceeds both seed ranges.} \\ \addlinespace[2pt]
Learned-table search $-$ catalogue-table search & $-$0.00161 & [$-$0.00605, $+$0.00122] & 13 / 11 / 12 & 0.839 / 0.900 \\*
\multicolumn{5}{@{}p{\textwidth}@{}}{All targets have runs met 84 versus 78, with 9 versus 3 met by one variant only. Seed ranges are 0.00065 and 0.00000. The difference exceeds both seed ranges.} \\ \addlinespace[2pt]
Learned-table search $-$ catalogue-table search & $-$0.00344 & [$-$0.01212, $+$0.00183] & 10 / 7 / 1 & 0.629 / 0.747 \\*
\multicolumn{5}{@{}p{\textwidth}@{}}{Blend targets have runs met 32 versus 24, with 9 versus 1 met by one variant only. Seed ranges are 0.00098 and 0.00000. The difference exceeds both seed ranges.} \\ \addlinespace[2pt]
Learned-table search $-$ categorical-encoding search & $-$0.00414 & [$-$0.00996, $-$0.00023] & 20 / 7 / 9 & 0.019 / 0.032 \\*
\multicolumn{5}{@{}p{\textwidth}@{}}{All targets have runs met 84 versus 79, with 10 versus 5 met by one variant only. Seed ranges are 0.00065 and 0.00168. The difference exceeds both seed ranges.} \\ \addlinespace[2pt]
Learned-table search $-$ categorical-encoding search & $-$0.00857 & [$-$0.02044, $-$0.00096] & 15 / 3 / 0 & 0.008 / 0.006 \\*
\multicolumn{5}{@{}p{\textwidth}@{}}{Blend targets have runs met 32 versus 25, with 10 versus 3 met by one variant only. Seed ranges are 0.00098 and 0.00350. The difference exceeds both seed ranges.} \\ \addlinespace[2pt]
Categorical-encoding search $-$ catalogue-table search & $+$0.00253 & [$-$0.00132, $+$0.00821] & 10 / 16 / 10 & 0.327 / 0.165 \\*
\multicolumn{5}{@{}p{\textwidth}@{}}{All targets have runs met 79 versus 78, with 8 versus 7 met by one variant only. Seed ranges are 0.00168 and 0.00000. The difference exceeds both seed ranges.} \\ \addlinespace[2pt]
Categorical-encoding search $-$ catalogue-table search & $+$0.00513 & [$-$0.00268, $+$0.01644] & 7 / 11 / 0 & 0.481 / 0.154 \\*
\multicolumn{5}{@{}p{\textwidth}@{}}{Blend targets have runs met 25 versus 24, with 8 versus 7 met by one variant only. Seed ranges are 0.00350 and 0.00000. The difference exceeds both seed ranges.} \\ \addlinespace[2pt]
Learned tables for ranking only $-$ catalogue-table search & $-$0.00132 & [$-$0.00684, $+$0.00246] & 13 / 11 / 12 & 0.839 / 0.790 \\*
\multicolumn{5}{@{}p{\textwidth}@{}}{All targets have runs met 84 versus 78, with 8 versus 2 met by one variant only. Seed ranges are 0.00039 and 0.00000. The difference exceeds both seed ranges.} \\ \addlinespace[2pt]
Learned tables for ranking only $-$ catalogue-table search & $-$0.00224 & [$-$0.01325, $+$0.00536] & 9 / 8 / 1 & 1.000 / 1.000 \\*
\multicolumn{5}{@{}p{\textwidth}@{}}{Blend targets have runs met 30 versus 24, with 8 versus 2 met by one variant only. Seed ranges are 0.00085 and 0.00000. The difference exceeds both seed ranges.} \\ \addlinespace[2pt]
Learned tables for refinement only $-$ catalogue-table search & $+$0.00076 & [$-$0.00025, $+$0.00211] & 5 / 10 / 21 & 0.302 / 0.188 \\*
\multicolumn{5}{@{}p{\textwidth}@{}}{All targets have runs met 80 versus 78, with 4 versus 2 met by one variant only. Seed ranges are 0.00100 and 0.00000. The difference does not exceed both seed ranges.} \\ \addlinespace[2pt]
Learned tables for refinement only $-$ catalogue-table search & $+$0.00136 & [$-$0.00067, $+$0.00405] & 5 / 8 / 5 & 0.581 / 0.340 \\*
\multicolumn{5}{@{}p{\textwidth}@{}}{Blend targets have runs met 26 versus 24, with 4 versus 2 met by one variant only. Seed ranges are 0.00208 and 0.00000. The difference does not exceed both seed ranges.} \\ \addlinespace[2pt]
Learned-table search $-$ uncorrected-proposal control & $-$0.00796 & [$-$0.01338, $-$0.00355] & 20 / 0 / 16 & $<$0.001 / $<$0.001 \\*
\multicolumn{5}{@{}p{\textwidth}@{}}{All targets have runs met 84 versus 65, with 20 versus 1 met by one variant only. Seed ranges are 0.00065 and 0.00194. The difference exceeds both seed ranges.} \\ \addlinespace[2pt]
Learned-table search $-$ uncorrected-proposal control & $-$0.01411 & [$-$0.02447, $-$0.00557] & 16 / 0 / 2 & $<$0.001 / $<$0.001 \\*
\multicolumn{5}{@{}p{\textwidth}@{}}{Blend targets have runs met 32 versus 17, with 16 versus 1 met by one variant only. Seed ranges are 0.00098 and 0.00366. The difference exceeds both seed ranges.} \\ \addlinespace[2pt]
Learned-table search $-$ no-residual control & $-$0.00736 & [$-$0.01263, $-$0.00318] & 20 / 0 / 16 & $<$0.001 / $<$0.001 \\*
\multicolumn{5}{@{}p{\textwidth}@{}}{All targets have runs met 84 versus 65, with 20 versus 1 met by one variant only. Seed ranges are 0.00065 and 0.00249. The difference exceeds both seed ranges.} \\ \addlinespace[2pt]
Learned-table search $-$ no-residual control & $-$0.01291 & [$-$0.02304, $-$0.00488] & 16 / 0 / 2 & $<$0.001 / $<$0.001 \\*
\multicolumn{5}{@{}p{\textwidth}@{}}{Blend targets have runs met 32 versus 17, with 16 versus 1 met by one variant only. Seed ranges are 0.00098 and 0.00477. The difference exceeds both seed ranges.} \\ \addlinespace[2pt]
Learned-table search $-$ pattern-search refinement & $-$0.00991 & [$-$0.01644, $-$0.00466] & 20 / 0 / 16 & $<$0.001 / $<$0.001 \\*
\multicolumn{5}{@{}p{\textwidth}@{}}{All targets have runs met 84 versus 59, with 25 versus 0 met by one variant only. Seed ranges are 0.00065 and 0.00476. The difference exceeds both seed ranges.} \\ \addlinespace[2pt]
Learned-table search $-$ pattern-search refinement & $-$0.01825 & [$-$0.03095, $-$0.00782] & 16 / 0 / 2 & $<$0.001 / $<$0.001 \\*
\multicolumn{5}{@{}p{\textwidth}@{}}{Blend targets have runs met 32 versus 11, with 21 versus 0 met by one variant only. Seed ranges are 0.00098 and 0.00749. The difference exceeds both seed ranges.} \\ \addlinespace[2pt]
Learned-table search $-$ three architectures with learned tables & $-$0.00738 & [$-$0.01240, $-$0.00331] & 23 / 3 / 10 & $<$0.001 / $<$0.001 \\*
\multicolumn{5}{@{}p{\textwidth}@{}}{All targets have runs met 84 versus 71, with 17 versus 4 met by one variant only. Seed ranges are 0.00065 and 0.00408. The difference exceeds both seed ranges.} \\ \addlinespace[2pt]
Learned-table search $-$ three architectures with learned tables & $-$0.01034 & [$-$0.01964, $-$0.00321] & 14 / 2 / 2 & 0.004 / 0.002 \\*
\multicolumn{5}{@{}p{\textwidth}@{}}{Blend targets have runs met 32 versus 28, with 7 versus 3 met by one variant only. Seed ranges are 0.00098 and 0.01252. The difference does not exceed both seed ranges.} \\ \addlinespace[2pt]
Learned-table search $-$ three-architecture catalogue search & $-$0.00371 & [$-$0.00903, $+$0.00039] & 15 / 9 / 12 & 0.307 / 0.317 \\*
\multicolumn{5}{@{}p{\textwidth}@{}}{All targets have runs met 84 versus 84, with 5 versus 5 met by one variant only. Seed ranges are 0.00065 and 0.00000. The difference exceeds both seed ranges.} \\ \addlinespace[2pt]
Learned-table search $-$ three-architecture catalogue search & $-$0.00733 & [$-$0.01754, $+$0.00061] & 11 / 6 / 1 & 0.332 / 0.306 \\*
\multicolumn{5}{@{}p{\textwidth}@{}}{Blend targets have runs met 32 versus 30, with 5 versus 3 met by one variant only. Seed ranges are 0.00098 and 0.00000. The difference exceeds both seed ranges.} \\ \addlinespace[2pt]
Three-architecture catalogue search $-$ catalogue-table search & $+$0.00210 & [$-$0.00059, $+$0.00598] & 3 / 5 / 28 & 0.727 / 0.461 \\*
\multicolumn{5}{@{}p{\textwidth}@{}}{All targets have runs met 84 versus 78, with 6 versus 0 met by one variant only. Seed ranges are 0.00000 and 0.00000. The difference exceeds both seed ranges.} \\ \addlinespace[2pt]
Three-architecture catalogue search $-$ catalogue-table search & $+$0.00388 & [$-$0.00143, $+$0.01165] & 3 / 3 / 12 & 1.000 / 0.688 \\*
\multicolumn{5}{@{}p{\textwidth}@{}}{Blend targets have runs met 30 versus 24, with 6 versus 0 met by one variant only. Seed ranges are 0.00000 and 0.00000. The difference exceeds both seed ranges.} \\ \addlinespace[2pt]
First batch of 3 $-$ Learned-table search & $+$0.00046 & [$-$0.00032, $+$0.00138] & 4 / 10 / 22 & 0.180 / 0.104 \\*
\multicolumn{5}{@{}p{\textwidth}@{}}{All targets have runs met 83 versus 84, with 2 versus 3 met by one variant only. Seed ranges are 0.00063 and 0.00065. The difference does not exceed both seed ranges.} \\ \addlinespace[2pt]
First batch of 3 $-$ Learned-table search & $+$0.00066 & [$-$0.00090, $+$0.00245] & 4 / 7 / 7 & 0.549 / 0.320 \\*
\multicolumn{5}{@{}p{\textwidth}@{}}{Blend targets have runs met 32 versus 32, with 2 versus 2 met by one variant only. Seed ranges are 0.00177 and 0.00098. The difference does not exceed both seed ranges.} \\ \addlinespace[2pt]
First batch of 6 $-$ Learned-table search & $-$0.00014 & [$-$0.00078, $+$0.00041] & 6 / 5 / 25 & 1.000 / 1.000 \\*
\multicolumn{5}{@{}p{\textwidth}@{}}{All targets have runs met 87 versus 84, with 3 versus 0 met by one variant only. Seed ranges are 0.00193 and 0.00065. The difference does not exceed both seed ranges.} \\ \addlinespace[2pt]
First batch of 6 $-$ Learned-table search & $-$0.00051 & [$-$0.00173, $+$0.00057] & 6 / 3 / 9 & 0.508 / 0.496 \\*
\multicolumn{5}{@{}p{\textwidth}@{}}{Blend targets have runs met 35 versus 32, with 3 versus 0 met by one variant only. Seed ranges are 0.00311 and 0.00098. The difference does not exceed both seed ranges.} \\ \addlinespace[2pt]
First batch of 15 $-$ Learned-table search & $+$0.00085 & [$+$0.00037, $+$0.00143] & 0 / 10 / 26 & 0.002 / 0.002 \\*
\multicolumn{5}{@{}p{\textwidth}@{}}{All targets have runs met 79 versus 84, with 0 versus 5 met by one variant only. Seed ranges are 0.00113 and 0.00065. The difference does not exceed both seed ranges.} \\ \addlinespace[2pt]
First batch of 15 $-$ Learned-table search & $+$0.00169 & [$+$0.00074, $+$0.00284] & 0 / 10 / 8 & 0.002 / 0.002 \\*
\multicolumn{5}{@{}p{\textwidth}@{}}{Blend targets have runs met 27 versus 32, with 0 versus 5 met by one variant only. Seed ranges are 0.00336 and 0.00098. The difference does not exceed both seed ranges.} \\ \addlinespace[2pt]
1 MLP $-$ Learned-table search & $+$0.00088 & [$-$0.00044, $+$0.00250] & 11 / 14 / 11 & 0.690 / 0.791 \\*
\multicolumn{5}{@{}p{\textwidth}@{}}{All targets have runs met 80 versus 84, with 5 versus 9 met by one variant only. Seed ranges are 0.00292 and 0.00065. The difference does not exceed both seed ranges.} \\ \addlinespace[2pt]
1 MLP $-$ Learned-table search & $+$0.00219 & [$-$0.00028, $+$0.00533] & 7 / 10 / 1 & 0.629 / 0.459 \\*
\multicolumn{5}{@{}p{\textwidth}@{}}{Blend targets have runs met 28 versus 32, with 4 versus 8 met by one variant only. Seed ranges are 0.00609 and 0.00098. The difference does not exceed both seed ranges.} \\ \addlinespace[2pt]
9 MLPs $-$ Learned-table search & $+$0.00061 & [$-$0.00115, $+$0.00331] & 13 / 10 / 13 & 0.678 / 0.482 \\*
\multicolumn{5}{@{}p{\textwidth}@{}}{All targets have runs met 82 versus 84, with 2 versus 4 met by one variant only. Seed ranges are 0.00058 and 0.00065. The difference does not exceed both seed ranges.} \\ \addlinespace[2pt]
9 MLPs $-$ Learned-table search & $+$0.00129 & [$-$0.00203, $+$0.00663] & 10 / 7 / 1 & 0.629 / 0.459 \\*
\multicolumn{5}{@{}p{\textwidth}@{}}{Blend targets have runs met 29 versus 32, with 1 versus 4 met by one variant only. Seed ranges are 0.00177 and 0.00098. The difference does not exceed both seed ranges.} \\ \addlinespace[2pt]
Pattern-search starts with corrected refinement $-$ retrieval with pattern search & $-$0.01215 & [$-$0.01829, $-$0.00722] & 26 / 2 / 8 & $<$0.001 / $<$0.001 \\*
\multicolumn{5}{@{}p{\textwidth}@{}}{All targets have runs met 45 versus 23, with 22 versus 0 met by one variant only. Seed ranges are 0.00000 and 0.00180. The difference exceeds both seed ranges.} \\ \addlinespace[2pt]
Pattern-search starts with corrected refinement $-$ retrieval with pattern search & $-$0.01885 & [$-$0.03044, $-$0.00964] & 17 / 0 / 1 & $<$0.001 / $<$0.001 \\*
\multicolumn{5}{@{}p{\textwidth}@{}}{Blend targets have runs met 18 versus 2, with 16 versus 0 met by one variant only. Seed ranges are 0.00000 and 0.00357. The difference exceeds both seed ranges.} \\ \addlinespace[2pt]
Three-architecture catalogue search $-$ pattern-search starts with corrected refinement & $-$0.01039 & [$-$0.02120, $-$0.00086] & 18 / 11 / 7 & 0.265 / 0.024 \\*
\multicolumn{5}{@{}p{\textwidth}@{}}{All targets have runs met 84 versus 45, with 42 versus 3 met by one variant only. Seed ranges are 0.00000 and 0.00000. The difference exceeds both seed ranges.} \\ \addlinespace[2pt]
Three-architecture catalogue search $-$ pattern-search starts with corrected refinement & $-$0.00825 & [$-$0.02908, $+$0.00962] & 9 / 9 / 0 & 1.000 / 0.766 \\*
\multicolumn{5}{@{}p{\textwidth}@{}}{Blend targets have runs met 30 versus 18, with 15 versus 3 met by one variant only. Seed ranges are 0.00000 and 0.00000. The difference exceeds both seed ranges.} \\ \addlinespace[2pt]
\end{longtable}\endgroup

\begingroup\fontsize{7}{8.5}\selectfont
\setlength{\tabcolsep}{2pt}
\setlength{\LTleft}{0pt}\setlength{\LTright}{0pt plus 1fill}
\setlength{\LTcapwidth}{\textwidth}

\begin{longtable}{@{}L{32mm}C{16mm}C{29mm}C{24mm}C{20mm}@{}}
\caption{Paired success-rate contrasts over all test targets and over blends. Differences subtract the mean of \(b\) from that of \(a\). \emph{Exceeds both seed ranges} indicates that the absolute difference exceeds both seed ranges. The 95\% intervals bootstrap paired target means. The exact sign test uses paired target means and omits ties. Boundary intervals carry no information about new targets.}\label{tab:contrasts_success_rate}\\

\toprule
Contrast (\(a-b\)) & Difference & 95\% interval & Targets better / worse / tied & Sign \(p\) \\ \midrule\endfirsthead
\multicolumn{5}{l}{Table \thetable\ (continued)}\\
\toprule
Contrast (\(a-b\)) & Difference & 95\% interval & Targets better / worse / tied & Sign \(p\) \\ \midrule\endhead
\midrule\multicolumn{5}{r}{Continued on next page}\\\endfoot
\bottomrule\endlastfoot
Learned-table search $-$ retrieval with pattern search & $+$0.565 & [$+$0.417, $+$0.704] & 25 / 0 / 11 & $<$0.001 \\*
\multicolumn{5}{@{}p{\textwidth}@{}}{All targets have runs met 84 versus 23, with 61 versus 0 met by one variant only. Seed ranges are 0.139 and 0.028. The difference exceeds both seed ranges.} \\ \addlinespace[2pt]
Learned-table search $-$ retrieval with pattern search & $+$0.556 & [$+$0.370, $+$0.741] & 14 / 0 / 4 & $<$0.001 \\*
\multicolumn{5}{@{}p{\textwidth}@{}}{Blend targets have runs met 32 versus 2, with 30 versus 0 met by one variant only. Seed ranges are 0.222 and 0.056. The difference exceeds both seed ranges.} \\ \addlinespace[2pt]
Learned-table search $-$ catalogue-table search & $+$0.056 & [$-$0.019, $+$0.139] & 6 / 2 / 28 & 0.289 \\*
\multicolumn{5}{@{}p{\textwidth}@{}}{All targets have runs met 84 versus 78, with 9 versus 3 met by one variant only. Seed ranges are 0.139 and 0.000. The difference does not exceed both seed ranges.} \\ \addlinespace[2pt]
Learned-table search $-$ catalogue-table search & $+$0.148 & [$+$0.019, $+$0.296] & 6 / 1 / 11 & 0.125 \\*
\multicolumn{5}{@{}p{\textwidth}@{}}{Blend targets have runs met 32 versus 24, with 9 versus 1 met by one variant only. Seed ranges are 0.222 and 0.000. The difference does not exceed both seed ranges.} \\ \addlinespace[2pt]
Learned-table search $-$ categorical-encoding search & $+$0.046 & [$-$0.037, $+$0.120] & 9 / 3 / 24 & 0.146 \\*
\multicolumn{5}{@{}p{\textwidth}@{}}{All targets have runs met 84 versus 79, with 10 versus 5 met by one variant only. Seed ranges are 0.139 and 0.167. The difference does not exceed both seed ranges.} \\ \addlinespace[2pt]
Learned-table search $-$ categorical-encoding search & $+$0.130 & [$-$0.019, $+$0.259] & 9 / 2 / 7 & 0.065 \\*
\multicolumn{5}{@{}p{\textwidth}@{}}{Blend targets have runs met 32 versus 25, with 10 versus 3 met by one variant only. Seed ranges are 0.222 and 0.333. The difference does not exceed both seed ranges.} \\ \addlinespace[2pt]
Categorical-encoding search $-$ catalogue-table search & $+$0.009 & [$-$0.083, $+$0.102] & 5 / 5 / 26 & 1.000 \\*
\multicolumn{5}{@{}p{\textwidth}@{}}{All targets have runs met 79 versus 78, with 8 versus 7 met by one variant only. Seed ranges are 0.167 and 0.000. The difference does not exceed both seed ranges.} \\ \addlinespace[2pt]
Categorical-encoding search $-$ catalogue-table search & $+$0.019 & [$-$0.167, $+$0.222] & 5 / 5 / 8 & 1.000 \\*
\multicolumn{5}{@{}p{\textwidth}@{}}{Blend targets have runs met 25 versus 24, with 8 versus 7 met by one variant only. Seed ranges are 0.333 and 0.000. The difference does not exceed both seed ranges.} \\ \addlinespace[2pt]
Learned tables for ranking only $-$ catalogue-table search & $+$0.056 & [$-$0.019, $+$0.139] & 4 / 1 / 31 & 0.375 \\*
\multicolumn{5}{@{}p{\textwidth}@{}}{All targets have runs met 84 versus 78, with 8 versus 2 met by one variant only. Seed ranges are 0.083 and 0.000. The difference does not exceed both seed ranges.} \\ \addlinespace[2pt]
Learned tables for ranking only $-$ catalogue-table search & $+$0.111 & [$-$0.037, $+$0.278] & 4 / 1 / 13 & 0.375 \\*
\multicolumn{5}{@{}p{\textwidth}@{}}{Blend targets have runs met 30 versus 24, with 8 versus 2 met by one variant only. Seed ranges are 0.167 and 0.000. The difference does not exceed both seed ranges.} \\ \addlinespace[2pt]
Learned tables for refinement only $-$ catalogue-table search & $+$0.019 & [$-$0.037, $+$0.074] & 3 / 1 / 32 & 0.625 \\*
\multicolumn{5}{@{}p{\textwidth}@{}}{All targets have runs met 80 versus 78, with 4 versus 2 met by one variant only. Seed ranges are 0.111 and 0.000. The difference does not exceed both seed ranges.} \\ \addlinespace[2pt]
Learned tables for refinement only $-$ catalogue-table search & $+$0.037 & [$-$0.074, $+$0.148] & 3 / 1 / 14 & 0.625 \\*
\multicolumn{5}{@{}p{\textwidth}@{}}{Blend targets have runs met 26 versus 24, with 4 versus 2 met by one variant only. Seed ranges are 0.222 and 0.000. The difference does not exceed both seed ranges.} \\ \addlinespace[2pt]
Learned-table search $-$ uncorrected-proposal control & $+$0.176 & [$+$0.093, $+$0.278] & 11 / 0 / 25 & $<$0.001 \\*
\multicolumn{5}{@{}p{\textwidth}@{}}{All targets have runs met 84 versus 65, with 20 versus 1 met by one variant only. Seed ranges are 0.139 and 0.028. The difference exceeds both seed ranges.} \\ \addlinespace[2pt]
Learned-table search $-$ uncorrected-proposal control & $+$0.278 & [$+$0.130, $+$0.444] & 9 / 0 / 9 & 0.004 \\*
\multicolumn{5}{@{}p{\textwidth}@{}}{Blend targets have runs met 32 versus 17, with 16 versus 1 met by one variant only. Seed ranges are 0.222 and 0.056. The difference exceeds both seed ranges.} \\ \addlinespace[2pt]
Learned-table search $-$ no-residual control & $+$0.176 & [$+$0.093, $+$0.278] & 11 / 0 / 25 & $<$0.001 \\*
\multicolumn{5}{@{}p{\textwidth}@{}}{All targets have runs met 84 versus 65, with 20 versus 1 met by one variant only. Seed ranges are 0.139 and 0.028. The difference exceeds both seed ranges.} \\ \addlinespace[2pt]
Learned-table search $-$ no-residual control & $+$0.278 & [$+$0.130, $+$0.444] & 9 / 0 / 9 & 0.004 \\*
\multicolumn{5}{@{}p{\textwidth}@{}}{Blend targets have runs met 32 versus 17, with 16 versus 1 met by one variant only. Seed ranges are 0.222 and 0.056. The difference exceeds both seed ranges.} \\ \addlinespace[2pt]
Learned-table search $-$ pattern-search refinement & $+$0.231 & [$+$0.148, $+$0.315] & 16 / 0 / 20 & $<$0.001 \\*
\multicolumn{5}{@{}p{\textwidth}@{}}{All targets have runs met 84 versus 59, with 25 versus 0 met by one variant only. Seed ranges are 0.139 and 0.167. The difference exceeds both seed ranges.} \\ \addlinespace[2pt]
Learned-table search $-$ pattern-search refinement & $+$0.389 & [$+$0.241, $+$0.537] & 13 / 0 / 5 & $<$0.001 \\*
\multicolumn{5}{@{}p{\textwidth}@{}}{Blend targets have runs met 32 versus 11, with 21 versus 0 met by one variant only. Seed ranges are 0.222 and 0.278. The difference exceeds both seed ranges.} \\ \addlinespace[2pt]
Learned-table search $-$ three architectures with learned tables & $+$0.120 & [$+$0.037, $+$0.204] & 10 / 2 / 24 & 0.039 \\*
\multicolumn{5}{@{}p{\textwidth}@{}}{All targets have runs met 84 versus 71, with 17 versus 4 met by one variant only. Seed ranges are 0.139 and 0.028. The difference does not exceed both seed ranges.} \\ \addlinespace[2pt]
Learned-table search $-$ three architectures with learned tables & $+$0.074 & [$-$0.019, $+$0.185] & 4 / 1 / 13 & 0.375 \\*
\multicolumn{5}{@{}p{\textwidth}@{}}{Blend targets have runs met 32 versus 28, with 7 versus 3 met by one variant only. Seed ranges are 0.222 and 0.056. The difference does not exceed both seed ranges.} \\ \addlinespace[2pt]
Learned-table search $-$ three-architecture catalogue search & 0.000 & [$-$0.074, $+$0.074] & 4 / 3 / 29 & 1.000 \\*
\multicolumn{5}{@{}p{\textwidth}@{}}{All targets have runs met 84 versus 84, with 5 versus 5 met by one variant only. Seed ranges are 0.139 and 0.000. The difference does not exceed both seed ranges.} \\ \addlinespace[2pt]
Learned-table search $-$ three-architecture catalogue search & $+$0.037 & [$-$0.093, $+$0.167] & 4 / 2 / 12 & 0.688 \\*
\multicolumn{5}{@{}p{\textwidth}@{}}{Blend targets have runs met 32 versus 30, with 5 versus 3 met by one variant only. Seed ranges are 0.222 and 0.000. The difference does not exceed both seed ranges.} \\ \addlinespace[2pt]
Three-architecture catalogue search $-$ catalogue-table search & $+$0.056 & [0.000, $+$0.139] & 2 / 0 / 34 & 0.500 \\*
\multicolumn{5}{@{}p{\textwidth}@{}}{All targets have runs met 84 versus 78, with 6 versus 0 met by one variant only. Seed ranges are 0.000 and 0.000. The difference exceeds both seed ranges.} \\ \addlinespace[2pt]
Three-architecture catalogue search $-$ catalogue-table search & $+$0.111 & [0.000, $+$0.278] & 2 / 0 / 16 & 0.500 \\*
\multicolumn{5}{@{}p{\textwidth}@{}}{Blend targets have runs met 30 versus 24, with 6 versus 0 met by one variant only. Seed ranges are 0.000 and 0.000. The difference exceeds both seed ranges.} \\ \addlinespace[2pt]
First batch of 3 $-$ Learned-table search & $-$0.009 & [$-$0.037, $+$0.019] & 1 / 2 / 33 & 1.000 \\*
\multicolumn{5}{@{}p{\textwidth}@{}}{All targets have runs met 83 versus 84, with 2 versus 3 met by one variant only. Seed ranges are 0.111 and 0.139. The difference does not exceed both seed ranges.} \\ \addlinespace[2pt]
First batch of 3 $-$ Learned-table search & 0.000 & [$-$0.056, $+$0.056] & 1 / 1 / 16 & 1.000 \\*
\multicolumn{5}{@{}p{\textwidth}@{}}{Blend targets have runs met 32 versus 32, with 2 versus 2 met by one variant only. Seed ranges are 0.167 and 0.222. The difference does not exceed both seed ranges.} \\ \addlinespace[2pt]
First batch of 6 $-$ Learned-table search & $+$0.028 & [0.000, $+$0.056] & 3 / 0 / 33 & 0.250 \\*
\multicolumn{5}{@{}p{\textwidth}@{}}{All targets have runs met 87 versus 84, with 3 versus 0 met by one variant only. Seed ranges are 0.139 and 0.139. The difference does not exceed both seed ranges.} \\ \addlinespace[2pt]
First batch of 6 $-$ Learned-table search & $+$0.056 & [0.000, $+$0.111] & 3 / 0 / 15 & 0.250 \\*
\multicolumn{5}{@{}p{\textwidth}@{}}{Blend targets have runs met 35 versus 32, with 3 versus 0 met by one variant only. Seed ranges are 0.222 and 0.222. The difference does not exceed both seed ranges.} \\ \addlinespace[2pt]
First batch of 15 $-$ Learned-table search & $-$0.046 & [$-$0.093, $-$0.009] & 0 / 4 / 32 & 0.125 \\*
\multicolumn{5}{@{}p{\textwidth}@{}}{All targets have runs met 79 versus 84, with 0 versus 5 met by one variant only. Seed ranges are 0.083 and 0.139. The difference does not exceed both seed ranges.} \\ \addlinespace[2pt]
First batch of 15 $-$ Learned-table search & $-$0.093 & [$-$0.185, $-$0.019] & 0 / 4 / 14 & 0.125 \\*
\multicolumn{5}{@{}p{\textwidth}@{}}{Blend targets have runs met 27 versus 32, with 0 versus 5 met by one variant only. Seed ranges are 0.111 and 0.222. The difference does not exceed both seed ranges.} \\ \addlinespace[2pt]
1 MLP $-$ Learned-table search & $-$0.037 & [$-$0.074, 0.000] & 1 / 5 / 30 & 0.219 \\*
\multicolumn{5}{@{}p{\textwidth}@{}}{All targets have runs met 80 versus 84, with 5 versus 9 met by one variant only. Seed ranges are 0.083 and 0.139. The difference does not exceed both seed ranges.} \\ \addlinespace[2pt]
1 MLP $-$ Learned-table search & $-$0.074 & [$-$0.148, 0.000] & 1 / 5 / 12 & 0.219 \\*
\multicolumn{5}{@{}p{\textwidth}@{}}{Blend targets have runs met 28 versus 32, with 4 versus 8 met by one variant only. Seed ranges are 0.167 and 0.222. The difference does not exceed both seed ranges.} \\ \addlinespace[2pt]
9 MLPs $-$ Learned-table search & $-$0.019 & [$-$0.056, $+$0.009] & 1 / 3 / 32 & 0.625 \\*
\multicolumn{5}{@{}p{\textwidth}@{}}{All targets have runs met 82 versus 84, with 2 versus 4 met by one variant only. Seed ranges are 0.056 and 0.139. The difference does not exceed both seed ranges.} \\ \addlinespace[2pt]
9 MLPs $-$ Learned-table search & $-$0.056 & [$-$0.111, 0.000] & 0 / 3 / 15 & 0.250 \\*
\multicolumn{5}{@{}p{\textwidth}@{}}{Blend targets have runs met 29 versus 32, with 1 versus 4 met by one variant only. Seed ranges are 0.056 and 0.222. The difference does not exceed both seed ranges.} \\ \addlinespace[2pt]
Pattern-search starts with corrected refinement $-$ retrieval with pattern search & $+$0.204 & [$+$0.093, $+$0.333] & 8 / 0 / 28 & 0.008 \\*
\multicolumn{5}{@{}p{\textwidth}@{}}{All targets have runs met 45 versus 23, with 22 versus 0 met by one variant only. Seed ranges are 0.000 and 0.028. The difference exceeds both seed ranges.} \\ \addlinespace[2pt]
Pattern-search starts with corrected refinement $-$ retrieval with pattern search & $+$0.296 & [$+$0.111, $+$0.500] & 6 / 0 / 12 & 0.031 \\*
\multicolumn{5}{@{}p{\textwidth}@{}}{Blend targets have runs met 18 versus 2, with 16 versus 0 met by one variant only. Seed ranges are 0.000 and 0.056. The difference exceeds both seed ranges.} \\ \addlinespace[2pt]
Three-architecture catalogue search $-$ pattern-search starts with corrected refinement & $+$0.361 & [$+$0.194, $+$0.528] & 14 / 1 / 21 & $<$0.001 \\*
\multicolumn{5}{@{}p{\textwidth}@{}}{All targets have runs met 84 versus 45, with 42 versus 3 met by one variant only. Seed ranges are 0.000 and 0.000. The difference exceeds both seed ranges.} \\ \addlinespace[2pt]
Three-architecture catalogue search $-$ pattern-search starts with corrected refinement & $+$0.222 & [0.000, $+$0.444] & 5 / 1 / 12 & 0.219 \\*
\multicolumn{5}{@{}p{\textwidth}@{}}{Blend targets have runs met 30 versus 18, with 15 versus 3 met by one variant only. Seed ranges are 0.000 and 0.000. The difference exceeds both seed ranges.} \\ \addlinespace[2pt]
\end{longtable}\endgroup

\begingroup\scriptsize\setlength{\tabcolsep}{1pt}
\setlength{\LTleft}{0pt}\setlength{\LTright}{0pt plus 1fill}\setlength{\LTcapwidth}{\textwidth}
\begin{longtable}{@{}L{39mm}C{14mm}C{24mm}C{24mm}C{22mm}@{}}
\caption{Test-set results by seed, in order 0, 1 and 2. Each seed contributes 36 runs, including 18 blend runs. The catalogue-only searches and the retrieval control are deterministic}
\label{tab:per_seed}\\
\toprule
Variant & \shortstack{Runs met\\all (36)} & \shortstack{Runs met\\blend (18)} & \shortstack{Design error\\all} & \shortstack{Design error\\blend} \\
\midrule\endfirsthead
\multicolumn{5}{@{}l}{Table~\thetable\ (continued)}\\
\toprule
Variant & \shortstack{Runs met\\all (36)} & \shortstack{Runs met\\blend (18)} & \shortstack{Design error\\all} & \shortstack{Design error\\blend} \\
\midrule\endhead
\bottomrule\endfoot
\shortstack[l]{Learned-table\\search} & 27, 26, 31 & 10, 9, 13 & 0.00645, 0.00669, 0.00604 & 0.01004, 0.01103, 0.01033 \\
\shortstack[l]{Catalogue-table\\search} & 26, 26, 26 & 8, 8, 8 & 0.00800, 0.00800, 0.00800 & 0.01391, 0.01391, 0.01391 \\
\shortstack[l]{Categorical-encoding\\search} & 24, 25, 30 & 6, 7, 12 & 0.00948, 0.01116, 0.01095 & 0.01681, 0.02031, 0.01998 \\
\shortstack[l]{Uncorrected-proposal\\control} & 22, 22, 21 & 6, 6, 5 & 0.01342, 0.01427, 0.01536 & 0.02283, 0.02440, 0.02649 \\
\shortstack[l]{Three-architecture\\catalogue search} & 28, 28, 28 & 10, 10, 10 & 0.01010, 0.01010, 0.01010 & 0.01779, 0.01779, 0.01779 \\
First batch of 6 architectures & 28, 27, 32 & 11, 10, 14 & 0.00698, 0.00672, 0.00505 & 0.01104, 0.01087, 0.00794 \\
\shortstack[l]{Retrieval with\\pattern search} & 8, 8, 7 & 1, 1, 0 & 0.03223, 0.03195, 0.03375 & 0.04409, 0.04351, 0.04708 \\
Random control & 1, 1, 0 & 0, 0, 0 & 0.15070, 0.15388, 0.12440 & 0.19539, 0.20991, 0.15091 \\
\end{longtable}\endgroup

\begingroup\fontsize{7}{8.5}\selectfont\setlength{\tabcolsep}{1pt}

\setlength{\LTleft}{0pt}\setlength{\LTright}{0pt plus 1fill}\begin{longtable}{@{}L{31mm}C{13mm}C{17mm}C{17mm}C{13mm}C{17mm}C{17mm}@{}}
\caption{Verified successes at relative tolerances of 2\%, 3\% and 5\%. The first panel re-scores designs returned under the 2\% stop rule. The second panel gives matched-stop-rule runs, with re-scored counts in parentheses at 3\% and 5\%. Every row uses at most 30 simulator calls per target and zero language-model calls}\label{tab:tolerance_levels}\\
\toprule
 & \multicolumn{3}{c}{All targets (108 runs)} & \multicolumn{3}{c}{Blend targets (54 runs)} \\
\cmidrule(lr){2-4}\cmidrule(lr){5-7}
Variant & 2\,\% & 3\,\% & 5\,\% & 2\,\% & 3\,\% & 5\,\% \\
\midrule\endfirsthead
\multicolumn{7}{l}{Table \thetable\ (continued)}\\
\toprule
 & \multicolumn{3}{c}{All targets (108 runs)} & \multicolumn{3}{c}{Blend targets (54 runs)} \\
\cmidrule(lr){2-4}\cmidrule(lr){5-7}
Variant & 2\,\% & 3\,\% & 5\,\% & 2\,\% & 3\,\% & 5\,\% \\
\midrule\endhead\bottomrule\endfoot

\shortstack[l]{Learned-table\\search} & 84 & 94 & 100 & 32 & 42 & 47 \\
\shortstack[l]{Catalogue-table\\search} & 78 & 99 & 105 & 24 & 45 & 51 \\
\shortstack[l]{Categorical-encoding\\search} & 79 & 91 & 98 & 25 & 37 & 44 \\
\shortstack[l]{Uncorrected-proposal\\control} & 65 & 75 & 87 & 17 & 27 & 36 \\
\shortstack[l]{Three-architecture\\catalogue search} & 84 & 96 & 99 & 30 & 42 & 45 \\
\shortstack[l]{Pattern-search\\starts with corrected\\refinement} & 45 & 51 & 69 & 18 & 24 & 36 \\
\shortstack[l]{Retrieval with\\pattern search} & 23 & 29 & 46 & 2 & 5 & 16 \\
Retrieval control & 0 & 0 & 6 & 0 & 0 & 3 \\
Random control & 2 & 3 & 7 & 0 & 1 & 4 \\
\addlinespace[4pt]
\multicolumn{7}{@{}p{\textwidth}@{}}{\textit{Matched stop rules (re-scored counts in parentheses)}} \\
\multicolumn{7}{l}{\textit{Test set}} \\
Learned-table search & 84 & 92 (94) & 102 (100) & 32 & 40 (42) & 49 (47) \\
Catalogue-table search & 78 & 93 (99) & 105 (105) & 24 & 42 (45) & 51 (51) \\
\shortstack[l]{Three-architecture\\catalogue search} & 84 & 96 (96) & 99 (99) & 30 & 42 (42) & 45 (45) \\
Bracketing search & 72 & 84 (87) & 96 (96) & 21 & 30 (33) & 42 (42) \\
\shortstack[l]{Bayesian\\optimisation} & 71 & 87 (87) & 101 (101) & 21 & 33 (33) & 47 (47) \\
Nelder--Mead search & 69 & 75 (75) & 90 (90) & 18 & 24 (24) & 36 (36) \\
\shortstack[l]{Retrieval with\\pattern search} & 23 & 29 (29) & 46 (46) & 2 & 5 (5) & 16 (16) \\
\multicolumn{7}{l}{\textit{Fresh draw}} \\
Learned-table search & 94 & 98 (99) & 101 (103) & 43 & 47 (48) & 50 (52) \\
Catalogue-table search & 87 & 90 (93) & 102 (102) & 36 & 39 (42) & 51 (51) \\
\shortstack[l]{Three-architecture\\catalogue search} & 93 & 93 (93) & 99 (99) & 45 & 45 (45) & 48 (48) \\
Bracketing search & 87 & 90 (93) & 96 (96) & 36 & 39 (42) & 45 (45) \\
\shortstack[l]{Bayesian\\optimisation} & 84 & 94 (94) & 102 (102) & 34 & 43 (43) & 51 (51) \\
Nelder--Mead search & 78 & 78 (78) & 96 (96) & 30 & 30 (30) & 45 (45) \\
\shortstack[l]{Retrieval with\\pattern search} & 27 & 37 (37) & 62 (62) & 0 & 10 (10) & 26 (26) \\

\end{longtable}\endgroup

\begingroup\scriptsize\setlength{\tabcolsep}{1pt}
\setlength{\LTleft}{0pt}\setlength{\LTright}{0pt plus 1fill}\setlength{\LTcapwidth}{\textwidth}
\begin{longtable}{@{}L{29mm}C{29mm}C{16mm}C{16mm}C{15mm}C{17mm}@{}}
\caption{Development-set outcomes for 12 targets (3 TPMS, 3 strut and 6 blend) with one seed and at most 30 simulator calls per target. TPMS denotes triply periodic minimal surface}
\label{tab:dev_subset}\\
\toprule
Variant & \shortstack{Targets met\\(TPMS / strut\\/ blend)} & \shortstack{Design error\\all} & \shortstack{Design error\\blend} & \shortstack{Calls to\\success} & \shortstack{Simulator\\calls per\\target} \\
\midrule\endfirsthead
\multicolumn{6}{@{}l}{Table~\thetable\ (continued)}\\
\toprule
Variant & \shortstack{Targets met\\(TPMS / strut\\/ blend)} & \shortstack{Design error\\all} & \shortstack{Design error\\blend} & \shortstack{Calls to\\success} & \shortstack{Simulator\\calls per\\target} \\
\midrule\endhead
\bottomrule\endfoot
\multicolumn{6}{l}{\textit{Reference}} \\*
\shortstack[l]{Learned-table\\search} & 10/12 (3 / 3 / 4) & 0.00607 & 0.00986 & 11.17 & 15.17 \\
\addlinespace[2pt]
\multicolumn{6}{l}{\textit{Learned component}} \\*
\shortstack[l]{Catalogue-table\\search} & 11/12 (3 / 3 / 5) & 0.00944 & 0.01649 & 8.17 & 13.67 \\
\shortstack[l]{Categorical-encoding\\search} & 11/12 (3 / 3 / 5) & 0.01149 & 0.01996 & 12.58 & 17.00 \\
\addlinespace[2pt]
\multicolumn{6}{l}{\textit{Role of the learned tables}} \\*
\shortstack[l]{Learned tables for\\ranking only} & 10/12 (3 / 3 / 4) & 0.00602 & 0.00976 & 10.83 & 15.00 \\
\shortstack[l]{Learned tables for\\refinement only} & 11/12 (3 / 3 / 5) & 0.00944 & 0.01649 & 10.33 & 15.75 \\
\addlinespace[2pt]
\multicolumn{6}{l}{\textit{Simulator feedback}} \\*
\shortstack[l]{Uncorrected-proposal\\control} & 8/12 (3 / 2 / 3) & 0.02395 & 0.04001 & 14.75 & 18.33 \\
No-residual control & 8/12 (3 / 2 / 3) & 0.02346 & 0.03904 & 14.75 & 18.33 \\
Pattern-search refinement & 8/12 (3 / 2 / 3) & 0.01471 & 0.02188 & 17.50 & 21.00 \\
\addlinespace[2pt]
\multicolumn{6}{l}{\textit{Architectures searched}} \\*
\shortstack[l]{Three architectures\\with learned tables} & 11/12 (3 / 3 / 5) & 0.00636 & 0.01043 & 6.83 & 7.92 \\
\shortstack[l]{Three-architecture\\catalogue search} & 11/12 (3 / 3 / 5) & 0.00972 & 0.01704 & 5.08 & 6.58 \\
First batch of 3 architectures & 11/12 (3 / 3 / 5) & 0.00636 & 0.01043 & 7.67 & 8.92 \\
First batch of 6 architectures & 11/12 (3 / 3 / 5) & 0.00580 & 0.00932 & 9.17 & 11.67 \\
First batch of 15 architectures & 10/12 (3 / 3 / 4) & 0.00807 & 0.01385 & 13.25 & 19.33 \\
\addlinespace[2pt]
\multicolumn{6}{l}{\textit{Ensemble size}} \\*
1 MLP & 10/12 (3 / 3 / 4) & 0.00708 & 0.01188 & 10.75 & 15.00 \\
9 MLPs & 10/12 (3 / 3 / 4) & 0.01279 & 0.02331 & 10.33 & 15.00 \\
\addlinespace[2pt]
\multicolumn{6}{l}{\textit{Catalogue starts}} \\*
\shortstack[l]{Pattern-search\\starts with corrected\\refinement} & 4/12 (2 / 2 / 0) & 0.02171 & 0.03188 & 22.33 & 19.67 \\
\shortstack[l]{Retrieval with\\pattern search} & 5/12 (2 / 3 / 0) & 0.02477 & 0.04299 & 22.58 & 30.00 \\
\addlinespace[2pt]
\multicolumn{6}{l}{\textit{Controls}} \\*
Retrieval control & 0/12 (0 / 0 / 0) & 0.07154 & 0.07280 & 31.00 & 0.00 \\
Random control & 0/12 (0 / 0 / 0) & 0.16747 & 0.13537 & 31.00 & 30.00 \\
\end{longtable}\endgroup

\begingroup\scriptsize\setlength{\tabcolsep}{1pt}
\setlength{\LTleft}{0pt}\setlength{\LTright}{0pt plus 1fill}\setlength{\LTcapwidth}{\textwidth}
\begin{longtable}{@{}lcccccc@{}}
\caption{Five-fold cross-validation on 358 catalogue designs with zero additional simulator calls. Recall at rank \(k\) is the share of held-out designs whose source architecture is in the first \(k\) ranks. Entries show means with ranges over three model seeds in parentheses. RMSE is root-mean-square error in log properties. The properties are axial Young's modulus \(E_{11}\), Voigt bulk modulus \(K_{\mathrm{bulk}}\) and axial conductivity \(k_{11}\). Catalogue-fitted tables can predict a held-out design only if its architecture retains another design in the training fold}
\label{tab:offline_recall}\\
\toprule
 & \multicolumn{3}{c}{Recall of the true architecture} & \multicolumn{3}{c}{Held-out RMSE ($\log_{10}$)} \\
\cmidrule(lr){2-4}\cmidrule(lr){5-7}
Model & Rank 3 & Rank 10 & Rank 20 & $E_{11}$ & $K_\mathrm{bulk}$ & $k_{11}$ \\
\midrule\endfirsthead
\multicolumn{7}{@{}l}{Table~\thetable\ (continued)}\\
\toprule
 & \multicolumn{3}{c}{Recall of the true architecture} & \multicolumn{3}{c}{Held-out RMSE ($\log_{10}$)} \\
\cmidrule(lr){2-4}\cmidrule(lr){5-7}
Model & Rank 3 & Rank 10 & Rank 20 & $E_{11}$ & $K_\mathrm{bulk}$ & $k_{11}$ \\
\midrule\endhead
\bottomrule\endfoot
\multicolumn{7}{l}{\textit{All designs (358)}} \\*
Learned tables & 0.301 (0.031) & 0.519 (0.011) & 0.703 (0.022) & 0.144 & 0.136 & 0.093 \\
\shortstack[l]{Categorical-encoding\\tables} & 0.334 (0.020) & 0.550 (0.031) & 0.661 (0.025) & 0.143 & 0.138 & 0.097 \\
Catalogue-fitted tables & 0.388 (---) & 0.486 (---) & 0.542 (---) & 0.316 & 0.308 & 0.211 \\
Nearest catalogue design & 0.034 (---) & 0.137 (---) & 0.268 (---) & -- & -- & -- \\
\addlinespace
\multicolumn{7}{l}{\textit{TPMS designs (128)}} \\*
Learned tables & 0.378 (0.062) & 0.633 (0.023) & 0.706 (0.008) & 0.132 & 0.120 & 0.098 \\
\shortstack[l]{Categorical-encoding\\tables} & 0.370 (0.055) & 0.646 (0.016) & 0.703 (0.016) & 0.131 & 0.119 & 0.098 \\
Catalogue-fitted tables & 0.578 (---) & 0.711 (---) & 0.750 (---) & 0.133 & 0.119 & 0.098 \\
Nearest catalogue design & 0.000 (---) & 0.039 (---) & 0.203 (---) & -- & -- & -- \\
\addlinespace
\multicolumn{7}{l}{\textit{Strut designs (80)}} \\*
Learned tables & 0.504 (0.038) & 0.696 (0.050) & 0.754 (0.012) & 0.167 & 0.143 & 0.130 \\
\shortstack[l]{Categorical-encoding\\tables} & 0.592 (0.025) & 0.721 (0.012) & 0.762 (0.000) & 0.167 & 0.141 & 0.129 \\
Catalogue-fitted tables & 0.662 (---) & 0.750 (---) & 0.800 (---) & 0.165 & 0.139 & 0.128 \\
Nearest catalogue design & 0.113 (---) & 0.375 (---) & 0.588 (---) & -- & -- & -- \\
\addlinespace
\multicolumn{7}{l}{\textit{Blend designs (150)}} \\*
Learned tables & 0.127 (0.040) & 0.327 (0.020) & 0.673 (0.067) & 0.140 & 0.146 & 0.058 \\
\shortstack[l]{Categorical-encoding\\tables} & 0.167 (0.013) & 0.378 (0.080) & 0.571 (0.053) & 0.140 & 0.151 & 0.072 \\
Catalogue-fitted tables & 0.080 (---) & 0.153 (---) & 0.227 (---) & 0.469 & 0.464 & 0.306 \\
Nearest catalogue design & 0.020 (---) & 0.093 (---) & 0.153 (---) & -- & -- & -- \\
\addlinespace
\multicolumn{7}{l}{\textit{Held-out blend designs whose architecture has no design in the training fold (14)}} \\*
Learned tables & 0.167 (0.071) & 0.429 (0.000) & 0.690 (0.214) & 0.175 & 0.127 & 0.062 \\
\shortstack[l]{Categorical-encoding\\tables} & 0.119 (0.071) & 0.167 (0.071) & 0.238 (0.071) & 0.123 & 0.117 & 0.067 \\
Catalogue-fitted tables & 0.000 (---) & 0.000 (---) & 0.000 (---) & -- & -- & -- \\
Nearest catalogue design & 0.000 (---) & 0.000 (---) & 0.000 (---) & -- & -- & -- \\
\end{longtable}\endgroup

The offline diagnostic measures recovery of a held-out design's source architecture. The share recovered in the first ten ranks is 0.519 for learned tables with the shared encoding and 0.550 for categorical-encoding tables across all catalogue designs. On blends, the corresponding values are 0.327 and 0.378. At rank 20 the order reverses for blends (0.673 and 0.571). Held-out root-mean-square errors over all catalogue designs range from 0.093 to 0.144 in log units across responses and the two encodings, well above the tolerance of 0.00860. Figure~\ref{fig:offline} shows the dependence on shortlist size. The verified search can succeed through any architecture. The offline ranking result leaves the source of the shared encoding's lower design error unresolved.

\par\noindent\begin{minipage}{\textwidth}
\centering
\includegraphics[width=\textwidth]{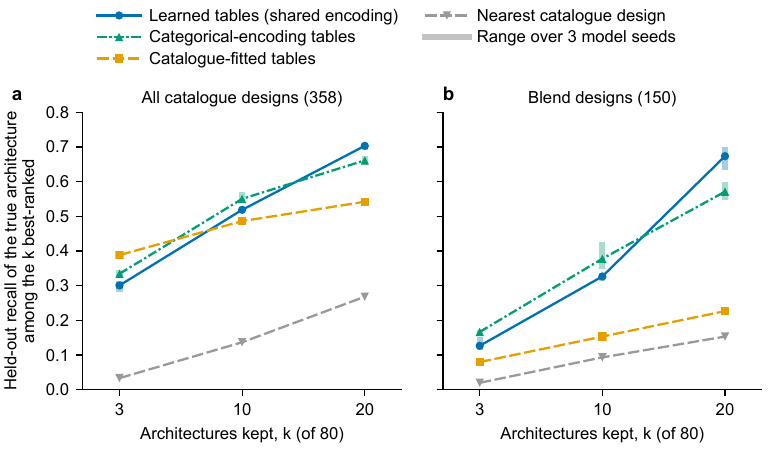}
\captionof{figure}{Offline cross-validated architecture recall. \textbf{a} All 358 catalogue designs. \textbf{b} The 150 blend designs. In five-fold cross-validation, each held-out design's properties define the target and recall measures whether its source architecture is among the best \(k\) of 80. Markers show means and bars span three model seeds. Catalogue-fitted tables and nearest-design retrieval are deterministic. This diagnostic uses existing responses and no simulator calls. At ranks three and ten the shared encoding has no recall advantage over categorical encoding. Held-out designs lie on the catalogue grid, so these results describe catalogue-like targets. No setting was selected using this diagnostic}
\label{fig:offline}
\end{minipage}\par

\begingroup\scriptsize\setlength{\tabcolsep}{2pt}
\setlength{\LTleft}{0pt}\setlength{\LTright}{0pt plus 1fill}\setlength{\LTcapwidth}{\textwidth}
\begin{longtable}{@{}L{26mm}C{10mm}C{23mm}C{25mm}C{17mm}C{17mm}@{}}
\caption{Primary comparison arms by target system. Each TPMS and strut block has 9 targets $\times$ three seeds, and the blend block has 18 $\times$ three. Means, ranges, intervals and failure convention follow Table~\ref{tab:main_comparison} of the article. TPMS denotes triply periodic minimal surface}\label{tab:system_comparison}\\

\toprule
Configuration & \shortstack{Runs\\met} & Success rate & Design error & \shortstack{Calls to\\success} & \shortstack{Simulator\\calls per\\target} \\
\midrule\endfirsthead
\multicolumn{6}{l}{Table~\thetable\ (continued)}\\
\toprule
Configuration & \shortstack{Runs\\met} & Success rate & Design error & \shortstack{Calls to\\success} & \shortstack{Simulator\\calls per\\target} \\
\midrule\endhead
\bottomrule\endfoot

\multicolumn{6}{l}{\textit{TPMS targets}} \\
Learned-table search & 25/27 & \centerentry{\shortstack{0.926 (0.111)\\{[0.778, 1.000]}}} & \centerentry{\shortstack{0.00463 (0.00220)\\{[0.00237, 0.00787]}}} & 7.37 (0.11) & 12.41 \\
Catalogue-table search & 27/27 & \centerentry{\shortstack{1.000 (---)\\{[1.000, 1.000]}}} & \centerentry{\shortstack{0.00419 (---)\\{[0.00277, 0.00561]}}} & 4.67 (---) & 11.22 \\
Bracketing search & 24/27 & 0.889 (---) & 0.00385 (---) & 5.78 (---) & 5.67 \\
Bayesian optimisation & 23/27 & 0.852 (0.111) & 0.00508 (0.00162) & 8.81 (5.67) & 8.67 \\
Nelder--Mead search & 24/27 & 0.889 (---) & 0.00642 (---) & 7.89 (---) & 7.78 \\
Retrieval with pattern search & 9/27 & \centerentry{\shortstack{0.333 (0.000)\\{[0.000, 0.667]}}} & \centerentry{\shortstack{0.02038 (0.00000)\\{[0.01086, 0.03142]}}} & 23.56 (0.22) & 30.00 \\
Retrieval control & 0/27 & \centerentry{\shortstack{0.000 (---)\\{[0.000, 0.000]}}} & \centerentry{\shortstack{0.03111 (---)\\{[0.02499, 0.03841]}}} & 31.00 (---) & 0.00 \\
Random control & 0/27 & \centerentry{\shortstack{0.000 (0.000)\\{[0.000, 0.000]}}} & \centerentry{\shortstack{0.08799 (0.00754)\\{[0.06518, 0.11356]}}} & 31.00 (0.00) & 30.00 \\
\addlinespace[3pt]
\multicolumn{6}{l}{\textit{Strut targets}} \\
Learned-table search & 27/27 & \centerentry{\shortstack{1.000 (0.000)\\{[1.000, 1.000]}}} & \centerentry{\shortstack{$<10^{-4}$ (0.00000)\\{[$<10^{-4}$, $<10^{-4}$]}}} & 2.96 (0.67) & 10.56 \\
Catalogue-table search & 27/27 & \centerentry{\shortstack{1.000 (---)\\{[1.000, 1.000]}}} & \centerentry{\shortstack{$<10^{-4}$ (---)\\{[$<10^{-4}$, $<10^{-4}$]}}} & 4.33 (---) & 12.22 \\
Bracketing search & 27/27 & 1.000 (---) & $<10^{-4}$ (---) & 1.56 (---) & 1.56 \\
Bayesian optimisation & 27/27 & 1.000 (0.000) & $<10^{-4}$ (0.00000) & 1.11 (0.00) & 1.11 \\
Nelder--Mead search & 27/27 & 1.000 (---) & $<10^{-4}$ (---) & 3.00 (---) & 3.00 \\
Retrieval with pattern search & 12/27 & \centerentry{\shortstack{0.444 (0.000)\\{[0.111, 0.778]}}} & \centerentry{\shortstack{0.02040 (0.00009)\\{[0.00709, 0.03467]}}} & 21.22 (0.22) & 29.56 \\
Retrieval control & 0/27 & \centerentry{\shortstack{0.000 (---)\\{[0.000, 0.000]}}} & \centerentry{\shortstack{0.04244 (---)\\{[0.03290, 0.05484]}}} & 31.00 (---) & 0.00 \\
Random control & 2/27 & \centerentry{\shortstack{0.074 (0.111)\\{[0.000, 0.185]}}} & \centerentry{\shortstack{0.11318 (0.02386)\\{[0.07357, 0.15645]}}} & 30.04 (2.33) & 30.00 \\
\addlinespace[3pt]
\multicolumn{6}{l}{\textit{Blend targets}} \\
Learned-table search & 32/54 & \centerentry{\shortstack{0.593 (0.222)\\{[0.407, 0.778]}}} & \centerentry{\shortstack{0.01046 (0.00098)\\{[0.00730, 0.01426]}}} & 21.20 (0.61) & 22.57 \\
Catalogue-table search & 24/54 & \centerentry{\shortstack{0.444 (---)\\{[0.222, 0.667]}}} & \centerentry{\shortstack{0.01391 (---)\\{[0.00725, 0.02528]}}} & 23.22 (---) & 24.00 \\
Bracketing search & 21/54 & 0.389 (---) & 0.02084 (---) & 21.06 (---) & 20.44 \\
Bayesian optimisation & 21/54 & 0.389 (0.111) & 0.01661 (0.00281) & 25.02 (2.94) & 24.41 \\
Nelder--Mead search & 18/54 & 0.333 (---) & 0.02932 (---) & 24.50 (---) & 23.83 \\
Retrieval with pattern search & 2/54 & \centerentry{\shortstack{0.037 (0.056)\\{[0.000, 0.111]}}} & \centerentry{\shortstack{0.04489 (0.00357)\\{[0.02702, 0.06663]}}} & 30.56 (0.67) & 29.33 \\
Retrieval control & 0/54 & \centerentry{\shortstack{0.000 (---)\\{[0.000, 0.000]}}} & \centerentry{\shortstack{0.07027 (---)\\{[0.04046, 0.10730]}}} & 31.00 (---) & 0.00 \\
Random control & 0/54 & \centerentry{\shortstack{0.000 (0.000)\\{[0.000, 0.000]}}} & \centerentry{\shortstack{0.18540 (0.05899)\\{[0.09822, 0.28935]}}} & 31.00 (0.00) & 30.00 \\
\addlinespace[3pt]

\end{longtable}\endgroup

\begingroup\scriptsize\setlength{\tabcolsep}{1pt}
\setlength{\LTleft}{0pt}\setlength{\LTright}{0pt plus 1fill}\setlength{\LTcapwidth}{\textwidth}
\begin{longtable}{@{}p{37mm}p{70mm}l@{}}
\caption{Search and model constants. Varied entries were sensitivity analyses on the test targets. Fixed values were neither tuned nor sensitivity-tested. The call allowance and success tolerance are benchmark requirements}
\label{tab:constants}\\
\toprule
Setting & Value & Status \\
\midrule\endfirsthead
\multicolumn{3}{@{}l}{Table~\thetable\ (continued)}\\
\toprule
Setting & Value & Status \\
\midrule\endhead
\bottomrule\endfoot
First-batch architectures & 10 (3, 6, 15 in analysis) & Varied \\
Networks in ensemble & 3 (1, 9 in analysis) & Varied \\
Maximum from one system in first batch & 6 & Fixed \\
Later-batch allocation & 3 refinements and 2 new proposals & Fixed \\
Initial window half-width & One quarter of each parameter range & Fixed \\
Window contraction & Half-width halved per additional valid observation & Fixed \\
Window floors & 0.003 density, 0.01 blend weight & Fixed \\
Close-match switch & Best observed error below $2\tau$ & Fixed \\
Architectures refined after switch & Up to 2 & Fixed \\
Release after consecutive misses & 3 & Fixed \\
Drop without valid observation & 2 invalid observations & Fixed \\
Invalid-query exclusion radius & Up to one tenth of the axis span & Fixed \\
System-shift eligibility & At least 3 first valid observations & Fixed \\
System-shift estimate & Componentwise median of first residuals & Fixed \\
System-shift acceptance & Leave-one-out mean maximum residual decreases & Fixed \\
Residual-fit row multiplier & $1/\sqrt{\|u\|_2^2+10^{-8}}$ & Fixed \\
Pure-table grid & Up to 201 density points & Fixed \\
Blend-table grid & 101 density $\times$ 41 weight points & Fixed \\
Pure-grid extension & Half a catalogue spacing, clipped to bounds & Fixed \\
Design precision and local grid step & Multiples of 0.001 & Fixed \\
Network hidden layers and width & 3 layers $\times$ 128 units & Fixed \\
Training steps and optimizer & 3,000 full-batch Adam steps & Fixed \\
Learning rate and weight decay & $0.003$ with cosine decay, $10^{-5}$ & Fixed \\
\end{longtable}\endgroup

\begingroup\fontsize{7}{8.5}\selectfont\setlength{\tabcolsep}{1pt}
\setlength{\LTleft}{0pt}\setlength{\LTright}{0pt plus 1fill}\setlength{\LTcapwidth}{\textwidth}

\begin{longtable}{@{}L{24mm}C{11mm}C{18mm}C{15mm}C{18mm}C{12mm}C{12mm}C{15mm}@{}}
\caption{Unchanged configurations on additional sets, each 36 targets $\times$ three seeds and at most 30 search calls per target. System counts are TPMS / strut / blend (out of 27 / 27 / 54). Searches use $32^3$ responses for every set. The last column re-scores returned designs under the 2\% stop rule. Ranges describe seed sensitivity and dashes denote deterministic configurations. Deterministic targets met out of 36 equal the reported runs met divided by three. All configurations use zero language-model calls. TPMS denotes triply periodic minimal surface}\label{tab:additional_sets}\\

\toprule
Configuration & Met & \shortstack{Met by\\system} & \shortstack{Design\\error} & \shortstack{Success rate\\(range)} & \shortstack{Calls to\\success} & \shortstack{Simulator\\calls} & \shortstack{Met at\\3\% / 5\%} \\
\midrule\endfirsthead
\multicolumn{8}{l}{Table \thetable\ (continued)}\\
\toprule
Configuration & Met & \shortstack{Met by\\system} & \shortstack{Design\\error} & \shortstack{Success rate\\(range)} & \shortstack{Calls to\\success} & \shortstack{Simulator\\calls} & \shortstack{Met at\\3\% / 5\%} \\
\midrule\endhead
\bottomrule\endfoot

\multicolumn{8}{l}{\textit{Fresh reachable draw}} \\
\shortstack[l]{Learned-table\\search} & 94/108 & 24 / 27 / 43 & 0.01230 & 0.870 (0.056) & 10.63 & 14.76 & 99 / 103 \\
\shortstack[l]{Catalogue-table\\search} & 87/108 & 24 / 27 / 36 & 0.01326 & 0.806 (---) & 11.58 & 16.06 & 93 / 102 \\
\shortstack[l]{Three-architecture\\catalogue search} & 93/108 & 21 / 27 / 45 & 0.01389 & 0.861 (---) & 8.67 & 9.56 & 93 / 99 \\
Bracketing search & 87/108 & 24 / 27 / 36 & 0.01121 & 0.806 (---) & 8.83 & 8.64 & 93 / 96 \\
\shortstack[l]{Bayesian\\optimisation} & 84/108 & 23 / 27 / 34 & 0.01337 & 0.778 (0.083) & 11.52 & 11.30 & 94 / 102 \\
Nelder--Mead search & 78/108 & 21 / 27 / 30 & 0.01138 & 0.722 (---) & 15.25 & 14.97 & 78 / 96 \\
\shortstack[l]{Retrieval with\\pattern search} & 27/108 & 12 / 15 / 0 & 0.02891 & 0.250 (0.000) & 26.14 & 29.56 & 37 / 62 \\
Retrieval control & 0/108 & 0 / 0 / 0 & 0.04587 & 0.000 (---) & 31.00 & 0.00 & 0 / 3 \\
Random control & 2/108 & 0 / 2 / 0 & 0.10821 & 0.019 (0.028) & 30.95 & 30.00 & 5 / 9 \\
\shortstack[l]{Pattern-search\\starts with\\corrected\\refinement} & 45/108 & 15 / 18 / 12 & 0.01973 & 0.417 (---) & 21.61 & 18.50 & 54 / 90 \\
\addlinespace[4pt]
\multicolumn{8}{l}{\textit{Interpolated responses}} \\
\shortstack[l]{Learned-table\\search} & 54/108 & 16 / 9 / 29 & 0.01758 & 0.500 (0.139) & 22.23 & 23.02 & 67 / 84 \\
\shortstack[l]{Catalogue-table\\search} & 54/108 & 12 / 12 / 30 & 0.02117 & 0.500 (---) & 22.72 & 23.53 & 66 / 84 \\
\shortstack[l]{Three-architecture\\catalogue search} & 51/108 & 9 / 12 / 30 & 0.01768 & 0.472 (---) & 20.89 & 20.69 & 75 / 90 \\
Bracketing search & 39/108 & 12 / 9 / 18 & 0.01743 & 0.361 (---) & 21.50 & 20.86 & 57 / 87 \\
\shortstack[l]{Bayesian\\optimisation} & 39/108 & 10 / 9 / 20 & 0.01771 & 0.361 (0.222) & 24.65 & 24.01 & 57 / 85 \\
Nelder--Mead search & 18/108 & 3 / 9 / 6 & 0.02406 & 0.167 (---) & 27.47 & 26.64 & 45 / 78 \\
\shortstack[l]{Retrieval with\\pattern search} & 4/108 & 0 / 3 / 1 & 0.03787 & 0.037 (0.028) & 30.44 & 29.78 & 13 / 36 \\
Retrieval control & 0/108 & 0 / 0 / 0 & 0.05321 & 0.000 (---) & 31.00 & 0.00 & 0 / 3 \\
Random control & 0/108 & 0 / 0 / 0 & 0.13079 & 0.000 (0.000) & 31.00 & 30.00 & 0 / 3 \\
\shortstack[l]{Pattern-search\\starts with\\corrected\\refinement} & 27/108 & 6 / 6 / 15 & 0.02387 & 0.250 (---) & 26.06 & 22.17 & 36 / 63 \\
\addlinespace[4pt]
\multicolumn{8}{l}{\textit{Fresh-reference $64^3$ responses}} \\
\shortstack[l]{Learned-table\\search} & 65/108 & 15 / 11 / 39 & 0.02060 & 0.602 (0.083) & 18.01 & 19.75 & 84 / 98 \\
\shortstack[l]{Catalogue-table\\search} & 66/108 & 15 / 12 / 39 & 0.01661 & 0.611 (---) & 18.83 & 20.11 & 81 / 96 \\
\shortstack[l]{Three-architecture\\catalogue search} & 69/108 & 15 / 15 / 39 & 0.02103 & 0.639 (---) & 15.28 & 14.11 & 84 / 96 \\
Bracketing search & 60/108 & 15 / 15 / 30 & 0.01763 & 0.556 (---) & 17.47 & 17.03 & 78 / 93 \\
\shortstack[l]{Bayesian\\optimisation} & 52/108 & 13 / 6 / 33 & 0.01751 & 0.481 (0.083) & 20.36 & 19.84 & 82 / 95 \\
Nelder--Mead search & 48/108 & 12 / 12 / 24 & 0.01690 & 0.444 (---) & 22.19 & 21.64 & 66 / 87 \\
\shortstack[l]{Retrieval with\\pattern search} & 12/108 & 6 / 3 / 3 & 0.03118 & 0.111 (0.000) & 28.29 & 29.44 & 30 / 64 \\
Retrieval control & 0/108 & 0 / 0 / 0 & 0.04536 & 0.000 (---) & 31.00 & 0.00 & 3 / 15 \\
Random control & 1/108 & 1 / 0 / 0 & 0.10301 & 0.009 (0.028) & 30.80 & 30.00 & 1 / 6 \\
\shortstack[l]{Pattern-search\\starts with\\corrected\\refinement} & 27/108 & 12 / 3 / 12 & 0.02433 & 0.250 (---) & 25.31 & 20.47 & 51 / 72 \\
\end{longtable}\endgroup

\par\noindent\begin{minipage}{\textwidth}
\centering
\scriptsize
\setlength{\tabcolsep}{1pt}
\captionof{table}{Resolution checks without further search. Mean absolute property shifts (percent) use distinct returned test geometries and are ordered top to bottom as axial Young's modulus \(E_{11}\), Voigt bulk modulus \(K_{\mathrm{bulk}}\) and axial conductivity \(k_{11}\). Count columns give, from top to bottom, 2\% / 3\% / 5\% matches out of 108. Test-set targets remain at $32^3$ for both returned-design evaluations. Fresh-set references and returned designs are evaluated at the same grid. No returned design is invalid at $64^3$. Convergence of $64^3$ is untested}
\label{tab:resolution}
\begin{tabular*}{\linewidth}{@{\extracolsep{\fill}}L{24mm}C{10mm}C{21mm}C{17mm}C{18mm}C{17mm}C{17mm}@{}}
\toprule
Configuration & Designs & \shortstack{Mean absolute\\property shift\\(\%)} & \shortstack{Test\\$32^3$} & \shortstack{Test $64^3$\\vs $32^3$} & \shortstack{Fresh\\$32^3$} & \shortstack{Fresh\\$64^3$} \\
\midrule
\shortstack[l]{Learned-table\\search} & 79 & \shortstack{3.19\\3.61\\1.94} & \shortstack{84\\94\\100} & \shortstack{12\\41\\64} & \shortstack{94\\99\\103} & \shortstack{57\\72\\91} \\ \addlinespace[3pt]
\shortstack[l]{Catalogue-table\\search} & 35 & \shortstack{3.22\\3.93\\1.87} & \shortstack{78\\99\\105} & \shortstack{18\\33\\66} & \shortstack{87\\93\\102} & \shortstack{60\\72\\90} \\ \addlinespace[3pt]
\shortstack[l]{Retrieval with\\pattern search} & 42 & \shortstack{3.70\\3.59\\2.45} & \shortstack{23\\29\\46} & \shortstack{11\\11\\35} & \shortstack{27\\37\\62} & \shortstack{9\\25\\52} \\ \addlinespace[3pt]
\bottomrule
\end{tabular*}
\end{minipage}\par\medskip

Pooling the 72 reachable targets of the test set and fresh draw gives a success-rate difference of 0.060 \([0.009,0.116]\), including a blend difference of 0.139 \([0.037,0.250]\). The overall sign test counts 11 targets better, four worse and 57 tied (\(p=0.12\)). The pooled mean design-error difference is \(-0.00128\) \([-0.00358,0.00023]\), and calls to success differ by \(-0.81\) \([-2.29,0.59]\). Thus the pooled success-rate interval favours this learned model, while the error and calls intervals include zero. These contrasts are post hoc and unadjusted for multiplicity, and the learned configuration had passed a development screen against the same comparator (see the article's benchmark comparisons), which can bias a smaller-set estimate upward. The pooled blend estimate of 0.139 lies above the confirmatory blend interval, while the pooled interval \([0.037,0.250]\) overlaps that interval. The confirmatory interval includes zero and its upper limit is below the pooled estimate. The confirmatory comparison therefore leaves a smaller effect unresolved. Table~\ref{tab:paired_additional} gives the paired outcomes for the additional sets.

\begingroup\fontsize{7}{8.5}\selectfont
\setlength{\tabcolsep}{2pt}
\setlength{\LTleft}{0pt}\setlength{\LTright}{0pt plus 1fill}
\setlength{\LTcapwidth}{\textwidth}

\begin{longtable}{@{}L{26mm}C{31mm}C{34mm}C{31mm}@{}}
\caption{Paired contrasts for the test set, additional sets and pooled reachable targets. Entries give mean difference, 95\% target-bootstrap interval and two-sided sign-test p value, with seeds averaged within target. Positive success and negative error or calls favour the first configuration. Targets are resampled within system and within set for pooling. These post hoc contrasts are unadjusted for multiplicity. Every search uses at most 30 simulator calls per target and zero language-model calls. Boundary intervals carry no information about performance on new targets.}\label{tab:paired_additional}\\

\toprule
System (targets) & Success rate & Design error & Calls to success \\ \midrule\endfirsthead
\multicolumn{4}{l}{Table \thetable\ (continued)}\\
\toprule
System (targets) & Success rate & Design error & Calls to success \\ \midrule\endhead
\midrule\multicolumn{4}{r}{Continued on next page}\\\endfoot
\bottomrule\endlastfoot
\multicolumn{4}{@{}p{\textwidth}@{}}{\textit{Test set. Learned-table search minus Catalogue-table search}} \\*
All (36) & \shortstack{+0.056\\{[-0.019, 0.139]}\\\(p=0.289\)} & \shortstack{-0.00161\\{[-0.00605, 0.00122]}\\\(p=0.839\)} & \shortstack{-0.68\\{[-2.95, 1.44]}\\\(p=0.701\)} \\ \addlinespace[3pt]
TPMS (9) & \shortstack{-0.074\\{[-0.222, 0.000]}\\\(p=1.000\)} & \shortstack{+0.00044\\{[-0.00237, 0.00355]}\\\(p=1.000\)} & \shortstack{+2.70\\{[-1.56, 7.41]}\\\(p=0.070\)} \\ \addlinespace[3pt]
Strut (9) & \shortstack{+0.000\\{[0.000, 0.000]}\\\text{---}} & \shortstack{+0.00000\\{[0.00000, 0.00000]}\\\text{---}} & \shortstack{-1.37\\{[-5.37, 1.48]}\\\(p=1.000\)} \\ \addlinespace[3pt]
Blend (18) & \shortstack{+0.148\\{[0.019, 0.296]}\\\(p=0.125\)} & \shortstack{-0.00344\\{[-0.01212, 0.00183]}\\\(p=0.629\)} & \shortstack{-2.02\\{[-5.50, 1.15]}\\\(p=0.424\)} \\ \addlinespace[3pt]
\multicolumn{4}{@{}p{\textwidth}@{}}{\textit{Test set. Learned-table search minus Bracketing search}} \\*
All (36) & \shortstack{+0.111\\{[0.000, 0.213]}\\\(p=0.021\)} & \shortstack{-0.00499\\{[-0.01257, 0.00015]}\\\(p=0.557\)} & \shortstack{+0.82\\{[-0.94, 2.78]}\\\(p=0.265\)} \\ \addlinespace[3pt]
TPMS (9) & \shortstack{+0.037\\{[0.000, 0.111]}\\\(p=1.000\)} & \shortstack{+0.00078\\{[-0.00177, 0.00307]}\\\(p=0.508\)} & \shortstack{+1.59\\{[0.22, 2.89]}\\\(p=0.180\)} \\ \addlinespace[3pt]
Strut (9) & \shortstack{+0.000\\{[0.000, 0.000]}\\\text{---}} & \shortstack{+0.00000\\{[0.00000, 0.00000]}\\\text{---}} & \shortstack{+1.41\\{[-0.67, 3.74]}\\\(p=0.375\)} \\ \addlinespace[3pt]
Blend (18) & \shortstack{+0.204\\{[-0.019, 0.407]}\\\(p=0.039\)} & \shortstack{-0.01038\\{[-0.02562, -0.00019]}\\\(p=0.143\)} & \shortstack{+0.15\\{[-3.11, 3.89]}\\\(p=1.000\)} \\ \addlinespace[3pt]
\multicolumn{4}{@{}p{\textwidth}@{}}{\textit{Test set. Learned-table search minus Bayesian optimisation}} \\*
All (36) & \shortstack{+0.120\\{[0.019, 0.222]}\\\(p=0.065\)} & \shortstack{-0.00319\\{[-0.00679, -0.00065]}\\\(p=0.006\)} & \shortstack{-1.81\\{[-4.57, 0.67]}\\\(p=1.000\)} \\ \addlinespace[3pt]
TPMS (9) & \shortstack{+0.074\\{[-0.185, 0.370]}\\\(p=1.000\)} & \shortstack{-0.00045\\{[-0.00292, 0.00235]}\\\(p=0.508\)} & \shortstack{-1.44\\{[-8.96, 3.78]}\\\(p=0.508\)} \\ \addlinespace[3pt]
Strut (9) & \shortstack{+0.000\\{[0.000, 0.000]}\\\text{---}} & \shortstack{+0.00000\\{[0.00000, 0.00000]}\\\text{---}} & \shortstack{+1.85\\{[0.37, 3.85]}\\\(p=0.125\)} \\ \addlinespace[3pt]
Blend (18) & \shortstack{+0.204\\{[0.056, 0.370]}\\\(p=0.070\)} & \shortstack{-0.00615\\{[-0.01319, -0.00137]}\\\(p=0.008\)} & \shortstack{-3.81\\{[-8.07, -0.09]}\\\(p=0.180\)} \\ \addlinespace[3pt]
\multicolumn{4}{@{}p{\textwidth}@{}}{\textit{Test set. Learned-table search minus Nelder--Mead search}} \\*
All (36) & \shortstack{+0.139\\{[0.028, 0.250]}\\\(p=0.065\)} & \shortstack{-0.00988\\{[-0.01744, -0.00384]}\\\(p=0.002\)} & \shortstack{-1.79\\{[-4.06, 0.61]}\\\(p=0.851\)} \\ \addlinespace[3pt]
TPMS (9) & \shortstack{+0.037\\{[0.000, 0.111]}\\\(p=1.000\)} & \shortstack{-0.00179\\{[-0.00379, 0.00034]}\\\(p=0.453\)} & \shortstack{-0.52\\{[-3.41, 1.48]}\\\(p=0.508\)} \\ \addlinespace[3pt]
Strut (9) & \shortstack{+0.000\\{[0.000, 0.000]}\\\text{---}} & \shortstack{+0.00000\\{[0.00000, 0.00000]}\\\text{---}} & \shortstack{-0.04\\{[-4.00, 2.74]}\\\(p=0.375\)} \\ \addlinespace[3pt]
Blend (18) & \shortstack{+0.259\\{[0.037, 0.481]}\\\(p=0.109\)} & \shortstack{-0.01886\\{[-0.03386, -0.00670]}\\\(p=0.002\)} & \shortstack{-3.30\\{[-7.20, 1.06]}\\\(p=0.057\)} \\ \addlinespace[3pt]
\multicolumn{4}{@{}p{\textwidth}@{}}{\textit{Fresh draw. Learned-table search minus Catalogue-table search}} \\*
All (36) & \shortstack{+0.065\\{[-0.009, 0.148]}\\\(p=0.453\)} & \shortstack{-0.00096\\{[-0.00189, -0.00020]}\\\(p=0.001\)} & \shortstack{-0.95\\{[-2.82, 0.82]}\\\(p=0.585\)} \\ \addlinespace[3pt]
TPMS (9) & \shortstack{+0.000\\{[0.000, 0.000]}\\\text{---}} & \shortstack{+0.00054\\{[-0.00029, 0.00191]}\\\(p=1.000\)} & \shortstack{+3.30\\{[1.04, 5.81]}\\\(p=0.289\)} \\ \addlinespace[3pt]
Strut (9) & \shortstack{+0.000\\{[0.000, 0.000]}\\\text{---}} & \shortstack{-0.00065\\{[-0.00194, 0.00000]}\\\(p=1.000\)} & \shortstack{-3.00\\{[-7.93, 0.67]}\\\(p=1.000\)} \\ \addlinespace[3pt]
Blend (18) & \shortstack{+0.130\\{[-0.019, 0.296]}\\\(p=0.453\)} & \shortstack{-0.00186\\{[-0.00353, -0.00068]}\\\(p=0.001\)} & \shortstack{-2.06\\{[-4.50, 0.59]}\\\(p=0.049\)} \\ \addlinespace[3pt]
\multicolumn{4}{@{}p{\textwidth}@{}}{\textit{Fresh draw. Learned-table search minus Bracketing search}} \\*
All (36) & \shortstack{+0.065\\{[-0.009, 0.148]}\\\(p=0.453\)} & \shortstack{+0.00109\\{[-0.00209, 0.00577]}\\\(p=0.307\)} & \shortstack{+1.80\\{[-0.13, 3.81]}\\\(p=0.122\)} \\ \addlinespace[3pt]
TPMS (9) & \shortstack{+0.000\\{[0.000, 0.000]}\\\text{---}} & \shortstack{+0.00833\\{[-0.00155, 0.02554]}\\\(p=1.000\)} & \shortstack{+3.41\\{[1.19, 5.89]}\\\(p=0.070\)} \\ \addlinespace[3pt]
Strut (9) & \shortstack{+0.000\\{[0.000, 0.000]}\\\text{---}} & \shortstack{+0.00000\\{[0.00000, 0.00000]}\\\text{---}} & \shortstack{+0.11\\{[-0.19, 0.48]}\\\(p=1.000\)} \\ \addlinespace[3pt]
Blend (18) & \shortstack{+0.130\\{[-0.019, 0.296]}\\\(p=0.453\)} & \shortstack{-0.00198\\{[-0.00495, 0.00028]}\\\(p=0.238\)} & \shortstack{+1.83\\{[-1.78, 5.76]}\\\(p=0.804\)} \\ \addlinespace[3pt]
\multicolumn{4}{@{}p{\textwidth}@{}}{\textit{Fresh draw. Learned-table search minus Bayesian optimisation}} \\*
All (36) & \shortstack{+0.093\\{[0.019, 0.157]}\\\(p=0.012\)} & \shortstack{-0.00107\\{[-0.00258, 0.00071]}\\\(p=0.009\)} & \shortstack{-0.89\\{[-2.52, 0.73]}\\\(p=1.000\)} \\ \addlinespace[3pt]
TPMS (9) & \shortstack{+0.037\\{[0.000, 0.111]}\\\(p=1.000\)} & \shortstack{+0.00045\\{[-0.00321, 0.00566]}\\\(p=0.727\)} & \shortstack{+1.70\\{[-1.52, 4.41]}\\\(p=0.125\)} \\ \addlinespace[3pt]
Strut (9) & \shortstack{+0.000\\{[0.000, 0.000]}\\\text{---}} & \shortstack{-0.00000\\{[-0.00000, 0.00000]}\\\text{---}} & \shortstack{+0.93\\{[0.15, 1.96]}\\\(p=0.125\)} \\ \addlinespace[3pt]
Blend (18) & \shortstack{+0.167\\{[0.019, 0.296]}\\\(p=0.021\)} & \shortstack{-0.00236\\{[-0.00459, 0.00019]}\\\(p=0.008\)} & \shortstack{-3.09\\{[-5.96, -0.20]}\\\(p=0.049\)} \\ \addlinespace[3pt]
\multicolumn{4}{@{}p{\textwidth}@{}}{\textit{Fresh draw. Learned-table search minus Nelder--Mead search}} \\*
All (36) & \shortstack{+0.148\\{[0.065, 0.250]}\\\(p=0.008\)} & \shortstack{+0.00093\\{[-0.00544, 0.01064]}\\\(p=0.152\)} & \shortstack{-4.62\\{[-7.13, -2.25]}\\\(p=0.016\)} \\ \addlinespace[3pt]
TPMS (9) & \shortstack{+0.111\\{[0.000, 0.333]}\\\(p=1.000\)} & \shortstack{+0.01522\\{[-0.00648, 0.05227]}\\\(p=1.000\)} & \shortstack{-2.59\\{[-9.93, 4.00]}\\\(p=0.727\)} \\ \addlinespace[3pt]
Strut (9) & \shortstack{+0.000\\{[0.000, 0.000]}\\\text{---}} & \shortstack{-0.00000\\{[-0.00000, 0.00000]}\\\text{---}} & \shortstack{-2.89\\{[-7.19, 0.11]}\\\(p=1.000\)} \\ \addlinespace[3pt]
Blend (18) & \shortstack{+0.241\\{[0.093, 0.407]}\\\(p=0.016\)} & \shortstack{-0.00576\\{[-0.01042, -0.00209]}\\\(p=0.096\)} & \shortstack{-6.50\\{[-9.43, -3.72]}\\\(p<0.001\)} \\ \addlinespace[3pt]
\multicolumn{4}{@{}p{\textwidth}@{}}{\textit{Interpolated responses. Learned-table search minus Catalogue-table search}} \\*
All (36) & \shortstack{-0.000\\{[-0.102, 0.102]}\\\(p=1.000\)} & \shortstack{-0.00359\\{[-0.01064, 0.00126]}\\\(p=1.000\)} & \shortstack{-0.49\\{[-2.85, 1.94]}\\\(p=0.839\)} \\ \addlinespace[3pt]
TPMS (9) & \shortstack{+0.148\\{[-0.148, 0.444]}\\\(p=0.625\)} & \shortstack{-0.00144\\{[-0.00354, 0.00037]}\\\(p=0.508\)} & \shortstack{-2.52\\{[-8.33, 3.48]}\\\(p=0.453\)} \\ \addlinespace[3pt]
Strut (9) & \shortstack{-0.111\\{[-0.259, 0.000]}\\\(p=0.500\)} & \shortstack{+0.00068\\{[-0.00133, 0.00279]}\\\(p=0.688\)} & \shortstack{+1.07\\{[-3.04, 5.74]}\\\(p=0.625\)} \\ \addlinespace[3pt]
Blend (18) & \shortstack{-0.019\\{[-0.148, 0.111]}\\\(p=1.000\)} & \shortstack{-0.00680\\{[-0.02109, 0.00290]}\\\(p=0.815\)} & \shortstack{-0.26\\{[-3.30, 2.94]}\\\(p=1.000\)} \\ \addlinespace[3pt]
\multicolumn{4}{@{}p{\textwidth}@{}}{\textit{Interpolated responses. Learned-table search minus Bracketing search}} \\*
All (36) & \shortstack{+0.139\\{[0.019, 0.259]}\\\(p=0.118\)} & \shortstack{+0.00015\\{[-0.00288, 0.00354]}\\\(p=0.007\)} & \shortstack{+0.73\\{[-2.02, 3.66]}\\\(p=1.000\)} \\ \addlinespace[3pt]
TPMS (9) & \shortstack{+0.148\\{[-0.148, 0.444]}\\\(p=0.625\)} & \shortstack{-0.00222\\{[-0.00413, -0.00053]}\\\(p=0.039\)} & \shortstack{+0.81\\{[-5.11, 7.85]}\\\(p=1.000\)} \\ \addlinespace[3pt]
Strut (9) & \shortstack{-0.000\\{[-0.111, 0.111]}\\\(p=1.000\)} & \shortstack{-0.00116\\{[-0.00246, -0.00015]}\\\(p=0.062\)} & \shortstack{+2.30\\{[-0.19, 5.74]}\\\(p=0.625\)} \\ \addlinespace[3pt]
Blend (18) & \shortstack{+0.204\\{[0.037, 0.389]}\\\(p=0.180\)} & \shortstack{+0.00199\\{[-0.00387, 0.00870]}\\\(p=0.481\)} & \shortstack{-0.09\\{[-4.46, 4.31]}\\\(p=0.774\)} \\ \addlinespace[3pt]
\multicolumn{4}{@{}p{\textwidth}@{}}{\textit{Interpolated responses. Learned-table search minus Bayesian optimisation}} \\*
All (36) & \shortstack{+0.139\\{[0.046, 0.241]}\\\(p=0.049\)} & \shortstack{-0.00012\\{[-0.00807, 0.00862]}\\\(p=0.175\)} & \shortstack{-2.42\\{[-4.75, -0.13]}\\\(p=0.169\)} \\ \addlinespace[3pt]
TPMS (9) & \shortstack{+0.222\\{[0.000, 0.407]}\\\(p=0.219\)} & \shortstack{-0.00187\\{[-0.00444, 0.00040]}\\\(p=0.508\)} & \shortstack{-3.15\\{[-6.11, 0.07]}\\\(p=0.125\)} \\ \addlinespace[3pt]
Strut (9) & \shortstack{+0.000\\{[-0.111, 0.111]}\\\(p=1.000\)} & \shortstack{-0.00077\\{[-0.00173, 0.00056]}\\\(p=0.289\)} & \shortstack{-1.30\\{[-6.70, 3.56]}\\\(p=1.000\)} \\ \addlinespace[3pt]
Blend (18) & \shortstack{+0.167\\{[0.019, 0.333]}\\\(p=0.180\)} & \shortstack{+0.00108\\{[-0.01494, 0.01889]}\\\(p=0.815\)} & \shortstack{-2.61\\{[-6.09, 0.85]}\\\(p=0.607\)} \\ \addlinespace[3pt]
\multicolumn{4}{@{}p{\textwidth}@{}}{\textit{Interpolated responses. Learned-table search minus Nelder--Mead search}} \\*
All (36) & \shortstack{+0.333\\{[0.222, 0.454]}\\\(p<0.001\)} & \shortstack{-0.00648\\{[-0.01261, -0.00129]}\\\(p=0.001\)} & \shortstack{-5.24\\{[-7.70, -2.79]}\\\(p<0.001\)} \\ \addlinespace[3pt]
TPMS (9) & \shortstack{+0.481\\{[0.222, 0.741]}\\\(p=0.031\)} & \shortstack{-0.00670\\{[-0.01287, -0.00262]}\\\(p=0.039\)} & \shortstack{-9.63\\{[-15.33, -4.44]}\\\(p=0.016\)} \\ \addlinespace[3pt]
Strut (9) & \shortstack{-0.000\\{[-0.111, 0.111]}\\\(p=1.000\)} & \shortstack{-0.00083\\{[-0.00190, 0.00004]}\\\(p=0.219\)} & \shortstack{+0.74\\{[-0.37, 2.52]}\\\(p=1.000\)} \\ \addlinespace[3pt]
Blend (18) & \shortstack{+0.426\\{[0.241, 0.611]}\\\(p<0.001\)} & \shortstack{-0.00919\\{[-0.02128, 0.00102]}\\\(p=0.096\)} & \shortstack{-6.04\\{[-10.00, -1.94]}\\\(p=0.003\)} \\ \addlinespace[3pt]
\multicolumn{4}{@{}p{\textwidth}@{}}{\textit{\(64^3\) responses. Learned-table search minus Catalogue-table search}} \\*
All (36) & \shortstack{-0.009\\{[-0.102, 0.093]}\\\(p=1.000\)} & \shortstack{+0.00399\\{[-0.00122, 0.01313]}\\\(p=1.000\)} & \shortstack{-0.82\\{[-3.19, 1.68]}\\\(p=0.134\)} \\ \addlinespace[3pt]
TPMS (9) & \shortstack{+0.000\\{[0.000, 0.000]}\\\text{---}} & \shortstack{+0.01799\\{[0.00002, 0.05321]}\\\(p=0.219\)} & \shortstack{-1.81\\{[-3.93, 0.00]}\\\(p=0.250\)} \\ \addlinespace[3pt]
Strut (9) & \shortstack{-0.037\\{[-0.333, 0.296]}\\\(p=1.000\)} & \shortstack{-0.00123\\{[-0.00374, 0.00114]}\\\(p=0.375\)} & \shortstack{+0.22\\{[-4.89, 5.81]}\\\(p=1.000\)} \\ \addlinespace[3pt]
Blend (18) & \shortstack{-0.000\\{[-0.111, 0.111]}\\\(p=1.000\)} & \shortstack{-0.00040\\{[-0.00231, 0.00112]}\\\(p=1.000\)} & \shortstack{-0.85\\{[-4.56, 3.28]}\\\(p=0.424\)} \\ \addlinespace[3pt]
\multicolumn{4}{@{}p{\textwidth}@{}}{\textit{\(64^3\) responses. Learned-table search minus Bracketing search}} \\*
All (36) & \shortstack{+0.046\\{[-0.046, 0.148]}\\\(p=0.727\)} & \shortstack{+0.00298\\{[-0.00183, 0.01141]}\\\(p=0.265\)} & \shortstack{+0.54\\{[-2.24, 3.20]}\\\(p=0.541\)} \\ \addlinespace[3pt]
TPMS (9) & \shortstack{+0.000\\{[0.000, 0.000]}\\\text{---}} & \shortstack{+0.01631\\{[-0.00102, 0.04907]}\\\(p=1.000\)} & \shortstack{+0.85\\{[-1.00, 3.41]}\\\(p=0.625\)} \\ \addlinespace[3pt]
Strut (9) & \shortstack{-0.148\\{[-0.370, 0.000]}\\\(p=0.500\)} & \shortstack{+0.00017\\{[-0.00112, 0.00179]}\\\(p=1.000\)} & \shortstack{+3.00\\{[-2.67, 9.07]}\\\(p=1.000\)} \\ \addlinespace[3pt]
Blend (18) & \shortstack{+0.167\\{[0.019, 0.352]}\\\(p=0.219\)} & \shortstack{-0.00229\\{[-0.00413, -0.00070]}\\\(p=0.238\)} & \shortstack{-0.85\\{[-5.48, 3.30]}\\\(p=1.000\)} \\ \addlinespace[3pt]
\multicolumn{4}{@{}p{\textwidth}@{}}{\textit{\(64^3\) responses. Learned-table search minus Bayesian optimisation}} \\*
All (36) & \shortstack{+0.120\\{[0.028, 0.231]}\\\(p=0.146\)} & \shortstack{+0.00309\\{[-0.00209, 0.01224]}\\\(p=0.099\)} & \shortstack{-2.35\\{[-5.07, 0.12]}\\\(p=1.000\)} \\ \addlinespace[3pt]
TPMS (9) & \shortstack{+0.074\\{[-0.074, 0.259]}\\\(p=1.000\)} & \shortstack{+0.01772\\{[-0.00094, 0.05316]}\\\(p=1.000\)} & \shortstack{-2.59\\{[-7.81, 1.59]}\\\(p=1.000\)} \\ \addlinespace[3pt]
Strut (9) & \shortstack{+0.185\\{[0.000, 0.407]}\\\(p=0.250\)} & \shortstack{-0.00147\\{[-0.00355, -0.00000]}\\\(p=0.625\)} & \shortstack{-4.85\\{[-9.96, 0.00]}\\\(p=0.250\)} \\ \addlinespace[3pt]
Blend (18) & \shortstack{+0.111\\{[-0.037, 0.278]}\\\(p=0.688\)} & \shortstack{-0.00194\\{[-0.00377, -0.00014]}\\\(p=0.096\)} & \shortstack{-0.98\\{[-5.19, 2.61]}\\\(p=0.607\)} \\ \addlinespace[3pt]
\multicolumn{4}{@{}p{\textwidth}@{}}{\textit{\(64^3\) responses. Learned-table search minus Nelder--Mead search}} \\*
All (36) & \shortstack{+0.157\\{[0.019, 0.296]}\\\(p=0.146\)} & \shortstack{+0.00371\\{[-0.00445, 0.01818]}\\\(p=0.004\)} & \shortstack{-4.19\\{[-7.28, -1.01]}\\\(p=0.007\)} \\ \addlinespace[3pt]
TPMS (9) & \shortstack{+0.111\\{[0.000, 0.333]}\\\(p=1.000\)} & \shortstack{+0.02537\\{[-0.00517, 0.08260]}\\\(p=0.219\)} & \shortstack{-2.70\\{[-7.00, 0.22]}\\\(p=0.625\)} \\ \addlinespace[3pt]
Strut (9) & \shortstack{-0.037\\{[-0.333, 0.296]}\\\(p=1.000\)} & \shortstack{-0.00040\\{[-0.00288, 0.00142]}\\\(p=0.625\)} & \shortstack{+2.11\\{[-4.93, 10.30]}\\\(p=1.000\)} \\ \addlinespace[3pt]
Blend (18) & \shortstack{+0.278\\{[0.093, 0.481]}\\\(p=0.070\)} & \shortstack{-0.00507\\{[-0.00761, -0.00297]}\\\(p=0.001\)} & \shortstack{-8.07\\{[-12.72, -3.54]}\\\(p=0.007\)} \\ \addlinespace[3pt]
\multicolumn{4}{@{}p{\textwidth}@{}}{\textit{Test set and fresh draw pooled. Learned-table search minus Catalogue-table search}} \\*
All (72) & \shortstack{+0.060\\{[0.009, 0.116]}\\\(p=0.118\)} & \shortstack{-0.00128\\{[-0.00358, 0.00023]}\\\(p=0.016\)} & \shortstack{-0.81\\{[-2.29, 0.59]}\\\(p=1.000\)} \\ \addlinespace[3pt]
TPMS (18) & \shortstack{-0.037\\{[-0.111, 0.000]}\\\(p=1.000\)} & \shortstack{+0.00049\\{[-0.00106, 0.00218]}\\\(p=1.000\)} & \shortstack{+3.00\\{[0.57, 5.61]}\\\(p=0.021\)} \\ \addlinespace[3pt]
Strut (18) & \shortstack{+0.000\\{[0.000, 0.000]}\\\text{---}} & \shortstack{-0.00032\\{[-0.00097, 0.00000]}\\\(p=1.000\)} & \shortstack{-2.19\\{[-5.31, 0.43]}\\\(p=0.754\)} \\ \addlinespace[3pt]
Blend (36) & \shortstack{+0.139\\{[0.037, 0.250]}\\\(p=0.057\)} & \shortstack{-0.00265\\{[-0.00709, 0.00017]}\\\(p=0.006\)} & \shortstack{-2.04\\{[-4.09, 0.00]}\\\(p=0.029\)} \\ \addlinespace[3pt]
\multicolumn{4}{@{}p{\textwidth}@{}}{\textit{Test set and fresh draw pooled. Learned-table search minus Bracketing search}} \\*
All (72) & \shortstack{+0.088\\{[0.023, 0.153]}\\\(p=0.013\)} & \shortstack{-0.00195\\{[-0.00611, 0.00165]}\\\(p=0.203\)} & \shortstack{+1.31\\{[0.00, 2.69]}\\\(p=0.044\)} \\ \addlinespace[3pt]
TPMS (18) & \shortstack{+0.019\\{[0.000, 0.056]}\\\(p=1.000\)} & \shortstack{+0.00456\\{[-0.00078, 0.01339]}\\\(p=0.607\)} & \shortstack{+2.50\\{[1.17, 3.91]}\\\(p=0.013\)} \\ \addlinespace[3pt]
Strut (18) & \shortstack{+0.000\\{[0.000, 0.000]}\\\text{---}} & \shortstack{+0.00000\\{[0.00000, 0.00000]}\\\text{---}} & \shortstack{+0.76\\{[-0.26, 1.93]}\\\(p=0.289\)} \\ \addlinespace[3pt]
Blend (36) & \shortstack{+0.167\\{[0.037, 0.296]}\\\(p=0.021\)} & \shortstack{-0.00618\\{[-0.01402, -0.00088]}\\\(p=0.041\)} & \shortstack{+0.99\\{[-1.47, 3.65]}\\\(p=1.000\)} \\ \addlinespace[3pt]
\multicolumn{4}{@{}p{\textwidth}@{}}{\textit{Test set and fresh draw pooled. Learned-table search minus Bayesian optimisation}} \\*
All (72) & \shortstack{+0.106\\{[0.046, 0.167]}\\\(p<0.001\)} & \shortstack{-0.00213\\{[-0.00410, -0.00056]}\\\(p<0.001\)} & \shortstack{-1.35\\{[-2.95, 0.14]}\\\(p=1.000\)} \\ \addlinespace[3pt]
TPMS (18) & \shortstack{+0.056\\{[-0.074, 0.204]}\\\(p=0.625\)} & \shortstack{+0.00000\\{[-0.00232, 0.00284]}\\\(p=0.332\)} & \shortstack{+0.13\\{[-3.83, 3.22]}\\\(p=0.077\)} \\ \addlinespace[3pt]
Strut (18) & \shortstack{+0.000\\{[0.000, 0.000]}\\\text{---}} & \shortstack{-0.00000\\{[-0.00000, 0.00000]}\\\text{---}} & \shortstack{+1.39\\{[0.46, 2.52]}\\\(p=0.008\)} \\ \addlinespace[3pt]
Blend (36) & \shortstack{+0.185\\{[0.083, 0.287]}\\\(p=0.001\)} & \shortstack{-0.00425\\{[-0.00798, -0.00145]}\\\(p<0.001\)} & \shortstack{-3.45\\{[-6.04, -1.06]}\\\(p=0.011\)} \\ \addlinespace[3pt]
\multicolumn{4}{@{}p{\textwidth}@{}}{\textit{Test set and fresh draw pooled. Learned-table search minus Nelder--Mead search}} \\*
All (72) & \shortstack{+0.144\\{[0.074, 0.218]}\\\(p<0.001\)} & \shortstack{-0.00447\\{[-0.00964, 0.00133]}\\\(p<0.001\)} & \shortstack{-3.20\\{[-4.90, -1.54]}\\\(p=0.048\)} \\ \addlinespace[3pt]
TPMS (18) & \shortstack{+0.074\\{[0.000, 0.204]}\\\(p=0.500\)} & \shortstack{+0.00671\\{[-0.00422, 0.02532]}\\\(p=0.581\)} & \shortstack{-1.56\\{[-5.48, 1.98]}\\\(p=0.332\)} \\ \addlinespace[3pt]
Strut (18) & \shortstack{+0.000\\{[0.000, 0.000]}\\\text{---}} & \shortstack{-0.00000\\{[-0.00000, 0.00000]}\\\text{---}} & \shortstack{-1.46\\{[-4.26, 0.81]}\\\(p=0.754\)} \\ \addlinespace[3pt]
Blend (36) & \shortstack{+0.250\\{[0.120, 0.389]}\\\(p=0.002\)} & \shortstack{-0.01231\\{[-0.02011, -0.00588]}\\\(p<0.001\)} & \shortstack{-4.90\\{[-7.31, -2.34]}\\\(p<0.001\)} \\ \addlinespace[3pt]
\end{longtable}\endgroup

\begingroup\fontsize{7}{8.5}\selectfont
\setlength{\tabcolsep}{2pt}
\setlength{\LTleft}{0pt}\setlength{\LTright}{0pt plus 1fill}
\setlength{\LTcapwidth}{\textwidth}

\begin{longtable}{@{}L{18mm}C{20mm}C{26mm}C{20mm}C{18mm}C{18mm}@{}}
\caption{Confirmatory results by system. Stochastic configurations use ten seeds. Deterministic configurations count each target once after identical two-seed runs. Success intervals bootstrap targets. Means include failures at 31 calls to success. All configurations use at most 30 simulator calls per target and zero language-model calls.}\label{tab:confirmatory}\\

\toprule
System & Met / runs & Success [95\% interval] & Design error & Calls to success & Simulator calls \\ \midrule\endfirsthead
\multicolumn{6}{l}{Table \thetable\ (continued)}\\
\toprule
System & Met / runs & Success [95\% interval] & Design error & Calls to success & Simulator calls \\ \midrule\endhead
\midrule\multicolumn{6}{r}{Continued on next page}\\\endfoot
\bottomrule\endlastfoot
\multicolumn{6}{@{}p{\textwidth}@{}}{\textit{Learned-table search}} \\*
All & 1687/2000 & \centerentry{\shortstack{0.844\\{[0.804, 0.881]}}} & 0.00669 & 10.82 & 15.27 \\ \addlinespace[4pt]
TPMS & 494/500 & \centerentry{\shortstack{0.988\\{[0.964, 1.000]}}} & 0.00343 & 5.64 & 11.33 \\ \addlinespace[4pt]
Strut & 490/500 & \centerentry{\shortstack{0.980\\{[0.942, 1.000]}}} & 0.00063 & 3.27 & 10.85 \\ \addlinespace[4pt]
Blend & 703/1000 & \centerentry{\shortstack{0.703\\{[0.626, 0.776]}}} & 0.01135 & 17.18 & 19.46 \\ \addlinespace[4pt]
\multicolumn{6}{@{}p{\textwidth}@{}}{\textit{Catalogue-table search}} \\*
All & 167/200 & \centerentry{\shortstack{0.835\\{[0.785, 0.880]}}} & 0.00667 & 10.61 & 15.43 \\ \addlinespace[4pt]
TPMS & 49/50 & \centerentry{\shortstack{0.980\\{[0.940, 1.000]}}} & 0.00376 & 4.96 & 11.90 \\ \addlinespace[4pt]
Strut & 50/50 & \centerentry{\shortstack{1.000\\{[1.000, 1.000]}}} & 0.00041 & 3.14 & 10.92 \\ \addlinespace[4pt]
Blend & 68/100 & \centerentry{\shortstack{0.680\\{[0.590, 0.770]}}} & 0.01127 & 17.17 & 19.44 \\ \addlinespace[4pt]
\multicolumn{6}{@{}p{\textwidth}@{}}{\textit{Three-architecture catalogue search}} \\*
All & 170/200 & \centerentry{\shortstack{0.850\\{[0.800, 0.895]}}} & 0.00745 & 8.40 & 9.05 \\ \addlinespace[4pt]
TPMS & 48/50 & \centerentry{\shortstack{0.960\\{[0.900, 1.000]}}} & 0.00536 & 3.24 & 3.72 \\ \addlinespace[4pt]
Strut & 49/50 & \centerentry{\shortstack{0.980\\{[0.940, 1.000]}}} & 0.00181 & 2.42 & 3.96 \\ \addlinespace[4pt]
Blend & 73/100 & \centerentry{\shortstack{0.730\\{[0.640, 0.810]}}} & 0.01132 & 13.97 & 14.26 \\ \addlinespace[4pt]
\multicolumn{6}{@{}p{\textwidth}@{}}{\textit{Bracketing search}} \\*
All & 157/200 & \centerentry{\shortstack{0.785\\{[0.735, 0.835]}}} & 0.00820 & 9.53 & 9.31 \\ \addlinespace[4pt]
TPMS & 50/50 & \centerentry{\shortstack{1.000\\{[1.000, 1.000]}}} & 0.00343 & 2.78 & 2.78 \\ \addlinespace[4pt]
Strut & 50/50 & \centerentry{\shortstack{1.000\\{[1.000, 1.000]}}} & 0.00067 & 2.02 & 2.02 \\ \addlinespace[4pt]
Blend & 57/100 & \centerentry{\shortstack{0.570\\{[0.470, 0.660]}}} & 0.01435 & 16.66 & 16.23 \\ \addlinespace[4pt]
\multicolumn{6}{@{}p{\textwidth}@{}}{\textit{Bayesian optimisation}} \\*
All & 1446/2000 & \centerentry{\shortstack{0.723\\{[0.682, 0.764]}}} & 0.00779 & 12.29 & 12.02 \\ \addlinespace[4pt]
TPMS & 445/500 & \centerentry{\shortstack{0.890\\{[0.810, 0.956]}}} & 0.00512 & 5.98 & 5.87 \\ \addlinespace[4pt]
Strut & 498/500 & \centerentry{\shortstack{0.996\\{[0.990, 1.000]}}} & 0.00077 & 2.36 & 2.35 \\ \addlinespace[4pt]
Blend & 503/1000 & \centerentry{\shortstack{0.503\\{[0.428, 0.578]}}} & 0.01264 & 20.42 & 19.92 \\ \addlinespace[4pt]
\multicolumn{6}{@{}p{\textwidth}@{}}{\textit{Nelder--Mead search}} \\*
All & 143/200 & \centerentry{\shortstack{0.715\\{[0.665, 0.765]}}} & 0.00928 & 14.15 & 13.87 \\ \addlinespace[4pt]
TPMS & 48/50 & \centerentry{\shortstack{0.960\\{[0.900, 1.000]}}} & 0.00419 & 6.38 & 6.34 \\ \addlinespace[4pt]
Strut & 49/50 & \centerentry{\shortstack{0.980\\{[0.940, 1.000]}}} & 0.00194 & 5.38 & 5.36 \\ \addlinespace[4pt]
Blend & 46/100 & \centerentry{\shortstack{0.460\\{[0.360, 0.560]}}} & 0.01550 & 22.43 & 21.89 \\ \addlinespace[4pt]
\multicolumn{6}{@{}p{\textwidth}@{}}{\textit{Retrieval with pattern search}} \\*
All & 545/2000 & \centerentry{\shortstack{0.273\\{[0.225, 0.320]}}} & 0.02313 & 25.73 & 29.48 \\ \addlinespace[4pt]
TPMS & 180/500 & \centerentry{\shortstack{0.360\\{[0.220, 0.500]}}} & 0.01587 & 23.79 & 29.28 \\ \addlinespace[4pt]
Strut & 340/500 & \centerentry{\shortstack{0.680\\{[0.540, 0.800]}}} & 0.00938 & 17.69 & 29.68 \\ \addlinespace[4pt]
Blend & 25/1000 & \centerentry{\shortstack{0.025\\{[0.000, 0.060]}}} & 0.03363 & 30.72 & 29.48 \\ \addlinespace[4pt]
\end{longtable}\endgroup

\begingroup\fontsize{7}{8.5}\selectfont
\setlength{\tabcolsep}{2pt}
\setlength{\LTleft}{0pt}\setlength{\LTright}{0pt plus 1fill}
\setlength{\LTcapwidth}{\textwidth}

\begin{longtable}{@{}L{26mm}C{31mm}C{34mm}C{31mm}@{}}
\caption{Confirmatory paired contrasts, formatted as in Table~\ref{tab:paired_additional}. The primary contrast is learned-table minus catalogue-table search, with success and design error overall and on blends as four declared endpoints. Other contrasts and calls are descriptive. All intervals and tests are unadjusted for multiplicity. Boundary intervals carry no information about new targets. Simulator allowance is 30 per target and language-model calls are zero.}\label{tab:paired_confirmatory}\\

\toprule
System (targets) & Success rate & Design error & Calls to success \\ \midrule\endfirsthead
\multicolumn{4}{l}{Table \thetable\ (continued)}\\
\toprule
System (targets) & Success rate & Design error & Calls to success \\ \midrule\endhead
\midrule\multicolumn{4}{r}{Continued on next page}\\\endfoot
\bottomrule\endlastfoot
\multicolumn{4}{@{}p{\textwidth}@{}}{\textit{Learned-table search minus Catalogue-table search}} \\*
All (200) & \shortstack{+0.008\\{[-0.027, 0.045]}\\\(p=0.233\)} & \shortstack{+0.00002\\{[-0.00104, 0.00099]}\\\(p=0.618\)} & \shortstack{+0.21\\{[-0.65, 1.07]}\\\(p=0.019\)} \\ \addlinespace[3pt]
TPMS (50) & \shortstack{+0.008\\{[-0.006, 0.030]}\\\(p=1.000\)} & \shortstack{-0.00032\\{[-0.00110, 0.00045]}\\\(p=0.878\)} & \shortstack{+0.68\\{[-0.37, 1.71]}\\\(p=0.004\)} \\ \addlinespace[3pt]
Strut (50) & \shortstack{-0.020\\{[-0.058, 0.000]}\\\(p=0.500\)} & \shortstack{+0.00022\\{[-0.00041, 0.00104]}\\\(p=1.000\)} & \shortstack{+0.13\\{[-1.12, 1.36]}\\\(p=0.248\)} \\ \addlinespace[3pt]
Blend (100) & \shortstack{+0.023\\{[-0.044, 0.093]}\\\(p=0.349\)} & \shortstack{+0.00008\\{[-0.00195, 0.00195]}\\\(p=0.475\)} & \shortstack{+0.01\\{[-1.49, 1.51]}\\\(p=0.826\)} \\ \addlinespace[3pt]
\multicolumn{4}{@{}p{\textwidth}@{}}{\textit{Learned-table search minus Three-architecture catalogue search}} \\*
All (200) & \shortstack{-0.007\\{[-0.043, 0.030]}\\\(p=0.054\)} & \shortstack{-0.00076\\{[-0.00241, 0.00074]}\\\(p=0.807\)} & \shortstack{+2.42\\{[1.40, 3.41]}\\\(p<0.001\)} \\ \addlinespace[3pt]
TPMS (50) & \shortstack{+0.028\\{[-0.004, 0.080]}\\\(p=1.000\)} & \shortstack{-0.00193\\{[-0.00519, 0.00007]}\\\(p=0.441\)} & \shortstack{+2.40\\{[1.11, 3.59]}\\\(p<0.001\)} \\ \addlinespace[3pt]
Strut (50) & \shortstack{+0.000\\{[-0.054, 0.058]}\\\(p=1.000\)} & \shortstack{-0.00118\\{[-0.00405, 0.00067]}\\\(p=1.000\)} & \shortstack{+0.85\\{[-0.91, 2.51]}\\\(p=0.036\)} \\ \addlinespace[3pt]
Blend (100) & \shortstack{-0.027\\{[-0.090, 0.037]}\\\(p=0.038\)} & \shortstack{+0.00003\\{[-0.00262, 0.00254]}\\\(p=0.841\)} & \shortstack{+3.21\\{[1.47, 4.95]}\\\(p<0.001\)} \\ \addlinespace[3pt]
\multicolumn{4}{@{}p{\textwidth}@{}}{\textit{Learned-table search minus Bracketing search}} \\*
All (200) & \shortstack{+0.059\\{[0.019, 0.099]}\\\(p=0.382\)} & \shortstack{-0.00151\\{[-0.00242, -0.00067]}\\\(p=0.121\)} & \shortstack{+1.29\\{[0.27, 2.32]}\\\(p<0.001\)} \\ \addlinespace[3pt]
TPMS (50) & \shortstack{-0.012\\{[-0.036, 0.000]}\\\(p=0.500\)} & \shortstack{-0.00000\\{[-0.00097, 0.00097]}\\\(p=0.542\)} & \shortstack{+2.86\\{[1.74, 4.01]}\\\(p<0.001\)} \\ \addlinespace[3pt]
Strut (50) & \shortstack{-0.020\\{[-0.058, 0.000]}\\\(p=0.500\)} & \shortstack{-0.00004\\{[-0.00084, 0.00086]}\\\(p=1.000\)} & \shortstack{+1.25\\{[0.06, 2.74]}\\\(p=0.185\)} \\ \addlinespace[3pt]
Blend (100) & \shortstack{+0.133\\{[0.059, 0.212]}\\\(p=0.126\)} & \shortstack{-0.00300\\{[-0.00473, -0.00145]}\\\(p=0.011\)} & \shortstack{+0.52\\{[-1.33, 2.35]}\\\(p=1.000\)} \\ \addlinespace[3pt]
\multicolumn{4}{@{}p{\textwidth}@{}}{\textit{Learned-table search minus Bayesian optimisation}} \\*
All (200) & \shortstack{+0.121\\{[0.083, 0.158]}\\\(p<0.001\)} & \shortstack{-0.00110\\{[-0.00194, -0.00020]}\\\(p<0.001\)} & \shortstack{-1.48\\{[-2.46, -0.51]}\\\(p=0.817\)} \\ \addlinespace[3pt]
TPMS (50) & \shortstack{+0.098\\{[0.040, 0.168]}\\\(p=0.002\)} & \shortstack{-0.00169\\{[-0.00266, -0.00075]}\\\(p=0.085\)} & \shortstack{-0.33\\{[-2.37, 1.49]}\\\(p=0.011\)} \\ \addlinespace[3pt]
Strut (50) & \shortstack{-0.016\\{[-0.056, 0.008]}\\\(p=1.000\)} & \shortstack{-0.00013\\{[-0.00069, 0.00063]}\\\(p=0.180\)} & \shortstack{+0.92\\{[-0.20, 2.25]}\\\(p=0.597\)} \\ \addlinespace[3pt]
Blend (100) & \shortstack{+0.200\\{[0.133, 0.266]}\\\(p<0.001\)} & \shortstack{-0.00129\\{[-0.00286, 0.00045]}\\\(p<0.001\)} & \shortstack{-3.24\\{[-4.85, -1.67]}\\\(p=0.073\)} \\ \addlinespace[3pt]
\multicolumn{4}{@{}p{\textwidth}@{}}{\textit{Learned-table search minus Nelder--Mead search}} \\*
All (200) & \shortstack{+0.128\\{[0.079, 0.179]}\\\(p=0.004\)} & \shortstack{-0.00259\\{[-0.00402, -0.00115]}\\\(p<0.001\)} & \shortstack{-3.34\\{[-4.61, -2.10]}\\\(p=0.079\)} \\ \addlinespace[3pt]
TPMS (50) & \shortstack{+0.028\\{[-0.004, 0.080]}\\\(p=1.000\)} & \shortstack{-0.00076\\{[-0.00174, 0.00013]}\\\(p=0.878\)} & \shortstack{-0.74\\{[-2.78, 1.10]}\\\(p=0.104\)} \\ \addlinespace[3pt]
Strut (50) & \shortstack{+0.000\\{[-0.054, 0.058]}\\\(p=1.000\)} & \shortstack{-0.00131\\{[-0.00419, 0.00059]}\\\(p=1.000\)} & \shortstack{-2.11\\{[-4.47, 0.19]}\\\(p=1.000\)} \\ \addlinespace[3pt]
Blend (100) & \shortstack{+0.243\\{[0.152, 0.337]}\\\(p=0.002\)} & \shortstack{-0.00415\\{[-0.00665, -0.00154]}\\\(p<0.001\)} & \shortstack{-5.25\\{[-7.30, -3.22]}\\\(p<0.001\)} \\ \addlinespace[3pt]
\multicolumn{4}{@{}p{\textwidth}@{}}{\textit{Learned-table search minus Retrieval with pattern search}} \\*
All (200) & \shortstack{+0.571\\{[0.510, 0.632]}\\\(p<0.001\)} & \shortstack{-0.01644\\{[-0.01898, -0.01404]}\\\(p<0.001\)} & \shortstack{-14.91\\{[-16.38, -13.42]}\\\(p<0.001\)} \\ \addlinespace[3pt]
TPMS (50) & \shortstack{+0.628\\{[0.496, 0.758]}\\\(p<0.001\)} & \shortstack{-0.01244\\{[-0.01626, -0.00887]}\\\(p<0.001\)} & \shortstack{-18.15\\{[-21.39, -14.77]}\\\(p<0.001\)} \\ \addlinespace[3pt]
Strut (50) & \shortstack{+0.300\\{[0.164, 0.440]}\\\(p=0.001\)} & \shortstack{-0.00875\\{[-0.01335, -0.00464]}\\\(p=0.007\)} & \shortstack{-14.42\\{[-17.42, -11.35]}\\\(p<0.001\)} \\ \addlinespace[3pt]
Blend (100) & \shortstack{+0.678\\{[0.600, 0.753]}\\\(p<0.001\)} & \shortstack{-0.02228\\{[-0.02661, -0.01850]}\\\(p<0.001\)} & \shortstack{-13.54\\{[-15.45, -11.62]}\\\(p<0.001\)} \\ \addlinespace[3pt]
\multicolumn{4}{@{}p{\textwidth}@{}}{\textit{Catalogue-table search minus Bracketing search}} \\*
All (200) & \shortstack{+0.050\\{[0.015, 0.090]}\\\(p=0.021\)} & \shortstack{-0.00153\\{[-0.00267, -0.00058]}\\\(p<0.001\)} & \shortstack{+1.08\\{[0.12, 2.06]}\\\(p=0.358\)} \\ \addlinespace[3pt]
TPMS (50) & \shortstack{-0.020\\{[-0.060, 0.000]}\\\(p=1.000\)} & \shortstack{+0.00032\\{[-0.00052, 0.00122]}\\\(p=0.690\)} & \shortstack{+2.18\\{[0.94, 3.58]}\\\(p=0.041\)} \\ \addlinespace[3pt]
Strut (50) & \shortstack{+0.000\\{[0.000, 0.000]}\\\text{---}} & \shortstack{-0.00026\\{[-0.00075, 0.00010]}\\\(p=0.375\)} & \shortstack{+1.12\\{[-0.04, 2.52]}\\\(p=1.000\)} \\ \addlinespace[3pt]
Blend (100) & \shortstack{+0.110\\{[0.040, 0.180]}\\\(p=0.007\)} & \shortstack{-0.00309\\{[-0.00532, -0.00125]}\\\(p<0.001\)} & \shortstack{+0.51\\{[-1.16, 2.15]}\\\(p=1.000\)} \\ \addlinespace[3pt]
\multicolumn{4}{@{}p{\textwidth}@{}}{\textit{Catalogue-table search minus Bayesian optimisation}} \\*
All (200) & \shortstack{+0.112\\{[0.068, 0.157]}\\\(p<0.001\)} & \shortstack{-0.00112\\{[-0.00190, -0.00027]}\\\(p<0.001\)} & \shortstack{-1.68\\{[-2.79, -0.60]}\\\(p=0.275\)} \\ \addlinespace[3pt]
TPMS (50) & \shortstack{+0.090\\{[0.032, 0.160]}\\\(p=0.004\)} & \shortstack{-0.00136\\{[-0.00244, -0.00031]}\\\(p=0.044\)} & \shortstack{-1.02\\{[-2.56, 0.42]}\\\(p=0.099\)} \\ \addlinespace[3pt]
Strut (50) & \shortstack{+0.004\\{[0.000, 0.010]}\\\(p=0.500\)} & \shortstack{-0.00036\\{[-0.00085, 0.00006]}\\\(p=0.727\)} & \shortstack{+0.78\\{[-0.28, 2.01]}\\\(p=1.000\)} \\ \addlinespace[3pt]
Blend (100) & \shortstack{+0.177\\{[0.095, 0.259]}\\\(p<0.001\)} & \shortstack{-0.00137\\{[-0.00283, 0.00026]}\\\(p<0.001\)} & \shortstack{-3.25\\{[-5.22, -1.29]}\\\(p=0.661\)} \\ \addlinespace[3pt]
\multicolumn{4}{@{}p{\textwidth}@{}}{\textit{Catalogue-table search minus Nelder--Mead search}} \\*
All (200) & \shortstack{+0.120\\{[0.070, 0.170]}\\\(p<0.001\)} & \shortstack{-0.00261\\{[-0.00387, -0.00138]}\\\(p<0.001\)} & \shortstack{-3.54\\{[-4.72, -2.38]}\\\(p<0.001\)} \\ \addlinespace[3pt]
TPMS (50) & \shortstack{+0.020\\{[0.000, 0.060]}\\\(p=1.000\)} & \shortstack{-0.00043\\{[-0.00140, 0.00032]}\\\(p=0.581\)} & \shortstack{-1.42\\{[-3.18, 0.28]}\\\(p=0.189\)} \\ \addlinespace[3pt]
Strut (50) & \shortstack{+0.020\\{[0.000, 0.060]}\\\(p=1.000\)} & \shortstack{-0.00153\\{[-0.00421, -0.00001]}\\\(p=0.219\)} & \shortstack{-2.24\\{[-4.46, -0.14]}\\\(p=0.238\)} \\ \addlinespace[3pt]
Blend (100) & \shortstack{+0.220\\{[0.130, 0.320]}\\\(p<0.001\)} & \shortstack{-0.00424\\{[-0.00639, -0.00203]}\\\(p<0.001\)} & \shortstack{-5.26\\{[-7.12, -3.39]}\\\(p<0.001\)} \\ \addlinespace[3pt]
\end{longtable}\endgroup

\begingroup\scriptsize\setlength{\tabcolsep}{1pt}
\setlength{\LTleft}{0pt}\setlength{\LTright}{0pt plus 1fill}\setlength{\LTcapwidth}{\textwidth}
\begin{longtable}{@{}L{22mm}C{18mm}C{13mm}C{19mm}C{18mm}C{13mm}C{23mm}@{}}
\caption{Corrected minus uncorrected refinement proposal for both table types. The control keeps every decision rule of the search and takes the queried design from the uncorrected table in the same window. Runs met are given as corrected vs uncorrected. Differences in success rate and design error average seeds within target; the 95\% intervals resample targets 20,000 times within system. The sign test counts targets better / worse for the corrected proposal, with its two-sided \(p\) value in parentheses. Learned tables use three seeds on the 36-target sets and ten seeds on the confirmatory set; catalogue tables are deterministic and count once per target on the confirmatory set. The learned-table rows of the test set repeat the contrast of Tables~\ref{tab:contrasts_design_error} and \ref{tab:contrasts_success_rate}. All other uncorrected-proposal runs were added after the confirmatory plan and are not among its declared configurations. All runs use at most 30 simulator calls per target and zero language-model calls}
\label{tab:correction_control}\\
\toprule
Tables, targets & Runs met & \shortstack{Success\\difference} & 95\% interval & \shortstack{Targets better\\/ worse (\(p\))} & \shortstack{Error\\difference} & 95\% interval \\
\midrule\endfirsthead
\multicolumn{7}{@{}l}{Table~\thetable\ (continued)}\\
\toprule
Tables, targets & Runs met & \shortstack{Success\\difference} & 95\% interval & \shortstack{Targets better\\/ worse (\(p\))} & \shortstack{Error\\difference} & 95\% interval \\
\midrule\endhead
\bottomrule\endfoot
\multicolumn{7}{@{}l}{\textit{Test set (36 targets)}} \\*
Learned tables, all & 84/108 vs 65/108 & $+$0.176 & [$+$0.093, $+$0.278] & 11 / 0 ($<$0.001) & $-$0.00796 & [$-$0.01338, $-$0.00355] \\
Learned tables, blend & 32/54 vs 17/54 & $+$0.278 & [$+$0.130, $+$0.444] & 9 / 0 (0.004) & $-$0.01411 & [$-$0.02447, $-$0.00557] \\
Catalogue tables, all & 78/108 vs 57/108 & $+$0.194 & [$+$0.083, $+$0.333] & 7 / 0 (0.016) & $-$0.01885 & [$-$0.03589, $-$0.00593] \\
Catalogue tables, blend & 24/54 vs 12/54 & $+$0.222 & [$+$0.056, $+$0.444] & 4 / 0 (0.125) & $-$0.03436 & [$-$0.06829, $-$0.00868] \\
\addlinespace[2pt]
\multicolumn{7}{@{}l}{\textit{Fresh draw (36 targets)}} \\*
Learned tables, all & 94/108 vs 73/108 & $+$0.194 & [$+$0.102, $+$0.296] & 11 / 0 ($<$0.001) & $-$0.00428 & [$-$0.00727, $-$0.00190] \\
Learned tables, blend & 43/54 vs 25/54 & $+$0.333 & [$+$0.185, $+$0.500] & 10 / 0 (0.002) & $-$0.00762 & [$-$0.01339, $-$0.00291] \\
Catalogue tables, all & 87/108 vs 69/108 & $+$0.167 & [$+$0.056, $+$0.306] & 6 / 0 (0.031) & $-$0.00423 & [$-$0.00751, $-$0.00170] \\
Catalogue tables, blend & 36/54 vs 24/54 & $+$0.222 & [$+$0.056, $+$0.444] & 4 / 0 (0.125) & $-$0.00756 & [$-$0.01391, $-$0.00261] \\
\addlinespace[2pt]
\multicolumn{7}{@{}l}{\textit{Confirmatory set (200 targets)}} \\*
Learned tables, all & 1687/2000 vs 1330/2000 & $+$0.179 & [$+$0.142, $+$0.217] & 66 / 0 ($<$0.001) & $-$0.00722 & [$-$0.00998, $-$0.00491] \\
Learned tables, blend & 703/1000 vs 397/1000 & $+$0.306 & [$+$0.243, $+$0.369] & 59 / 0 ($<$0.001) & $-$0.01242 & [$-$0.01756, $-$0.00821] \\
Catalogue tables, all & 167/200 vs 126/200 & $+$0.205 & [$+$0.150, $+$0.260] & 41 / 0 ($<$0.001) & $-$0.00861 & [$-$0.01179, $-$0.00592] \\
Catalogue tables, blend & 68/100 vs 37/100 & $+$0.310 & [$+$0.220, $+$0.400] & 31 / 0 ($<$0.001) & $-$0.01424 & [$-$0.02020, $-$0.00944] \\
\addlinespace[2pt]
\end{longtable}\endgroup

The two added coverage analyses ask whether a sparser catalogue would change the confirmatory result (Table~\ref{tab:coverage}). Among the 100 confirmatory blend targets, the learned-minus-catalogue success difference is 0.037 \([-0.050,0.143]\) for the 30 targets whose reference architecture has one or two catalogue designs. It is 0.010 \([-0.080,0.102]\) for the 49 targets with three designs and 0.033 \([-0.171,0.238]\) for the 21 targets with four. With the thinned catalogue, learned-table search meets 493 of 600 runs and catalogue-table search meets 163 of 200 targets. The difference is 0.007 \([-0.032,0.047]\) overall and 0.033 \([-0.033,0.103]\) on blends. Neither analysis detects a learned-table advantage. The thinning roughly halves the pure-architecture catalogue and leaves the blend catalogue unchanged, so it does not test sparser blend coverage.

\tablestart
\par\noindent\begin{minipage}{\textwidth}
\centering
\fontsize{7}{8.5}\selectfont
\setlength{\tabcolsep}{2pt}
\captionof{table}{Learned-table minus catalogue-table search under reduced catalogue coverage, on the 200 confirmatory targets. The thinned catalogue keeps every second density of each pure architecture, starting from the lowest, and all blend designs; learned tables are trained on it and catalogue tables are fitted to it. Runs met are given for learned-table search (three seeds in the first two blocks, ten in the third) and for deterministic catalogue-table search, which counts once per target. Differences average seeds within target. The 95\% intervals resample targets 20,000 times, within system in the first two blocks and within the row in the third. These analyses were added after the confirmatory plan and are not among its declared endpoints. All runs use at most 30 simulator calls per target and zero language-model calls}
\label{tab:coverage}
\begin{tabular*}{\linewidth}{@{\extracolsep{\fill}}L{28mm}C{8mm}C{22mm}C{24mm}C{15mm}C{22mm}@{}}
\toprule
Targets & \(n\) &\shortstack{Learned tables,\\runs met} & \shortstack{Catalogue tables,\\targets met} & \shortstack{Success\\difference} & 95\% interval \\
\midrule
\multicolumn{6}{@{}l}{\textit{Thinned catalogue (259 designs), learned-table seeds 0--2}} \\
All targets & 200 & 493/600 & 163/200 & $+$0.007 & [$-$0.032, $+$0.047] \\
TPMS targets & 50 & 143/150 & 48/50 & $-$0.007 & [$-$0.060, $+$0.040] \\
Strut targets & 50 & 139/150 & 48/50 & $-$0.033 & [$-$0.100, $+$0.013] \\
Blend targets & 100 & 211/300 & 67/100 & $+$0.033 & [$-$0.033, $+$0.103] \\
\addlinespace[2pt]
\multicolumn{6}{@{}l}{\textit{Full catalogue (358 designs), learned-table seeds 0--2}} \\
All targets & 200 & 500/600 & 167/200 & $-$0.002 & [$-$0.042, $+$0.038] \\
TPMS targets & 50 & 147/150 & 49/50 & $+$0.000 & [$+$0.000, $+$0.000] \\
Strut targets & 50 & 147/150 & 50/50 & $-$0.020 & [$-$0.053, $+$0.000] \\
Blend targets & 100 & 206/300 & 68/100 & $+$0.007 & [$-$0.070, $+$0.087] \\
\addlinespace[2pt]
\multicolumn{6}{@{}l}{\textit{Full catalogue, blend targets by catalogue designs of the reference architecture, ten seeds}} \\
One or two designs & 30 & 251/300 & 24/30 & $+$0.037 & [$-$0.050, $+$0.143] \\
Three designs & 49 & 325/490 & 32/49 & $+$0.010 & [$-$0.080, $+$0.102] \\
Four designs & 21 & 127/210 & 12/21 & $+$0.033 & [$-$0.171, $+$0.238] \\
\bottomrule
\end{tabular*}
\end{minipage}\par\medskip

\section*{Additional checks on the confirmatory targets}

On the same 200 confirmatory targets, we test whether search outcomes respond to large changes in table quality and whether learning helps when the catalogue omits architectures. We first reran the 2,400 original learned-table and catalogue-table runs and obtained identical outcomes. These controls keep the search, 30-call budget and 2\% stop rule fixed.

The degraded table uses one least-squares line per system to predict log responses from log density. It gives all architectures in a system the same prediction at a given density. The oracle table interpolates simulator responses on a coarse grid with 21 densities per pure architecture and 11 densities by nine weights per blend architecture. Constructing it used 6,048 simulator calls outside every search budget, including 4,767 newly solved designs and 1,281 served from the solver cache; 1,238 designs were rejected. Only the oracle uses these extra simulations.

The held-out-architecture test removes all 38 catalogue designs of one randomly chosen subset of 14 blend architectures, leaving 320 designs, and rebuilds both table types. Catalogue tables give no prediction for the removed architectures. Learned-table controls use seeds 0--2. Deterministic configurations count once per target. We planned these controls after seeing the confirmatory results and before running the controls. Tables~\ref{tab:table-controls} and \ref{tab:table-control-contrasts} report the outcomes and paired target-bootstrap intervals.

\begingroup\fontsize{7}{8.5}\selectfont
\setlength{\tabcolsep}{2pt}
\setlength{\LTleft}{0pt}\setlength{\LTright}{0pt plus 1fill}
\setlength{\LTcapwidth}{\textwidth}

\begin{longtable}{@{}L{18mm}C{20mm}C{26mm}C{20mm}C{18mm}C{18mm}@{}}
\caption{Table-quality controls and held-out architectures on the 200 confirmatory targets (50 TPMS, 50 strut, 100 blend). All searches keep the same schedule, at most 30 simulator calls, a 2\% stop rule and the \(32^3\) simulator, with zero language-model calls. Learned arms use seeds 0--2; deterministic arms count once. Intervals bootstrap targets within system. The degraded table uses one density fit per system, shared by all its architectures. Only the oracle uses extra simulator data (6,048 construction calls outside search budgets). The held-out test removes 38 designs of 14 blend architectures, leaving 320 designs, and rebuilds both table types. These controls were planned after the confirmatory results were known and are not declared endpoints. All 2,400 existing runs were first reproduced identically. TPMS denotes triply periodic minimal surface}\label{tab:table-controls}\\
\toprule
System & Met / runs & Success [95\% interval] & Design error & Calls to success & Simulator calls \\ \midrule\endfirsthead
\multicolumn{6}{l}{Table \thetable\ (continued)}\\\toprule
System & Met / runs & Success [95\% interval] & Design error & Calls to success & Simulator calls \\ \midrule\endhead
\midrule\multicolumn{6}{r}{Continued on next page}\\\endfoot
\bottomrule\endlastfoot
\multicolumn{6}{@{}p{\textwidth}@{}}{\textit{Degraded table}} \\*
All & 72/200 & \centerentry{\shortstack{0.360\\{[0.295, 0.420]}}} & 0.02054 & 25.98 & 26.32 \\ \addlinespace[4pt]
TPMS & 27/50 & \centerentry{\shortstack{0.540\\{[0.400, 0.680]}}} & 0.01254 & 23.40 & 24.32 \\ \addlinespace[4pt]
Strut & 26/50 & \centerentry{\shortstack{0.520\\{[0.380, 0.660]}}} & 0.01668 & 23.50 & 24.70 \\ \addlinespace[4pt]
Blend & 19/100 & \centerentry{\shortstack{0.190\\{[0.120, 0.270]}}} & 0.02647 & 28.52 & 28.13 \\ \addlinespace[4pt]
\multicolumn{6}{@{}p{\textwidth}@{}}{\textit{Catalogue tables, full catalogue}} \\*
All & 167/200 & \centerentry{\shortstack{0.835\\{[0.785, 0.880]}}} & 0.00667 & 10.61 & 15.43 \\ \addlinespace[4pt]
TPMS & 49/50 & \centerentry{\shortstack{0.980\\{[0.940, 1.000]}}} & 0.00376 & 4.96 & 11.90 \\ \addlinespace[4pt]
Strut & 50/50 & \centerentry{\shortstack{1.000\\{[1.000, 1.000]}}} & 0.00041 & 3.14 & 10.92 \\ \addlinespace[4pt]
Blend & 68/100 & \centerentry{\shortstack{0.680\\{[0.590, 0.770]}}} & 0.01127 & 17.17 & 19.44 \\ \addlinespace[4pt]
\multicolumn{6}{@{}p{\textwidth}@{}}{\textit{Learned tables, full catalogue, seeds 0--2}} \\*
All & 500/600 & \centerentry{\shortstack{0.833\\{[0.792, 0.873]}}} & 0.00672 & 10.95 & 15.34 \\ \addlinespace[4pt]
TPMS & 147/150 & \centerentry{\shortstack{0.980\\{[0.940, 1.000]}}} & 0.00361 & 5.74 & 11.50 \\ \addlinespace[4pt]
Strut & 147/150 & \centerentry{\shortstack{0.980\\{[0.947, 1.000]}}} & 0.00068 & 3.53 & 10.91 \\ \addlinespace[4pt]
Blend & 206/300 & \centerentry{\shortstack{0.687\\{[0.607, 0.763]}}} & 0.01129 & 17.26 & 19.48 \\ \addlinespace[4pt]
\multicolumn{6}{@{}p{\textwidth}@{}}{\textit{Catalogue tables, 14 blend architectures held out}} \\*
All & 166/200 & \centerentry{\shortstack{0.830\\{[0.785, 0.875]}}} & 0.00715 & 11.11 & 15.92 \\ \addlinespace[4pt]
TPMS & 49/50 & \centerentry{\shortstack{0.980\\{[0.940, 1.000]}}} & 0.00383 & 4.84 & 11.90 \\ \addlinespace[4pt]
Strut & 50/50 & \centerentry{\shortstack{1.000\\{[1.000, 1.000]}}} & 0.00041 & 2.96 & 10.92 \\ \addlinespace[4pt]
Blend & 67/100 & \centerentry{\shortstack{0.670\\{[0.580, 0.760]}}} & 0.01218 & 18.32 & 20.43 \\ \addlinespace[4pt]
\multicolumn{6}{@{}p{\textwidth}@{}}{\textit{Learned tables, 14 blend architectures held out}} \\*
All & 503/600 & \centerentry{\shortstack{0.838\\{[0.797, 0.878]}}} & 0.00621 & 11.24 & 15.48 \\ \addlinespace[4pt]
TPMS & 147/150 & \centerentry{\shortstack{0.980\\{[0.947, 1.000]}}} & 0.00342 & 5.86 & 11.42 \\ \addlinespace[4pt]
Strut & 147/150 & \centerentry{\shortstack{0.980\\{[0.940, 1.000]}}} & 0.00069 & 3.16 & 10.81 \\ \addlinespace[4pt]
Blend & 209/300 & \centerentry{\shortstack{0.697\\{[0.617, 0.773]}}} & 0.01037 & 17.97 & 19.85 \\ \addlinespace[4pt]
\multicolumn{6}{@{}p{\textwidth}@{}}{\textit{Oracle table}} \\*
All & 193/200 & \centerentry{\shortstack{0.965\\{[0.940, 0.990]}}} & 0.00317 & 3.79 & 11.21 \\ \addlinespace[4pt]
TPMS & 49/50 & \centerentry{\shortstack{0.980\\{[0.940, 1.000]}}} & 0.00228 & 2.24 & 10.48 \\ \addlinespace[4pt]
Strut & 49/50 & \centerentry{\shortstack{0.980\\{[0.940, 1.000]}}} & 0.00086 & 2.12 & 10.62 \\ \addlinespace[4pt]
Blend & 95/100 & \centerentry{\shortstack{0.950\\{[0.900, 0.990]}}} & 0.00477 & 5.40 & 11.86 \\ \addlinespace[4pt]
\end{longtable}\endgroup

\begingroup\fontsize{7}{8.5}\selectfont
\setlength{\tabcolsep}{2pt}
\setlength{\LTleft}{0pt}\setlength{\LTright}{0pt plus 1fill}
\setlength{\LTcapwidth}{\textwidth}

\begin{longtable}{@{}L{17mm}C{7mm}C{18mm}C{27mm}C{18mm}C{33mm}@{}}
\caption{Paired table-control contrasts on the confirmatory targets. Entries are the first arm minus the second, averaged over seeds within target; intervals are 95\% percentile target-bootstrap intervals (20,000 resamples within system, or within subgroup), unadjusted for multiplicity. Full-catalogue and held-out learned arms use seeds 0--2, and deterministic arms count once. These post-plan controls are not declared confirmatory endpoints. All searches have at most 30 simulator calls per target and zero language-model calls; the oracle's extra construction calls are stated in Table~\ref{tab:table-controls}. The four held-out endpoints (success rate and design error, all and blend) detect no difference. The subgroup of 22 held-out-reference targets leaves smaller benefits unresolved. One random subset of 14 architectures was tested; the planned 28-architecture level was not run because the earlier stages exceeded the plan's time limit. The coarse oracle table still misses seven targets. No intermediate degradation was tested.}\label{tab:table-control-contrasts}\\
\toprule
Targets & \(n\) & Success difference & 95\% interval & Error difference & 95\% interval \\ \midrule\endfirsthead
\multicolumn{6}{l}{Table \thetable\ (continued)}\\\toprule
Targets & \(n\) & Success difference & 95\% interval & Error difference & 95\% interval \\ \midrule\endhead
\midrule\multicolumn{6}{r}{Continued on next page}\\\endfoot
\bottomrule\endlastfoot
\multicolumn{6}{@{}p{\textwidth}@{}}{\textit{Degraded minus catalogue tables}} \\*
All & 200 & \(-0.475\) & \([-0.545, -0.405]\) & \(+0.01386\) & \([0.01132, 0.01654]\) \\
Blend & 100 & \(-0.490\) & \([-0.590, -0.380]\) & \(+0.01520\) & \([0.01192, 0.01870]\) \\
\multicolumn{6}{@{}p{\textwidth}@{}}{\textit{Oracle minus catalogue tables}} \\*
All & 200 & \(+0.130\) & \([0.085, 0.180]\) & \(-0.00351\) & \([-0.00487, -0.00238]\) \\
Blend & 100 & \(+0.270\) & \([0.190, 0.360]\) & \(-0.00650\) & \([-0.00915, -0.00434]\) \\
\multicolumn{6}{@{}p{\textwidth}@{}}{\textit{Learned minus catalogue tables, full catalogue, seeds 0--2}} \\*
All & 200 & \(-0.002\) & \([-0.042, 0.038]\) & \(+0.00004\) & \([-0.00095, 0.00094]\) \\
Blend & 100 & \(+0.007\) & \([-0.070, 0.087]\) & \(+0.00002\) & \([-0.00190, 0.00173]\) \\
\multicolumn{6}{@{}p{\textwidth}@{}}{\textit{Learned minus catalogue tables, 14 architectures held out}} \\*
All & 200 & \(+0.008\) & \([-0.028, 0.045]\) & \(-0.00094\) & \([-0.00248, 0.00044]\) \\
Blend & 100 & \(+0.027\) & \([-0.040, 0.097]\) & \(-0.00182\) & \([-0.00483, 0.00095]\) \\
\multicolumn{6}{@{}p{\textwidth}@{}}{\textit{Catalogue tables, held out minus full catalogue}} \\*
All & 200 & \(-0.005\) & \([-0.030, 0.015]\) & \(+0.00048\) & \([-0.00007, 0.00128]\) \\
Blend & 100 & \(-0.010\) & \([-0.050, 0.030]\) & \(+0.00092\) & \([-0.00017, 0.00253]\) \\
\multicolumn{6}{@{}p{\textwidth}@{}}{\textit{Learned tables, held out minus full catalogue}} \\*
All & 200 & \(+0.005\) & \([-0.020, 0.028]\) & \(-0.00051\) & \([-0.00149, 0.00026]\) \\
Blend & 100 & \(+0.010\) & \([-0.037, 0.057]\) & \(-0.00092\) & \([-0.00292, 0.00061]\) \\
\multicolumn{6}{@{}p{\textwidth}@{}}{\textit{Learned minus catalogue tables, held-out blend subgroups}} \\*
Held-out reference & 22 & \(+0.061\) & \([-0.030, 0.182]\) & \(-0.00504\) & \([-0.01320, 0.00017]\) \\
Other blends & 78 & \(+0.017\) & \([-0.064, 0.098]\) & \(-0.00091\) & \([-0.00413, 0.00212]\) \\
\end{longtable}\endgroup

\end{document}